\documentstyle[twoside,art12]{article}
\font\myaddressfont=cmcsc10
\input amssym.def
\input amssym.tex
 \newif\ifMarginNotes \MarginNotestrue
\def\mrgn#1{\ifMarginNotes\setbox0=\vtop{\hsize 6.75pc
   {\noindent\relax #1\par}}\leavevmode
   \vadjust{\dimen0=\dp0 \dimen1=\ht0\advance\dimen1 by .5ex
 \advance\dimen0 by -.5ex
  \kern-\dimen1\hbox{\kern\hsize\kern.5pc$\leftarrow$
  \box0}\kern-\dimen0}\fi}
\catcode`\@=11
\font\twelvemsb=msbm10 scaled 1200
\font\tenmsb=msbm10
\font\ninemsb=msbm7 scaled 1200
\newfam\msbfam
\textfont\msbfam=\twelvemsb  \scriptfont\msbfam=\tenmsb
  \scriptscriptfont\msbfam=\ninemsb
\def\msb@{\hexnumber@\msbfam}
\def\Bbb{\relax\ifmmode\let\next\Bbb@\else
 \def\next{\errmessage{Use \string\Bbb\space only in math
mode}}\fi\next}
\def\Bbb@#1{{\Bbb@@{#1}}}
\def\Bbb@@#1{\fam\msbfam#1}

\font\twelveeufm=eufm10 scaled 1200
\font\teneufm=eufm10

\font\seveneufm=eufm7

\newfam\eufmfam
\textfont\eufmfam=\twelveeufm
\scriptfont\eufmfam=\teneufm
\scriptscriptfont\eufmfam=\seveneufm
\def\frak{\relax\ifmmode\let\next\frak@\else
 \def\next{\errmessage{Use \string\frak\space only in math
mode}}\fi\next}
\def\frak@#1{{\frak@@{#1}}}
\def\frak@@#1{\fam\eufmfam#1}

\catcode`\@=12

\def\ts{\tilde s}
\def\ritem#1{\item[{\rm #1}]}
\def\eglo{\eta_{\rm global}}
\def\eloc{\eta_{\rm local}}
\def\hpo{{\cal H}_{p_1}}
\def\hx{{\hat X}}
\def\hpt{{\cal H}_{p_2}}
\def\tm{{\tilde M}}
\def\Ga{\Gamma}
\def\T{{\cal T}}
\def\bbt{{\Bbb T}}
\def\dq{\dot q}
\def\fz{{\frak Z}}
\def\La{\Lambda}
\def\ext{{\rm ext}}

\def\mbni#1{\medbreak\noindent{\bf #1}}
\def\vcv{\vert\!\vert\!\vert\cdot\vert\!\vert\!\vert}
\def\vv#1{\vert\!\vert\!\vert #1 \vert\!\vert\!\vert}
\def\Vrt#1{\Vert #1\Vert}
\def\la{\lambda}
\def\part{\partial}
\def\cs{{\cal S}}
\def\Mat{{\rm Mat}}
\def\tr{{\rm tr}\,}
\def\Tr{{\rm Tr}\,}
\def\llangle{\left\langle}
\def\jlo{{\rm JLO}}
\def\rrangle{\right\rangle}
\def\lra#1{\llangle #1\rrangle}
\def\A{{\frak A}}
\def\F{{\frak F}}
\def\fg{{\frak G}}
\def\jba{{\frak J_{\beta,\alpha}}}
\def\jd{{\frak J_\delta}}
\def\E{{\cal E}}
\def\J{{\cal J}}
\def\fj{{\frak J}}
\def\hensp#1{\enspace \hbox{#1}\enspace}
\def\N{{\cal N}}
\def\C{{\cal C}}
\def\D{{\cal D}}
\def\R{{\Bbb R}}
\def\B{{\cal B}}
\def\CH{{\cal H}}
\def\ep{\epsilon}
\def\el{\epsilon,\lambda}
\def\bh{{\cal B}({\cal H})}
\def\Z{{\Bbb Z}}
\def\clips{,\ldots}
\def\be{\begin{equation}}
\def\ee{\end{equation}}
\def\beq{\begin{eqnarray}}
\def\nn{\nonumber}
\def\eeq{\end{eqnarray}}
\def\l{\left}
\def\r{\right}
\def\ctbaj{\T(-\beta_j, \alpha_j)}
\def\etatot{{\eta_{\rm tot}}}
\def\etamin{\eloc}

\title{Quantum Harmonic Analysis\\
and Geometric Invariants\thanks{Supported
in part by the Department of Energy under Grant DE-FG02-94ER-25228
and by the National Science Foundation under Grant DMS-94-24344.}}
\author{Arthur Jaffe\\
Harvard University}
\date{September 1997}

\begin{document}
\hsize=7truein
\maketitle
\thispagestyle{empty}
\begin{abstract}
We develop two topics in parallel and show their inter-relation.  The first
centers on the notion of a fractional-differentiable structure on a
commutative or a non-commutative space.  
We call this study {\it quantum harmonic analysis}.
The second concerns homotopy invariants for these spaces and is 
an aspect of non-commutative geometry.

We study an algebra $\A$, which will be a Banach algebra with unit, 
represented as an algebra of operators on a Hilbert space $\CH$.
In order to obtain a geometric interpretation of $\A$, we define
a derivative on elements of $\A$. We do this 
in a Hilbert space context, taking $da$ as a
commutator $da = [Q,a]$.  Here $Q$ is a basic self-adjoint operator with
discrete spectrum, increasing sufficiently rapidly
that ${\rm exp}{(-\beta Q^2)}$ has a trace whenever $\beta>0$.  

We can define fractional differentiability 
of order $\mu$, with $0<\mu\le 1$, by the boundedness of 
$(Q^2 +I)^{\mu/2} a(Q^2+I)^{-\mu/2}$.
Alternatively we can require the boundedness of an appropriate smoothing 
(Bessel potential) of $da$.  We find that it is convenient to 
assume the boundedness of $(Q^2+I)^{-\beta/2} da (Q^2+I)^{-\alpha/2}$, 
where we choose $\alpha,\beta \ge 0$ such that $\alpha+\beta<1$.  
We show that this also ensures a fractional derivative of order $\mu=1-\beta$
in the first sense. 
We define a family of interpolation spaces $\fj_{\beta,\alpha}$.  
Each such space is a Banach algebra of operator, whose 
elements have a fractional derivative of order $\mu=1-\beta>0$.  

We concentrate on subalgebras $\A$ of $\fj_{\beta,\alpha}$ 
which have certain additional 
covariance properties under a group $\Z_2\times {\fg }$ acting on $\CH$
by a unitary representation $\gamma\times U(g)$.  
In addition, the derivative $Q$ is assumed to be ${\fg }$-invariant.
The geometric interpretation 
flows from the assumption 
that elements of $\A$ possess an arbitrarily small fractional 
derivative. 
We study homotopy invariants of $\A$ in 
terms of equivariant, entire cyclic cohomology. In fact, the existence of
a fractional derivative on $\A$ allows the construction of the  cochain
$\tau^\jlo$, which plays the role of the integral of differential forms.   We 
give a simple expression for a homotopy invariant $\fz^Q(a;g)$, determined 
by pairing $\tau^\jlo$, with a 
${\fg }$-invariant element $a \in\A$, such that $a$ is a square root 
of the identity.  This invariant is 
$\fz^Q(a;g)
={1\over \sqrt{\pi}} \int^\infty_{-\infty}e^{-t^2}\Tr\l(\gamma
U(g)ae^{-Q^2 +itda}\r)dt$. 

This representation of the pairing is 
reminiscent of the heat-kernel representation for an index.
In fact this quantity is an invariant, in the following sense. 
We isolate a simple condition on a family $Q(\la)$ of differentiations
that yields a continuously-differentiable family $\tau^\jlo(\la)$ of
cochains.   Since
$\fz^Q(a;g)$ need not be an integer, continuity of 
$\tau^\jlo(\la)$ in $\la$ is insufficient to prove the constancy of the 
pairing.  However the
existence of the derivative leads to the existence of the homotopy.  As
$\lra{\tau,a}$ vanishes for $\tau$ a coboundary, and as 
$d\tau^\jlo(\la)/d\la$ is a coboundary, our condition on
$Q(\la)$ ensures that
$\fz^{Q(\la)}(a;g)$ is independent of $\la$.  Hence it is 
a homotopy invariant.   

The theory of $\fz^Q(a;g)$ reduces to the study of the Radon
transform of sequences of certain functions.  The fractional differentiability
properties of elements of $\A$ translate into properties of the
asymptotics of the sequences of Radon transforms.    The condition that
$\tau^\jlo$ fit into the framework of entire cyclic cohomology translates
to the existence of some fractional derivative for functions in
the algebra under study, and in particular the assumption $\alpha+\beta<1$.  
Thus the study of fractionally-differentiable
structures dove-tails naturally with the theory of homotopy invariants.

In our study of quantum harmonic analysis, we
introduce spaces $\T(-\beta,\alpha)$ of operator-valued 
distributions.  These spaces are bounded, linear operators between
Sobolev spaces.  The elements of the interpolation
spaces, the Banach algebras $\fj_{\beta,\alpha}$, have derivatives 
$da$ which belong to the spaces $\T(-\beta,\alpha)$. 
For a certain range of $\beta$ and $\alpha$, we
extend the theory of the Radon transform from products
of regularized, bounded operators to products of regularized, 
operator-valued distributions.  

We sometimes wish to evaluate such an invariant at the endpoint of an
interval such as $\lambda \in (0,1]$, where $\fz^{Q(\lambda)}(a;g)$
becomes singular as $\lambda \rightarrow 0$.  We discuss in brief 
a procedure to regularize the endpoint, and a method to recover
$\fz^{Q(\lambda)}(a;g)$ fully from certain
partial information at the endpoint.

Finally, we generalize this approach to cover the case when $Q$ can be
split into the sum of ``independent'' parts $Q_1+Q_2$, such that 
$(Q_1+Q_2)^2 = (Q_1)^2+ (Q_2 )^2$.
Here we assume that $Q_1$ and $(Q_2 )^2$ are ${\fg }$-invariant, but 
not necessarily $Q_2$.  With further assumptions on $a$, the most important 
being that
$(Q_1)^2-(Q_2)^2$ commutes with $a$, we obtain a modified formula for an
invariant, namely
$\fz^{\{Q_j\}}(a;g)
={1\over \sqrt{\pi}} \int^\infty_{-\infty}e^{-t^2}
\Tr\l(\gamma U(g)ae^{-(Q_1^2 + Q_2^2)/2 +itd_1a} \r)dt$.

\end{abstract}

\newpage
\tableofcontents

\textwidth=7in\textheight=8.3in 
\textwidth=7in\textheight=8in 
\newpage
\section{Introduction}
We study  algebras $\A$ of bounded linear transformation on a Hilbert
space $\CH$.  Each  algebra describes the geometry of a  space, and we
study homotopy invariants of this space.  All this fits into Alain Connes'
formulation of non-commutative geometry \cite{C1}, as well as extensions 
of that theory studied by others.  
We emphasize here  the analytic aspects of this
theory, and in doing so we develop the relation between the details of the
analysis and the existence of invariants.

The algebra $\A$ will be a Banach algebra with a unit $I$, and 
with  a norm $\vcv$ that dominates the operator norm $\Vrt{\cdot}$ on $\bh$,
the algebra of all 
bounded linear transformations on $\CH$.  Thus for $a\in \A$,
\be
\Vrt{a}\le \vv{a}\;,
\ee
and the injection $\A\to \bh$ is continuous.  

In order to describe $\A$ and the norm $\vcv$, we need to define a
derivative on $\A$.  We do this in terms of a given, self-adjoint, linear
transformation 
$Q=Q^\ast$  on
$\CH$ with domain $\D$.  Assume that  the spectrum of $Q$ is discrete,
and that the eigenvalues increase sufficiently rapidly that $\exp(-sQ^2)$
has a trace for every $s>0$.    Connes calls the  condition  
$\Theta$-{\it summability} \cite{C2}.  The derivative
$da$ of an element of
$\A$ is defined by a commutator
\be
da = [Q,a]=Qa-aQ\;.
\ee
In general $da$ is not an element of $\bh$, but it is always given by a
(possibly unbounded) sesquilinear form on $\CH\times \CH$ with domain
$\D\times \D$.

We define Sobolev spaces  in \S V as the domain of fractional
powers of $Q$.  They are realized as a Gelfand family of Hilbert spaces, see
\cite{GV}.  Define $R=(Q^2 +I)^{-1/2}$ to be a smoothing operator of degree
$-1$.  The space $\CH_\mu$ is the domain of  the operator $R^{-\mu}$ of
degree
$\mu$.  One definition of order-$\mu$, bounded-differentiability of 
$a\in \bh$ is that
\be 
R^{-\mu}a R^\mu\in \bh\;.
\ee
In other words, (I.3) is the assertion that $a$ is a bounded, linear
transformation on $\CH_\mu$.  

Alternatively, because of the fundamental
nature of the differential $da$, we wish to pose differentiability
properties of $a$ in terms of properties of $da$.  The differentiable case 
for $a$ becomes the assumption that $da$ is a densely defined, 
bounded sesquilinear form on $\CH$, so $da$ uniquely determines a 
bounded, linear transformation in $\bh$.  For short, $da \in \bh$. 
Equivalently, $a$ is a bounded, linear transformation on $\CH_1$.
This assumption  that  $da \in \bh$ is made in all earlier work.  This 
ranges from the differentiable case studied in \cite{CM1}, to the smooth case
in \cite{JLO1}, where one assumes that for all $n$, $d^n\A\in\bh$.
(I am thankful to A. Connes for 
informing me that he and Moscovici have also considered 
certain algebras of pseudo-differential operartors in \cite{CM2}.)

Here we do not assume that $da$ is necessarily an element of $\bh$. Rather, 
we require that $da$ defines a bounded, linear transformation between
certain Sobolev spaces $\CH_p$. In \S V  we formulate the assumption
that $da$ belongs to an interpolation space of
fractionally-differentiable operators. Namely we assume that $da$ is
a bounded map
\be
da:\CH_\alpha \to \CH_{-\beta}
\ee
where $0\le \alpha,\beta$ and 
\be
\alpha+\beta<1\;.
\ee
We also write (I.4) as  $da \in \T(-\beta,\alpha)$,  where 
$\T(-\beta,\alpha)$ denotes the space of bounded, linear
maps from $\CH_\alpha$ to $\CH_{-\beta}$. 
The condition (I.5) is crucial to the resulting analysis.
We show that (I.4--5) 
ensures that each $a$ we consider has a fractional derivative of order
$\mu=1-\beta>0$.  Such an element $da$ is an operator-valued, generalized
function.  The differentiable case is $\beta=0$.

In \S V we define Banach algebras $\jba$ of elements in $\bh$ which
have the property (I.4) with $0\le \alpha,\beta$ and (I.5).  We call the
$\jba$ interpolation spaces and assume that
$$
\A\subset \jba\;,
$$
including the requirement that
$$
\Vrt{a}\le \Vrt{a}_{\jba} \le \vv{a}\;.
$$
We show in \S V.5 that if $a,b\in \jba$, then both $(da)b$ and
$a^\gamma(db)$ are bounded maps from $\CH_\alpha\to \CH_{-\beta}$.  We
also show that the graded Leibniz rule
\be
d(ab) = (da)b +a^\gamma(db)
\ee
extends from the differentiable case, to the case of interpolation spaces
$\jba$.  (Here $a^\gamma=a$ if also $a\in \A$, where (I.6) reduces to the
ordinary Leibniz rule.) These considerations lead us
to the notion of a fractionally-differentiable structure $\{\CH,Q,\gamma,
U(g),\A\}$ on
$\A$, see
\S VI. 

In addition to the differential acting from $\A$ to $\T(-\beta,\alpha)$,
we also assume that two groups act as automorphism groups of $\A$.  Each
group will act on $\CH$ as a 
strongly-continuous, unitary representation. We assume that 
this conjugation by this unitary defines a *-automorphism of $\A$.
The first group is $\Z_2$, which will be 
represented by a self-adjoint, unitary operator
$\gamma=\gamma^\ast = \gamma^{-1}$ in $\bh$.  The 
action on $\A$ is given by,
\be
a\to a^\gamma = \gamma a\gamma^{-1} = \gamma a \gamma\;,
\ee
and we assume that $\A$ is pointwise invariant: $a=a^\gamma$ for all 
$a\in \A$.  Furthermore we 
assume
that
\be
\gamma Q\gamma = Q^\gamma = -Q\;,
\ee
so
\be
(da)^\gamma = -d(a^\gamma)\;.
\ee
This structure is familiar from ordinary geometry.\footnote{In the case that
$\CH$ is the $L^2$-Hilbert
space of differential forms on a smooth, compact manifold ${\cal M}$, the 
$\Z_2$-grading
$\gamma$ may be taken to equal $(-1)^n$ on  the subspace of forms of
degree $n$.  One can take the elements of
$\A$ to be smooth forms of degree zero (functions on ${\cal M}$). In that case,
with $d_{{\ext}}$ the exterior derivative, 
a possible example of $Q$ is $Q= d^\ast_\ext + d_\ext$, see \S V.1. }
 
The second group ${\fg }$ is represented by the continuous unitary
representation $g\to U(g)$ on $\CH$ and by the automorphism
\be
a\to a^g = U(g)aU(g)^\ast 
\ee
of $\A$.  
We define $\A^{\fg } \subset \A$  as the subalgebra of $\A$ which is 
pointwise invariant
under the ${\fg }$, namely the subalgebra of $\A$ which commutes
with the representation $U(g)$,
\be
a^g=a \qquad {\rm for} \qquad a\in\A^{\fg }\;.
\ee

The group $U(g)$ is also assumed to commute with $\gamma$ and with
$Q$.
\be
\gamma^g = U(g)\gamma U(g)^\ast=\gamma\;,\quad Q^g = U(g)QU(g)^\ast
=Q\;.
\ee
The group ${\fg }$ may equal the identity. However, if ${\fg }$ is not trivial, then
it gives rise to an equivariant theory.

Before embarking on the analysis of interpolation spaces, we consider the
theory of continuous, multi-linear functionals\footnote{Such functionals often
arise in a purely mathematical setting: 
in analysis, in probability theory, or in geometry.  Furthermore, such 
functionals arise in mathematical physics:  in statistical physics, 
in quantum theory, in quantum field theory, 
and in string theory. Within each of these fields, it may be true that 
one can represent a particular functional as a well-defined integral over a 
function space.  When this is the case, the functional integral representation
provides a powerful tool in order to prove mathematical properties 
of the functional in question, as well as a tool to estimate the functional, 
or possibly to evaluate it in closed form. In particular, constructive 
quantum field theory provides many examples of this phenomenon.}
on $\A$.  Once $\A$ has a topology given by
the norm $\vcv$, we can define spaces of continuous, multilinear
functionals on $\A$.  These are the spaces of cochains, introduced in \S II. 
In particular we focus on the space of entire chains
\be
\C = \oplus^\infty_{n=0}\C_n
\ee
where an element $f_n\in \C_n$ is an $(n+1)$-linear functional on $\A$
which is also a functional on ${\fg }$.  We use a complex of cochains for which
$f_n(a_0\clips,a_n;g)=0$ whenever any $a_j=I$ for $j=1,2\clips,n$ (but not
$j=0$).  

The {\it entire} condition 
on the functionals in $\C$ pertains to uniform asymptotics on the norm
of $f_n$ as $n\to\infty$. 
The sequence
$f=\{f_n\}$ must satisfy
\be
n^{1/2}\vv{f_n}^{1/n}\to 0\;.
\ee
This condition will appear in a natural way in the construction that follows;
the entire condition will be just the asymptotics needed to establish 
convergence of various power series and the existence of generating functionals
that we define on elements of $\C$. 
This condition was introduced in \cite{C2},
in order to prove the existence of a ``normalization'' operator.  Because we
use a different complex of cochains, we do not require normalization, but
the entire condition remains a natural analytic assumption, useful for other
reasons.

In \S II we give some continuous, linear operators on the spaces of cochains
including the three fundamental coboundary operators $b,B$, and $\part =
b+B$.  The operator $b$ is the Hochschild coboundary operator, $B$ is the
Connes coboundary operator, and $\part=b+B$ is the coboundary operator of
entire cyclic cohomology satisfying
\be
\part : \C\to \C\;,\quad \part^2=0\;.
\ee

In \S III we study a natural pairing $\lra{\tau,a}$ between cochains
$\tau\in \C$ and elements  $a\in\A^{\fg }$, which satisfy $a^2=I$.
More generally, we may take $a^2=I$ where  
$a\in\Mat_m(\A^{\fg })$.  Here $\Mat_m(\A^{\fg })$ denotes 
the space of $m\times m$ matrices with matrix entries
$a_{ij}\in \A^{\fg }$.  We show 
in \S III that the requirements
\be
\lra{\part G,a}=0
\ee
 for all $G\in \C$ and for all $a\in\A^{\fg }$ for which $a^2=I$,
determines the respresentation for a possible pairing.  It is 
\be
\lra{\tau,a} = {1\over \sqrt{\pi}} \int^\infty_{-\infty}  e^{-t^2}
\J(t;a;g) dt\;,
\ee
where $\J(t;a;g)$ is a generating functional for the components of $\tau$,
defined as the alternating sum
\be
\J(t;a;g)=\sum^\infty_{n=0}(-t^2)^{n}\tr \tau_{2n} (a,a\clips,a;g)\;,
\ee
see Corollary III.5.

In \S IV we introduce a particular cochain, the JLO-cochain 
$$
\tau^\jlo=\{\tau^\jlo_n\}\in\C\;.
$$
The $n^{\rm th}$
component $\tau^\jlo_n$ of $\tau^\jlo$ is an $(n+1)$-multilinear functional
on
$\A$,
\be
\tau^\jlo_n (a_0\clips,a_n;g)=\lra{a_0,da_1\clips,da_n;g}\;.
\ee
Here the expectation $\lra{\enspace}$ is defined in \S IV and V in terms
of the Radon transform of the heat-kernel regularized operator
\be
X^\jlo (s) = a_0e^{-s_0Q^2}da_1 e^{-s_0Q^2} \cdots da_n e^{-s_nQ^2}\;,
\ee
restricted to the sector $s_j>0$, for $0\le j\le n$.
In this sector, with each  $s_j>0$, we show that $X^\jlo(s)$ is trace class.
In \S V--VI, we show that 
the definition of $\tau^\jlo$ extends to the
algebras $\A\subset \jba$ that we introduce here.  
The trace norm $||X^\jlo(s)||_1$ of $X^\jlo(s)$ may diverge
on the hyperplane $s_0+s_1+\cdots+s_n=1$ as any $s_j\to 0$. We will estimate
the rate of divergence in terms of $\alpha+\beta$, using the bound on the 
norm $||da||_{\T(-\beta,\alpha)}$ of $da$ in (I.4). 

We establish the existence and the properties of the Radon transform,
$\tau^\jlo_n = \Tr (\gamma U(g) (\int X^\jlo(s)ds) )$,
where the integral is taken over the hyperplane $s_0+s_1+\cdots+s_n=1$.
In this step, we use the assumption $\alpha+\beta<1$.  
Our estimates also allow us to justify  the interchange of  the integration 
and the trace in the expression defining $\tau^\jlo _n$,
$$
\int\l(\Tr\l(\gamma U(g)X^\jlo(s)\r)\r)ds 
= \Tr \l(\gamma U(g)\l(\int X^\jlo(s)ds\r)\r)\;.
$$
Thus we simply write
\be
\tau_n^\jlo(a_0,\cdots, a_n;g) =
\int_{s_j>0} {\rm Tr} \l(\gamma U(\theta) a_0e^{-s_0Q^2}
da_1e^{-s_1Q^2} \cdots da_ne^{-s_nQ^2}\r) \delta(s_0 + \cdots +s_n -1)
ds_0 ds_1 \cdots ds_n\;.
\ee

In Proposition VI.2 we prove that
\be
n^{1/2} \vv{\tau^\jlo_n}^{1/n} \le O\l(n^{-(1-\alpha-\beta)/2}\r)\;,
\ee
which yields the required asymptotics (I.14).  The behavior of (I.22) for 
large $n$ is dependent on the analysis in \S V--VI 
of the differentiability properties of elements of $\A$.
The importance of the order-$\mu$, fractional-differentiability of 
elements of $\A$ emerges once more.  We require that the order of 
fractional derivative 
$\mu=1-\beta$ is greater than $0$, which is part of the assumption (I.5).  
Our methods break 
down just at the point when elements of $\A$ have no fractional derivative,
or more precisely when $\alpha+\beta=1$.

Parenthetically, we remark that in 
\S V we introduce sets $\{x_0, x_1, \cdots, x_n\}$ of $n$, operator-valued
generalized functions, which we call sets of {\it vertices}.  
In our study of quantum harmonic analysis, we define expectations of 
such sets of operators, as a multilinear functional.
Suppose there are $\alpha_j, \beta_j \ge 0$, with 
$\beta_{n+1} = \beta_0$, and such that for $0 \le j\le n$, 
\be
x_j \in \ctbaj\;\quad {\rm with \ } \alpha_j + \beta_{j+1} < 2\;.
\ee
Then we call the set $\{x_0, x_1, \cdots, x_n\}$ 
a {\it regular set } of vertices.  
The conditions (I.23) require that every $\alpha_j,\beta_j<2$;
however certain configurations of $\alpha_j, \beta_j$ may allow a 
particular vertex $x_j$ in a regular set to have $\alpha_j + \beta_j$ 
close to 4.

We define the heat kernel regularization
$X(s)$ of a 
regular set $\{x_0, x_1, \cdots, x_n\}$ of vertices as a trace
class operator. For 
parameters $0 < s_j $, let 
\be  
X(s) = (I+Q^2)^{-\beta_0/2}x_0e^{-s_0Q^2} x_1 e^{-s_1Q^2}\cdots
x_ne^{-s_nQ^2}(I+Q^2)^{\beta_0/2}\;.
\ee
While this operator $X(s)$ is trace class, 
the trace norm on $\CH$ may diverge as $s_j\to 0$.  The operator $X(s)$
has a trace class Radon transform on the
hyperplane $s_0 +\cdots + s_n = 1$, and the operations of taking the 
Radon transform and the trace commute.  We write
\be
\langle x_0,x_1,\cdots, x_n;g \rangle_n = \int_{s_j>0} {\rm Tr}
\l(\gamma U(\theta)x_0e^{-s_0Q^2}x_1e^{-s_1Q^2}\cdots x_ne^{-s_nQ^2}\r)
\delta(1- \sum_{j=0}^n s_j)ds_0\cdots ds_n\;.
\ee
This functional extends by continuity from its definition 
on the space of bounded vertices $x_j\in\B(\CH)$ to
a multi-linear functional on the space of vertices 
$x_j\in\ctbaj$ in a regular set.  Furthermore 
we bound this expectation (I.25) in Corollary V.4 by 
\be
\l|\lra{x_0,x_1\clips,x_n;g}_n\r| \le \frac{m(\etamin)^{n+1}}
{\Ga (\etatot )} {\rm Tr}\l(e^{-Q^2/2}\r)
\l(\prod^n_{j=0} \Vrt{x_j}_{(-\beta_j,\alpha_j)}\r)\;,
\ee
where  $m(\etamin) < \infty$ is a constant. Here the 
exponents $\eta$ that characterize the behavior of the expectations are
defined by
$$
\eta_j = \frac12(2 - \alpha_j - \beta_{j+1})\;, \qquad
\etamin = {\rm min}_{0\le j\le n}\{\eta_j\}\;, \qquad {\rm and} \qquad
\etatot = \sum_{j=0}^n \eta_j\;.
$$
We say that $\etamin$ characterizes the local regularity of the 
expectations (I.25), while $\eglo={\etatot}/{(n+1)}$ 
characterizes the global regularity of sets of such expectations as
a function of $n$.

In \S VI we also return to the fact that the 
functional $\tau^\jlo$ is a cocycle in $\C$, namely
\be
\part \tau^\jlo=0\;.
\ee
This was previously known in the differentiable case, for which
$\alpha=\beta=0$.
In \S VI we also analyze the generating functional corresponding 
to $\tau^\jlo$, namely 
\be
\J^\jlo(t;a;g) = \Tr 
\l( \gamma U(g) ae^{-Q^2 +itda}\r)\;.
\ee
Define the functional $\fz(a;g)$ as the Gaussian transform
of $\J(t;a;g)$ evaluated at the origin,
\be
\fz(a;g)= {1\over \sqrt\pi} \int^\infty_{-\infty}e^{-t^2} \J(t;a;g)dt\;.
\ee
In the case that $\J(t;a;g)=\J^\jlo(t;a;g)$, we indicate the 
dependence of $\fz$ on $Q$. We have 
\be
\fz^{Q}(a;g)
={1\over \sqrt{\pi}} \int^\infty_{-\infty}e^{-t^2}\Tr\l(\gamma
U(g)ae^{-Q^2 +itda}\r)dt\;.
\ee
We prove in \S V--VI that the functionals (I.28--30) exist for all $a \in \A$,
when $\A$ is contained in one of the allowed interpolation space $\jba$. 

Thus we have  a representation for the pairing 
(I.17) in an extremely simple and elegant form. 
Let $\A^{\fg }$ denote the subset of $\A$ that is pointwise 
invariant under ${\fg }$.
For $a\in\A^{\fg }$, or more generally for $a\in {\rm Mat}_m(\A^{\fg })$, and  
also $a^2=I$, then  (I.17) equals  
\be
\lra{\tau^\jlo,a}=\fz^{Q}(a;g)
={1\over \sqrt{\pi}} \int^\infty_{-\infty}e^{-t^2}\Tr\l(\gamma
U(g)ae^{-Q^2 +itda}\r)dt\;.
\ee
In other words, $\fz^Q(a;g)$ can be written as a trace of a heat
kernel.\footnote{In the language of physics, $\fz^Q(a;g)$ is called a 
{\it partition function}.  Such an object often arises 
in statistical physics
or in quantum theory. The Laplace operator which generates the heat kernel
is $Q^2$, perturbed
by $itda$. Note that if $a=a^*$, then the perturbation is symmetric. In physics, 
the perturbation $da$ is sometimes called a {\it local source}.
We mentioned earlier that it might be the case that a functional 
${\rm Tr}(\gamma U(g)\cdot e^{-Q^2})$ could be represented as a 
functional integral, given by a measure $d\mu_g$.  If this is the case, and
if in addition both $a$ and $da$ can be realized as functions on path space,
then the representation (I.31) further simplifies. The Gaussian integral
can be carried out, giving $\fz^Q(a,g)=\int a{e^{-(da)^2/4}d\mu_g}$. 
}
For the case $a=I$, we have $da=0$, and  (I.31) reduces  
to the equivariant index
\be
\fz^{Q}(I;g)=\Tr \l(\gamma U(g)e^{-Q^2}\r)\;.
\ee
The expression for $\fz^Q(a;g)$ in (I.31) is fundamental, and it 
provides a generalization of the McKean-Singer heat-kernel 
representation of the index $\Tr(\gamma e^{-Q^2})$.  In this context, the
cochain $\tau^\jlo$ is the equivariant Chern character for the
fractionally-differentiable structure on $\A$.  

In \S VII we return to the question of the precise sense in which 
(I.31) provides a formula for an invariant. We study the variation of 
$\fz^{Q(\la)}(a;g)$ corresponding to a family of 
cochains $\tau^\jlo(\la)$ depending 
on a parameter $\la$.  This family arises from a family of differentiations
$Q(\la)$ of the form
\be
Q(\la) = Q+q(\la)\;.
\ee
We assume that $q(\la)$ is a bounded map
\be
q(\la):\CH_1\to \CH_0 \;,
\ee
with (an appropriate) norm less than 1.  This condition (I.34) can also be
described by saying that $q(\la)$ is a perturbation of $Q$ in the sense of T.
Kato, see \cite{K2}.  We also assume that there is a bounded map $\dq(\la)$,
continuous in
$\la$, from
$\CH_1$ to
$\CH_0$, such that in the space of bounded, linear maps from $\CH_1$ to
$\CH_0$, the difference quotient is norm convergent to the derivative,
\be 
\lim_{\la'\to\la} \l( {q(\la) - q(\la')\over \la-\la'} - \dq(\la)\r) =0\;.
\ee

These assumptions form the basis of our definition in \S VII of a regular
family $Q(\la)$.  We also define a corresponding regular family of
fractionally-differentiable structures $\{\CH,Q(\la),\gamma, U(g),\A\}$ on
$\A$.  Under these hypotheses we prove that 
\be
\la\mapsto \tau^\jlo(\la)
\ee
is a continuously differentiable function from an interval $\La\subset \R$ to
$\C$.  Here
$\C$ carries the natural topology defined in \S II. 

These simple conditions
(on $Q(\la)$ and its derivative) allow a complete analysis of the trace-class 
nature and differentiability in the appropriate Schatten norm of 
$\la \mapsto {\rm exp}{(-sQ(\la)^2)}$, for $s>0$. In fact our hypotheses 
appear to cover most applications. As a consequence, the derivative
$d\tau^\jlo(\la)/d\la$ can be computed by differentiating the expression
(I.21) under the integral, and under the trace.  
In \S VII we justify this interchange of limits for $a\in
\A\subset
\jba$ and for $q(\la)$ as above.

Calculation of the derivative shows that there is a cochain $h\in
\C$ with coboundary $\part h$ such that
\be
{d\tau^\jlo(\la)\over d\la} = \part h(\la)\;.
\ee
Integrating this relation we obtain
\be
\tau^\jlo(\la) = \tau^\jlo(\la') + \part H(\la,\la')\;,
\ee
where $H(\la,\la')\in \C$.  Since the pairing (I.16) vanishes on
coboundaries, $\lra{\part H,a}=0$. The linearity of the pairing in $\tau$
ensures
\be
\fz^{Q(\la)}(a;g)
= \fz^{Q(\la')}(a;g)\;.
\ee
In other words, $\fz^{Q(\la)}$ does not depend on $\la$, and so
it is a homotopy invariant.  
As a special case of this
result, we show that the definition of $\tau^\jlo_n$, involved choosing a
particular hyperplane 
$s_0+s_1+\cdots +s_n=\beta$ for the Radon transform of (I.19), gives a
pairing independent of
$\beta$.  But more generally, $\fz^Q$ remains unchanged under a regular
deformation of a parameter in a potential, of a metric, etc.

We comment in \S VIII on using the homotopy invariance of
$\fz(a;g)$ in various settings, as a tool to study or to compute
these quantities.  In particular, we study the possibility that the family
$\{\CH,Q(\la),\gamma,U(g), \A\}$ for $\la\in\La$, may have a singularity at
one endpoint of an interval $\La$.  We give a method to study such an
endpoint, by introducing a family $\tau^\jlo(\la,\ep)$ of approximations to
$\tau^\jlo(\la)$.  

In \S IX we generalize this approach to cover the case when $Q$ can be
split into the sum of ``independent'' parts $Q=(Q_1+Q_2)/\sqrt2$, such that 
$(Q_1+Q_2)^2 = (Q_1)^2+ (Q_2 )^2$.
We assume here that $Q_1$ and $(Q_2 )^2$, are ${\fg }$-invariant, but we do not
make that assumption about $Q_2$.
With further assumptions on $a$, the most important being that
$(Q_1)^2-(Q_2)^2$ commutes with $a$, we obtain a modified formula for 
a pairing.  Namely, with $d_1a=[Q,a]$ and $a^2=I$, the quantity
\be
\fz^{\{Q_j\}}(a;g)
={1\over \sqrt{\pi}} \int^\infty_{-\infty}e^{-t^2}
\Tr\l(\gamma U(g)ae^{-Q^2 +itd_1a} \r)dt\
\ee
is invariant under regular deformations of $Q_1(\la)$ and $Q_2(\la)$ which 
leave $Q_1^2-Q_2^2$ fixed.

\section{Cochains and Coboundary Operators}
\setcounter{equation}{0}
Functionals on $\A$ (maps from $\A$ to $\Bbb C$) play a central 
role in non-commutative geometry. 
These functionals may be integrals or traces and are called 
{\it expectations} of operators in $\A$. We introduce three spaces
$\N\subset\C\subset\D$ of multilinear functionals on $\A$ and use them 
below. 

In this paper, the fundamental space is the space
$\C$ of entire cochains.  The other spaces allow us to simplify the
discussion of certain linear transformations of $\C$ into $\C$.  Each of
these spaces are defined on
$\A$ and have an invariance under ${\fg }$, so we denote the spaces by
$\D=\D(\A;{\fg })$, $\C=\C(\A;{\fg })$, etc.  We sometimes suppress either $\A$ or
${\fg }$ in order to simplify notation.

A coboundary operator $\part $ is a continuous, linear transformation from
$\C$ to $\C$, with the property $\part^2=0$.  We study three coboundary
operators in $\C$: the Hochschild operator $b$, the Connes operator $B$,
and their sum $\part = b+B$ which is the coboundary operator in entire
cyclic cohomology.
\subsection{Spaces of Cochains}
\mbni{The Space $\D$.}  

Define $\D_n$ as a vector space of functionals on $\A^{n+1}\times {\fg }$, where
every $f_n\in \D_n$ is an $(n+1)$-continuous, multilinear functional on
$\A$, that is also a continuous function on ${\fg }$. We denote the values of
$f_n$ by
$f_n(a_0\clips,a_n;g)$, where $a_j\in \A$ and $g\in {\fg }$.  We assume that
$f_n$ is invariant on the diagonal, in the sense that
\be
f_n\l(
a^{g^{-1}}_0,a^{g^{-1}}_1\clips,a^{g^{-1}}_n;g\r)=f_n\l(a_0\clips,a_n;g\r)\;.
\ee
The norm $\vv{f_n}$ of $f_n$ is defined with respect to the 
norm $\vcv$ on $\A$ and the $\sup$ norm on ${\fg }$. Thus,
\be
\vv{f_n} 
 = \sup_{\vv{a_j}\le 1\atop g\in {\fg }} \l|f_n
(a_0\clips,a_n; g)\r|\;.  
\ee
Elements of $\D$ are sequences $f=\{f_n:f_n\in \D_n,n\in\Z_+\}$, which
also are assumed to satisfy the {\it entire} condition:
\be
\lim_{n\to \infty} n^{1/2}\vv{f_n}^{1/n} = 0\;.
\ee

When we write an identity $f=g$ in $\D$, this means $f_n(a_0\clips,a_n;g)
= g_n(a_0\clips,a_n;g)$ for each $n\in \Z_+$, for all $a_j\in \A$,
and all $g\in {\fg }$.  Likewise we interpret identities with the ${\fg }$-dependence
suppressed as holding pointwise in ${\fg }$.
\mbni{The Space $\C\subset \D$.}

The key property of the space of entire cochains $\C$ is that elements of
$\C$ vanish when evaluated on $I$ except in the $0^{\rm th}$ place.  More
precisely, the elements
$f\in
\C$ are those elements $f\in \D$ such that for every $n\ge 1$,
$f_n(a_0\clips,a_n;g)=0$ when any $a_j=I$, for any  $j\ge 1$ (but not
$j=0$).  

\mbni{The Space $\N\subset \C$.}

The subspace $\N$ of $\C$ is the annihilator of $I$.  In other words, 
$f\in \D$ is in $\N$ if each $f_n$ $(a_0\clips,a_n;g)$ vanishes whenever 
{\it any} $a_j=I$, for $0\le j\le n$.  Thus $f\in \C$ belongs to $\N$ if 
for every $n\in \Z_+$, $f_n(I,a_1\clips,a_n;g)=0$.

\subsection{Elementary Linear Transformations}
We define a number of bounded, linear transformations, $T:\D\to \D$.  In
other words, all these maps have domain $\D$ and range in $\D$.  We remark
below specifically  which of these operators also map $\C$ into $\C$, map 
$\C$ into $\N$, etc.

Let $\cs$ denote a generic linear transformation $\cs:\D\to \D$.  Since $\D$
is contained in the direct sum $\oplus_n \D_n$ and since $\cs$ is linear, it
is sufficient to define $\cs$ on each $\D_n$, $n\in \Z_+$.  We often denote
by $\cs_n$ the action of $\cs$ on $\D_n$.  In all the examples here, $\cs$ is
tridiagonal: the range of $\cs_n$ lies in $\D_{n-1}\oplus \D_n\oplus
\D_{n+1}$.   

\mbni{Cyclic Transposition.} $T:\N\to \N$.
\be
(Tf_n) (a_0\clips,a_n;g)=(-1)^n f_n\l(a_n^{g^{-1}},a_0\clips,a_{n-1};g\r)\;.
\ee
Note that as a
consequence of the invariance (II.1), it is true that \be
T^{n+1}=I\;.
\ee

\mbni{Cyclic Antisymmetrization.} $A:\N\to \N$.

The cyclic antisymmetrization $A_n$ on $\D_n$ is defined by
$A_n=\sum^n_{j=0} T^j_n$.  Then (II.5) ensures that for any $s\in \Z$, 
\be
A_n = \sum^n_{j=0} T_n^{j+s}\;  .
\ee
\mbni{Annihilation.} $U:\C\to \N$, $U:\N\to 0$.

The annihilation transformation  $U$   maps $\D_n$ into
$\D_{n-1}$.  It is defined by
\be
U_0f_0=0\;,\qquad (U_nf_n)(a_0\clips,a_{n-1})= f_n(I,a_0\clips,a_{n-1})\;.
\ee
As  $U$ acts on the first variable , $U:\C\to \N$ and $U:\N\to 0$.

\mbni{Creation.} $V$

The creation operator maps $\D_n$ to $\D_{n+1}$, according to the rule
\be
(V_nf_n)(a_0\clips,a_{n+1}) = f_n(a_0a_1,a_2\clips,a_{n+1})\;.
\ee
We also introduce $V(r): \D_n\to \D_{n+1}$, defined by conjugating $V$ 
by $T^r$, namely
$$
V(r)=T^{-r}VT^r\;.
$$
Then $V(0)=V$, and acting on $\D_n$,
\be
\l( V(r)_n f_n\r)(a_0\clips,a_{n+1}) = \l\{ \begin{array}{ll}
(-1)^rf_n(a_0\clips,a_ra_{r+1}\clips,a_{n+1})\;,& 0\le r\le n\;,\\
(-1)^{n+1} f_n\l(a^{g^{-1}}_{n+1} a_0,a_1\clips,a_n\r) \;, &
r=n+1.\end{array}\r. 
\ee
These definitions yield the following elementary identities:
\beq
U_{n+1} V_n&=&I_n\;,\\
U_{n+1}V(r)_n +V(r-1)_{n-1} U_n &=& 0\;,\quad 1\le r\le n\;, \\
U_{n+1} V(n+1)_n&=&-T_n\;, 
\eeq
and
\be
V(r)_{n+1}V(s)_n+V(s+1)_{n+1}V(r)_n=0\;,\quad 0\le r\le s\le n+1\;.
\ee
\subsection{Coboundary Operators}
\mbni{The Hochschild Coboundary.} $b:\C\to \C$.

The Hochschild coboundary operator $b$ maps $\D_n$ to $\D_{n+1}$.
It is defined by
\be
b_n = \sum^{n+1}_{r=0} V(r)_n\;.
\ee
Thus
\be
(b_nf_n) (a_0\clips,a_{n+1})  =  \sum^n_{j=0} (-1)^j
f_n(a_0\clips,a_ja_{j+1} \clips,a_{n+1})  
 + (-1)^{n+1} f_n \l(
a^{g^{-1}}_{n+1}a_0,a_1\clips,a_n\r)\;.
\ee
Note that
\beq
b_{n+1} b_n& =& \sum^{n+2}_{r=0} \sum^{n+1}_{s=0} V(r)_{n+1}
V(s)_n\nn\\
&=&
\sum_{0\le r\le s\le n+1} \l(V(r)_{n+1} V(s)_n + V(s+1)_{n+1}
V(r)_n\r)=0\;,
\eeq
where the last equality follows from (II.13).  

As a consequence of (II.16) we have established that $b$ is a {\it
coboundary operator} on $\D$.  In particular, we have proved on the identity
$\D$,
\be
b^2=0\;.
\ee

Finally we verify that $b$ acts on $\C$, namely $b:\C\to \C$, from 
which we conclude that $b$ is also a coboundary operator on $\C$, 
namely $b^2:\C\to 0$.  Assume that $f\in \C$; we need to show that 
$bf\in \C$.  Evaluate $(b_nf_n)(a_0\clips,a_{n+1})$ using (II.15), 
and also assume that $a_k=I$, for some fixed $k$ with $1\le k\le n+1$.  
Then
\be 
(b_nf_n)(a_0\clips,a_{n+1})  
= \l\{ \begin{array}{l}
\l( (-1)^{k-1} + (-1)^k\r) 
f_n\l(a_0\clips,a_{k-1},a_{k+1}\clips,a_{n+1}\r)\;, k\le n\\
\noalign{\vskip4pt}
\l( (-1)^n +(-1)^{n+1}\r) 
f_n\l(a_0\clips,a_{n}\r)\;, k=n+1\; . \\
\noalign{\vskip4pt}
\end{array}\r.
\ee 
In both cases the right hand side of (II.18) vanishes, so $b:\C\to \C$.

\mbni{The Connes Coboundary.} $B:\C\to\N$.

The Connes coboundary operator $B$  is defined on $\C$ by
\be
B=AU\;.
\ee
In particular, $B_n=A_{n-1}U_n$, and on $\C_n$ this can be written as 
\be
(B_nf_n)(a_0\clips,a_{n-1}) = \sum^{n-1}_{j=0} (-1)^{(n-1)j} f_n \l(
I,a^{g^{-1}}_{n-j} \clips,a^{g^{-1}}_{n-1}, a_0\clips,a_{n-j-1}\r)\;.
\ee
Since $U:\C\to \N$ and $A:\N\to \N$, it follows that $B:\C\to\N$.  But
$U:\N\to 0$.  Thus we have shown that $B$ is a coboundary operator on
$\C$, namely\footnote{Note that on the space $\D$, the Connes operator has
 an additional term, namely
$B=AU(I-T^{-1})$.  The term $UT^{-1}$ vanishes on $\C$.  It can be checked
that $(AU(I-T^{-1}))^2=0$ on $\D$, see for example \cite{C1}.}
\be
B^2:\C \to 0\;.
\ee

Finally we verify
that on $\D$, the two coboundary operators satisfy
\be
Bb+bB=0\;.
\ee
In fact
\beq
(Bb+bB)_n &=& B_{n+1}b_n +b_{n-1} B_n  \nn\\
&=& A_nU_{n+1}\sum^{n+1}_{r=0}V(r)_n+\sum^n_{r=0}
V(r)_{n-1}A_{n-1}U_n\; . 
\eeq
Using (II.10,12), the $r=0$ and $r=n+1$ terms in the first sum in (II.23) are
equal to $A_n(I_n-T_n)$.  By (II.5--6), this is zero.  For the remaining
terms in (II.23), we use (II.11) to obtain
\be
(Bb+bB)_n = -A_n \sum^{n-1}_{r=0} V(r)_{n-1}U_n
+\sum^n_{r=0}V(r)_{n-1}A_{n-1}U_n\;.
\ee
Again by (II.5--6),
\beq
\sum^{n-1}_{r=0} A_nV(r)_{n-1}U_n&=& \sum^n_{j=0}\sum^{n-1}_{r=0}
T^j_n V_{n-1}T^r_{n-1} U_n = A_n V_{n-1}A_{n-1} U_n\nn\\
&=& \sum^n_{r=0} V(r)_{n-1}A_{n-1}U_n\;.
\eeq
Hence we have shown that (II.24) vanishes and (II.22) holds.

\mbni{The Entire Coboundary.} $\partial:\C\to \C$.

The entire coboundary operator is the sum of $b$ and $B$.  Define
\be
\part =b+B\;.
\ee
Both $b$ and $B$ act on $\C$ and are coboundaries.  By (II.22),
\be
\part^2 = 0\;.
\ee
Alternatively, we could use the couboundary operator 
$\overline\part=b-B$ which also is nilpotent.

\mbni{The Cocycle Condition.} $\partial \tau =0$.

A cochain $\tau\in\C(\A)$ is a cocycle if 
\be
\partial\tau = 0\;.\nn
\ee

\mbni{Equivalence Classes of Cochains.} $[f]$.

We define an equivalence class of entire cochains, modulo coboundaries.  
An entire cochain
$f\in \C(\A)$ defines an equivalence class
$[f]$ by
\be
[f] = \{f +\part G:G\in \C(\A)\}\;.
\ee
Clearly the equivalence $[f]$ depends on the space $\C(\A)$ of allowed
cochains. 

\subsection{Convergence of Cochains}
We introduce here a topology on $\D$ or on $\C$.  Let us consider
a family of cochains indexed by a parameter $\la$ ranging over an index
set $\Lambda$.  (For example, $\Lambda$  might be $\Z_+$, an interval
$(\la_1,\la_2)$ on the real line, a subset of $\C$, etc.)  We use this 
structure to study analysis on $\C$ or on $\D$. 

Let $\F$ denote a family of cochains $f(\la)$,
\be
\F = \{f(\la):f(\la)\in \C(\hbox{or }\D), \la\in \Lambda\}\;.
\ee
We say that $\F\in \C$ (or $\D$) is {\it bounded} if there exists a sequence
$\alpha_n$ such that
\be
0\le \alpha_n\;,\quad n^{1/2} \alpha_n^{1/n}\to 0\; ,
\ee
and for $n\in \Z_+$,
\be
\sup_{\la\in \Lambda} \vv{f_n(\la)} \le \alpha_n\;.
\ee
We say that the family is {\it Cauchy} as $\la\to\la_0$ if $\F$ is bounded
in $\C$ (or $\D$) and for every $n$, 
\be
\lim_{\la,\la'\to \la_0} \vv{f_n(\la)-f_n(\la')}=0 \;.\
\ee
A family $\F$ that is Cauchy as $\la\to\la_0$ has a limit $f$ in $\C$ (or
$\D$); in this case, for all $n\in \Z_+$,
\be
\vv{f_n} \le \alpha_n\;,\hensp{and} \lim_{\la\to\la_0} \vv{f_n(\la)-f_n}
=0\;.
\ee
We write
\be
\lim_{\la\to \la_0} f(\la) = f\;.
\ee
The standard notions of $\F$ being closed or compact follow.

If $\Lambda$ is an open, real interval $(\la_1,\la_2)$, we say that $f(\la)$
is differentiable at $\la_0\in \Lambda$ if for  $\la\in
\Lambda$,
$$
\lim_{\la\to\la_0} {f(\la_0)-f(\la)\over \la_0-\la} = g
$$
exists.  Then $f'(\la_0) = g$, etc. 

\section{Pairing a Cochain}
\setcounter{equation}{0}
Let us define a pairing $\lra{\tau,a}$ between a cochain $\tau\in \C(\A)$, 
and a root of unity $a \in \A^{\fg }$, as a non-linear functional 
$\tau: \A^{\fg }\rightarrow {\Bbb C}$ which depends on $\tau$ only through its 
equivalence class $[\tau]$. 
Since $\lra{\tau,a}$ is linear in $\tau$, it must be true that 
\be
\lra{\part G, a} = 0
\ee
for all $G\in\C(\A)$ and all $a\in\A^G$.
This condition allows us to determine a pairing, at least for even elements
$\{\tau_{2n}\}\subset \C(\A)$.

\subsection{$\lra{\part G,a}=0$ Determines a Pairing}
There is a canonical way to pair a cochain $\tau\in \C(\A)$ with an element 
$a\in \A^{\fg }$, such that the result $\lra{\tau,a}$ is linear in $\tau$.  
We suppose that the dependence of $\lra{\tau,a}$ on $\tau_n$ is a linear 
function of either 
\be
\tau_n (a,a\clips,a)\hensp{or} \tau_n(I,a,a\clips,a)\;.
\ee
Under this assumption, the general form of $\lra{\tau,a}$ is 
determined by a sequence of
numerical coefficients $\alpha_n,\beta_n$ independent of $\tau$ and $a$,
namely
\be
\lra{\tau,a} =
\sum^\infty_{n=0}\alpha_n\tau_n(a,a\clips,a)+\sum^\infty_{n=1}
\beta_n\tau_n(I,a,a\clips,a) \;.
\ee
We use the requirement $\lra{\part G,a}=0$, along with the assumption that
the odd components of $\tau$ vanish, to limit the form of the pairing.

\mbni{Proposition III.1.} {\it Let $a\in \A^{\fg }$ satisfy $a^2=I$.  Consider a
pairing such that 
\be
\lra{\tau,a} = \sum^\infty_{n=0}
\alpha_{2n}\tau_{2n}(a,a\clips,a)+
\sum^\infty_{n=0}\beta_{2n}\tau_{2n}(I,a,a\clips,a)\;.
\ee
Suppose that 
$$
\lra{\tau,a} = 0
$$ 
whenever $\tau=\part G$ and $G\in \C(\A)$.  Then
\be
\lra{\tau,a} = \alpha_0 \sum^\infty_{n=0} \l( -{1\over 4}\r)^n 
{(2n)!\over n!} \tau_{2n} (a,a\clips,a)\;.
\ee  
}

\mbni{Remarks:} (1) We remark that it is no loss of generality to normalize 
the pairing (III.4) so that $\alpha_0=1$. With this normalization, 
$\langle\tau,I\rangle = \tau_0(I)$.

(2) Proposition III.1 replaces Connes ``normalization'' condition
\cite{C2}, which is unnecessary for this complex.  
Furthermore, Proposition III.1 
is dual to the related result of Getzler and Szenes \cite{GS}. 
These other results concern a pairing of $\tau$ with 
idempotents $p^2=p\in \A^{\fg }$, rather than a pairing with operators 
$a\in \A^{\fg }$ which are square roots of unity, $a^2=I$.  
There is a one-to-one correspondence between
square roots of unity $a$ and idempotents $p$ given by $a=2p-I$.
In terms of these variables, the pairing introduced by Connes, 
which we denote $\langle\tau , p\rangle^C$, equals our pairing for 
$a=I$.  In general, it is the average of our pairing and its value at 
$a=I$, namely
\be
\langle\tau , p\rangle^C = 
{{1} \over {2}} \langle\tau , a\rangle +
{{1} \over {2}} \langle\tau , I\rangle\;.
\ee

(3) In \S IV we introduce a cochain $\tau^\jlo\in \C(\A)$ which will be the
focus of much of the remainder of this paper.  This cochain automatically
has the property $\tau^\jlo_{2n+1}=0$ of the form (III.4)

\mbni{Proof.}  Let $\tau=\part G$.  For $a\in\A^{\fg }$, 
it is the case that $a=a^\gamma= a^{g^{-1}}= a^{g^{-1} \gamma}$. 
Assume $a^2=I$ and let $n$ be even. 
Then the following three relations hold:
\beq
b_{n-1}G_{n-1} (a,a\clips,a) &=& 
G_{n-1}(a^2,a\clips,a)+ G_{n-1}(a^{g^{-1}} a, a\clips,a)=
2 G_{n-1}(I,a\clips,a) \;,\nn\\
B_{n+1}G_{n+1} (a\clips,a) &=& 
A_n G_{n+1}(I, a\clips,a) =
(n+1)G_{n+1} (I,a\clips,a)\;,
\eeq
and
\be
\l(b_{n-1}G_{n-1}\r) (I,a\clips,a) =
2G_{n-1}(a\clips,a) \;.
\ee
As $B:\C\to \N$, we also have  $(B_{n+1}G_{n+1}) (I,a\clips,a)=0$ .  
Thus for $n\ge 1$,
\be
\tau_{2n} (a\clips,a) =2 G_{2n-1}(a,a\clips,a)+(2n+1)G_{2n+1}(I,a\clips,a)\;,
\ee
and
\be
\tau_{2n}(I,a\clips,a) = 2G_{2n-1} (a\clips,a) \;.
\ee
Inserting the identities (III.9--10) into (III.4) yields
\beq
\lra{\tau,a} &=& \sum^\infty_{n=0} \l( 2\alpha_{2n+2} 
+(2n+1)\alpha_{2n}\r) G_{2n+1} (1,a,a\clips,a) \nn\\
&& + \sum^\infty_{n=1}\beta_{2n}G_{2n-1} (a,a\clips,a)\;.
\;.
\eeq
The vanishing of (III.11) for all $G\in \C$  ensures
vanishing of the coefficients in (III.11).  Thus 
$\beta_{2n}=0$ for $n\ge 1$, and
$ (2n+1)\alpha_{2n}+2\alpha_{2n+2}=0$.  This recursion relation is satisfied
by
$\alpha_{2n} = (-1/4)^n (2n) !(n!)^{-1}\alpha_0$.  
Substituting the coefficients
$\alpha_{2n}$ and
$\beta_{2n}$ into (III.4) yields (III.5).

The space $\Mat_m(\A)$ is the space of $m\times m$ matrices with
entries in $\A$.  Elements $a\in \Mat_m(\A^{\fg })$ satisfying $a^2=I$ 
are matrices $a=\{a_{ij}\}$  
$$
\sum^m_{j=1}a_{ij} a_{jn} = \delta_{in}\;.$$  
The pairing $\langle \tau,a\rangle$ for $a\in \Mat_m(\A^{\fg })$, $a^2=I$, is
defined by
\be 
\langle \tau,a\rangle 
  =  \sum^\infty_{n=0} (-1/4)^n {(2n)!\over n!} \sum_{1\le
i_0,i_1\clips,i_{2n}\le m} \tau_{2n} \l(
a_{i_0i_1},a_{i_1i_2}\clips,a_{i_{2n}i_0}\r)\;. 
\ee 
We use a shorthand notation for (III.12), so $m\times m$ matrices which
enter $\tau$ are multiplied.  Thus we write (III.12) as
\be 
  \langle \tau,a\rangle  =  \sum^\infty_{n=0} (-1/4)^n {(2n)!\over n!} \tr
\tau_{2k} \l( a ,a\clips,a\r)\;. 
\ee 
The $\tr$ in (III.13) denotes the trace in the space of $m\times m$
matrices $\Mat_m(\A)$ with entries in $\A$.   
We summarize this discussion by stating

\mbni{Proposition III.2} {\it Let $\tau\in \C$ and $a \in
{\rm Mat}_m(\A^{\fg })$
with $a^2=I$.  Then the pairing (III.13) exists. 
Furthermore, $\lra{\tau, a}$ depends on $\tau$ only through
its class $[\tau]$, so 
$$
\langle \tau,a\rangle = \langle \tau+\part G,a\rangle\;,
$$
where $G\in \C(\A)$, and 
\be
\langle \part G,a\rangle =0\;.
\ee  
}

In the case $m=1$, the pairing reduces to (III.4) normalized so
$\alpha_0=1$. The proof of (III.13) reduces step by step to the proof
of (III.4).

\subsection{A Generating Functional $\J(z;a)$ for $\tau$}
Let us define a generating function $\J(z;a)$  for $\tau$ by 
\be
\J (z;a) = \sum^\infty_{n=0} (- z^2)^{n}\tr \tau_{2n} \l(
 a ,a\clips,a\r)\;.
\ee
As a consequence of the assumption that $\tau$ is an entire cochain, and
that
$\vv{a}\le M$, we have
\be
\l| \tau_n(a\clips,a) \r|^{1/n} \le o(n^{-1/2})\; . 
\ee
Hence the series (III.15) converges to define an entire function of 
$z$ of order at most two.

Let $h(z)$ denote an entire function of one variable,
\be
h(z) = \sum^\infty_{n=0} h_nz^n\; .
\ee
We say that $h\in \E_\eta$, $\eta\ge 0$, if
\be
n^\eta |h_n|^{1/n} =  o(1) \hensp{as}n\to \infty\;. 
\ee
A topology on $\E_\eta$, can be defined as follows:
A family of functions $\{h^{(\la)}\}\subset \E_\eta$, indexed by $\la$ in an
index set $\La$, is bounded in
$\E_\eta$ if there is a bound $n^{\eta}\l|h^{(\la)}_n\r|^{1/n} \le  o(1)$, where
$o(1)$ is independent of $\la$.  The sequence {\it converges} to $h\in
\E_\eta$ if it is bounded and as $\la\to \la_0$,  and if
$h^{(\la)}_n\to h_n$.  Remark that the space
$\E_0$ contains all entire functions.  

The operator $D=d/dz$ maps $\E_\eta$ into $\E_\eta$,
for all $\eta\ge 0$.  Likewise, multiplication by $z$ maps $\E_\eta$ to
$\E_\eta$.
Consider the operator
\be
R=\exp(D^2/4)=\sum^\infty_{n=0} \l({D^{2n}\over 4^nn!}\r)\;.
\ee
For bounded or $L^2$ functions, the operator $R$ is given by convolution
with a Gaussian kernel, and on those spaces it defines a contraction.  
\mbni{Lemma III.3.} {\it  The transformation $R$
is   defined on $\E_\eta$ for $\eta\ge 1/2$, and 
\be
R:\E_\eta \to \E_{\eta-\frac12}\;, \hensp{for} \eta\ge \frac12\;.
\ee
}

\mbni{Proof.} 
Let us first establish (III.20) in case $h(z)=h(-z)$.  Thus $h(z) =
\sum^\infty_{n=0} h_{2n} z^{2n}$.  Since
$$
Rz^{2n}  = \sum^n_{k=0} {(2n)!\over 4^k(2n-2k)!k!} z^{2n-2k}\;,
$$
we write $R$ in the form
\be
(Rh)(z)=\sum^\infty_{m=0}\beta_{2m}z^{2m}\;,
\ee
with
\be
\beta_{2m} = \sum^\infty_{n=m} {(2n)!h_{2n}\over 4^{n-m}(2m)!(n-m)!} \;.
\ee
Using the hypothesis $(2n)!^\eta |h_{2n}|\le o(1)^n$, we obtain
\beq
(2m)!^{(\eta-\frac12)} |\beta_{2m}| &\le & \sum^\infty_{n=m} o(1)^n
{(2n)!^{1-\eta} (2m)!^{\eta-\frac12}\over (2m)!(n-m)!}\;, \nn\\
&\le &\sum^\infty_{n=m} o(1)^n \le o(1)^m\; .
\eeq
Here we used $(2n)!^{1/2}(n-m)!^{-1}\le O(1)^nm!$, and $\eta\ge \frac12$. 
Hence $Rh\in \E_{\eta-\frac12}$, and we have established (III.20) for even
$h(z)$.  In general, $h\in \E_\eta$ can be written $h=h_e+zh_o$ where both
$h_e$ and $h_o$ are even elements of $\E_\eta$.  Since
$Rz=(z+2D)R$, the above analysis shows that (III.20) holds in general.

As an entire function of $z$, $\J(z;a)$ defined in (III.15) is an element of
$\E_{1/2}$.  This is the consequence of assumption (II.3) for entire
cochains.  Therefore we infer from (III.20) that $\J$ is in the domain of
$R$, and that $(R\J)(z;a)$ is an entire function of $z$.  As a consequence, we
obtain a simple representation for the pairing (III.13).

\subsection{The Pairing Expressed in Terms of $\J(z;a)$  }
We now express the pairing $\langle \tau,p\rangle$ in
terms of the generating functional $\J(z;a)$ of (III.15).  This leads
us to the Gaussian transform $\fz(a;g)$ of $\J(z;a)$, 
evaluated at the origin.

\mbni{Proposition III.4.} {\it Let $\tau$ denote an entire cochain for $\A$
and let $a\in \Mat_m(\A^{\fg })$ satisfy $a^2=I$.  Then the pairing (III.13) can
be expressed in terms of the generating functional $\J$ as
\be
\fz(a;g) = \langle \tau,a\rangle 
= {1\over  \sqrt\pi} \int^\infty_{-\infty} e^{-t^2} \J(t;a)  dt\;.
\ee
}
\mbni{Proof.}  Since $\J(z;a)$ is even, and an element of $\E_{1/2}$, we
have from (III.21-22) that
\be
(R\J) (0;a)=\sum^\infty_{n=0} (-1/4)^n {(2n)!\over n!} \tr \tau_{2n}
( a ,a\clips,a)\;.
\ee
Comparing with (III.13), we find
$$
(R\J)\l( 0;a  \r) = \langle \tau,a\rangle  \;.
$$
As remarked above, the operator $R$ can be expressed as convolution by a
Gaussian.  In particular,
\be
(Rf)(s) = {1\over \sqrt \pi} \int^\infty_{-\infty} 
e^{-(s-t)^2} f(t)\, dt\;,
\ee
which also can be seen from
\be
{1\over \sqrt\pi} \int^\infty_{-\infty} e^{-t^2} t^{2n}dt = {(2n)!\over
n!4^n} \;.
\ee
Hence we have the representation (III.24).

\subsection{Pairing for Families}
In \S II.2 we introduced the notion of a family $\F=\{\tau(\la\}\subset \C$
of cochains depending on a parameter $\la$ belonging to an index set $\La$.
An important consequence of the topology introduced for $\C$, is that the
pairing $\langle \tau(\la),a\rangle$ of a family inherits the convergence
properties from $\C$.  Associated with the family $\{\tau(\la\}\subset
\C$, we have a family of generating functions $\{\J(\la)\}$ defined by
\be
\J(\la) = \J(z;a;\la) = \sum^\infty_{n=0}
(-z^2)^{n}\tau_{2n}(\la)( a ,a\clips,a;g) = \sum^\infty_{n=0}
\J_n(\la)z^n\;.
\ee 
For each $a\in \A$, and $\la\in \La$, the function
$\J(\cdot;a;\la)$ is a function in $\E_{1/2}$, as defined in (III.18).  
We consider
$\{\J(\la)\}\subset \E_{1/2}$ as a family of functions in $\E_{1/2}$
parameterized by $\la\in \La$.

\mbni{Proposition III.6.} {\it Let the family $\{\tau(\la)\}\subset \C$ be
bounded, continuous at $\la_0$, or differentiable at $\la_0$ in the sense of
\S II.2 as a map from $\La$ to $\C$.  Then
\begin{itemize}
\ritem{(i)} The family $\{\J(\la)\}\subset \E_{1/2}$ is respectively:
bounded, continuous at $\la_0$, or differentiable at $\la_0$ in the sense of
a family in
$\E_{1/2}$.
\ritem{(ii)} For $a^2=I \in {\rm Mat}_m(\A)$, the pairing of $\tau(\la)$ with
$a$ of (III.13) or of Corollary III.3 defines the pairing function
\be
\langle \tau(\la),a\rangle\;.
\ee
As a function of the variable $\la$, $\lra{\tau(\la),a}$ is respectively:
bounded uniformly for $\la\in \La$, continuous at $\la_0$, or differentiable
at
$\la_0$.
\end{itemize}
}

\mbni{Proof.} If the family defined by $\tau(\la)$ is bounded, then the
bound (II.30--31) ensures that the coefficients $\J_n(\la)$ in (III.28)
satisfy
\be
\sup_{\la\in \La} |\J_n(\la)|\le \alpha_n 
\ee
where $n^{1/2}\alpha_n^{1/n}\to 0$ as $n\to \infty$.  Thus $\{\J(\la)\}$ is
a bounded family in $\E_{1/2}$.  Likewise, convergence of the uniformly
bounded coefficients $\J_n(\la)$ as $\la\to\la_0$, ensures convergence of
$\J(\la)$ to a function $\J\in \E_{1/2}$.   This is a consequence of the uniform
bound on $\J(\la)$ as a function of $z$,  and  of the
Vitali convergence theorem for holomorphic functions.  Likewise, if the
difference quotient
$(\tau(\la_0)-\tau(\la))/(\la_0-\la)$ converges in $\C$ as $\la\to \la_0$,
this means that $(\J(\la_0)-\J(\la))/(\la-\la_0)$ converges in $\E_{1/2}$. 
This proves part (i) of the proposition.  The proof of part (ii) is similar.  It
is a consequence of the uniform bound (II.30--31), along with the continuity
or differentiability.  The uniform bound (II.30--31) ensures that the sum
\be
\lra{\tau(\la),a} =   \sum^\infty_{n=0} (-1/4)^n {(2n)!\over
n!}
\tr
\tau_{2n}(\la) \l(a,a\clips,a\r) 
\ee
converges uniformly for $\la\in \La$ and is bounded by the constant $M$
defined by
\be
M=\sum^\infty_{n=0}\l(\sup_{i,j} \vv{a_{ij}} m\r)^{2n+1} 
{(2n)!\over n!} \alpha_{2n}\;.
\ee
Thus $\lra{\tau(\la),a}$ is a
uniformly bounded function.  If $\tau(\la)$ converges as $\la\to \la_0$, 
then $\vv{\tau_n(\la)-\tau_n(\la')}$ is Cauchy for 
each $n$ as $\la, \la'\to\la_0$.  Define
$\lra{\tau(\la),a}_N$ as (III.31), but 
with the sum over $n$ limited to $n\le N$.  We write
\be
| \lra{\tau(\la),a} -\lra{\tau(\la'),a}|  
\le   |\lra{\tau(\la),a} -\lra{\tau(\la),a}_N|  
  + |\lra{\tau(\la),a}_N-\lra{\tau(\la'),a}_N| 
  + |\lra{\tau(\la'),a}_N -\lra{\tau(\la'),a}|\;.
\ee 
The convergence of (III.32) ensures that there exists $N_0<\infty$ such that
for $N>N_0$, 
$|\lra{\tau(\la),a}-\lra{\tau(\la),a}_N| <\ep$ uniformly in $\la\in \La$.  
On the other hand, the fact that $\tau_n$ is Cauchy insures that for 
$N>N_0$ and fixed, we have 
$|\lra{\tau(\la),a}-\lra{\tau(\la'),a}|<\ep$ for $\la, \la'$ arbitrarily
close to $\la_0$. Thus 
$\lra{\tau(\la),a}$ converges as $\la\to \la_0$.  The argument that
differentiability of $\tau(\la)$ in $\C$ ensures differentiability of
$\lra{\tau(\la),a}$ is similar, so we omit the details.  This completes the
proof of the proposition.

\section{The JLO Cochain (Differentiable Case)}
\subsection{Heat Kernel Regularization and the Radon Transform}
The description of the JLO cochain requires some more structure, in
addition to  the algebra $\A$ and the functionals $\C$ on $\A$.  We
begin by introducing a Hilbert space $\CH$ and the algebra $\B(\CH)$ of
bounded linear transformations on $\CH$.  We represent $\A$ as a
subalgebra of $\B(\CH)$.  We let $\Vert\cdot\Vert$ denote the norm on
$\CH$ and $\vcv$ the inherited norm on $\A$.  We think of $\A$ as the
non-commutative generalization of an algebra of functions.  In this section
we formulate the case that $\A$ is an algebra of differentiable functions;
thus we call it the differentiable case.  We also need to define a derivative
on $\A$.  
This is natural for a differential geometric interpretation of
non-commutative geometry.
Some other operators play a special role.  They are the following:

\mbni{$\Z_2$-Grading $\gamma$:}

A $\Z_2$-grading $\gamma\in \B(\CH)$ is a self-adjoint, unitary
$\gamma=\gamma^\ast =
\gamma^{-1}$.  The grading $\gamma$ determines a decomposition of $\CH$,
$\CH=\CH_+\oplus
\CH_-$, into the $\pm$ eigenspaces of $\gamma$.  The orthogonal
projections of $\CH$ onto $\CH_\pm$ are $P_\pm = \frac12 (I\pm
\gamma)$, and $P_+ + P_- = I$.  We assume that
$\A$ is pointwise invariant under $\gamma$, 
\setcounter{equation}{0}
\be
\gamma a = a\gamma\;,\quad a\in \A\; . 
\ee
In general, we denote the action of $\gamma$ as
\be
b^\gamma = \gamma b\gamma^{-1}=\gamma b \gamma\;,\quad b\in \B(\CH)
\; .
\ee

\mbni{Dirac Operator $Q$:}

The operator\footnote{In the physics literature, the square root $Q$ of the 
energy operator is called the {\it supercharge}.  
The Laplacian $H=Q^2$ is called
the energy, or Hamiltonian.  The relation between supersymmetry in physics
and index theory was observed by Witten \cite{W1} in the context of the
index of the exterior differential. In quantum field theory, the supercharge
also operator has the interpretation of a Dirac operator on loop space.} 
$Q=Q^\ast$ is an (unbounded) operator on
$\CH$ whose square
$H=Q^2$ has the interpretation of Laplacian.  It is assumed that $Q$ and
$\gamma$ anticommute,
\be
\gamma Q + Q\gamma=0\;.
\ee
Let $Q_\pm = QP_\pm$.  Then $Q=Q_++Q_-$, where $Q^\ast_+=Q_-$ and
$Q^2_+ = Q^2_-=0$.  It follows that $H=Q_+Q_-+Q_-Q_+$. We also assume
that for $0<\beta$,
\be
{\rm Tr} \l(e^{-\beta Q^2}\r) <\infty\;.
\ee
The condition (IV.4) is called
$\Theta$-summability by Connes \cite{C2}.  At least in the case 
$\alpha=\beta=0$,   it can be replaced by a more general   condition, called
KMS, see \cite{K1,JLO2}.

\mbni{Derivation $d$.}

The operator $Q$ defines a graded derivation $d$.  For operators $b\in
\B(\CH)$ in the domain of $d$,
\be
db = Qb-b^\gamma Q\;.
\ee
We assume that all $a\in \A$ are in the domain of $d$, and that the bilinear
form $da=Qa-aQ$ defined on $\D(Q)\times \D(Q)$ uniquely determines a
bounded linear operator
$$
da  \in \B(\CH)\;.$$
We assume that for all $a,b\in \A$,
\be
d(ab)=(da)b+a(db)\; . 
\ee
Note that for $b$ in the domain of $d^2$, or as a sesquilinear form,
\be
d^2 b=[Q^2,b]= Q^2b -bQ^2\;.
\ee
In \S V we elaborate the properties of the domain of $d$.

\mbni{Symmetry Group ${\fg }$:}

Equivariance arises through the existence of a continuous, unitary
representation
$U(g)$ of a compact Lie group
${\fg }$ on
$\CH$.  We assume that
\be
U(g)\gamma = \gamma U(g)\;,\quad U(g)Q = QU(g)\;.
\ee
We let
\be
b^g = U(g) b U(g)^\ast \;,\quad b\in \B(\CH)\;,
\ee
denote the automorphism of $\B(\CH)$ induced by ${\fg }$.  We assume that
${\fg }$ acts on $\A$, namely that for all $a\in \A$, $a^g\in \A$. The group
${\fg }$ may be trivial, in which case the cochains are no longer functions on
${\fg }$.  This is the ordinary (rather than equivariant) theory.  (The
representation
$U(g)$ acting on $\CH$ has no relation with, and should not be confused
with,
$U$ in (II.7) which acts on $\D$.)

\mbni{Heat Kernel Regularization:}

Consider $(n+1)$ operators $b_j\in \bh, j=0,1\clips,n$.  Define the operator
valued function $X(s)$ on $\R^{n+1}$ by
\be
X(s) = \l\{ \begin{array}{l}
b_0e^{-s_0Q^2}b_1 e^{-s_1Q^2} \cdots b_ne^{-s_nQ^2}\;, \hensp{every}
s_j>0\\
0\;, \hensp{any} s_j\le 0\end{array}\r. \;.
\ee
The operator $X(s)$ is the {\it heat-kernel regularized density} of the
ordered set operators $\{b_0,b_1\clips,b_n\}$. See also \cite{J1}.  Note that
$X(s)$ is an $(n+1)$-multilinear function on the set $\bh$.  We call
$\{b_0\clips,b_n\}$ the set of {\it vertices}  of $X(s)$. We  
sometimes use $X$ to denote the set of vertices,
$$
X=\{b_0,b_1\clips,b_n\}\;,
$$
as well as the heat kernel regularization, at least 
in cases where no  confusion can  arise.

\mbni{The Radon Transform:}
We also consider the Radon transform $\hx(\beta)$ of $X(s)$ corresponding
to the hyperplane $s_0+s_1+\cdots +s_n=\beta > 0$.  In other words
\be
\hx(\beta) = \int X(s)\, d^{n}s(\beta)\;.
\ee 
Here $d^{n}s(\beta)$ 
denotes Lebesque measure on
$\R^{n+1}$ restricted to the hyperplane $\sum^n_{j=0} s_j=\beta$.
Explicitly, this measure is 
\be
d^{n}s(\beta)=\delta(s_0+\cdots + s_n - \beta)ds_0ds_1\cdots ds_n\;,
\ee
where $\delta$ denotes the Dirac measure.  
We shall see in (V.20,24), that the 
total measure of $d^{n}s(\beta)$, integrated over the positive 
``quadrant'' $\R_+^{n+1}$, equals $\beta^n/n!$. 

The heat-kernel-regularized density $X(s)$ and its Radon transform
$\hx(\beta)$   form the basic objects in  the geometric theory we develop
here.  We refer to both as {\it heat-kernel regularizations} of
$\{b_0\clips,b_n\}$.  It often turns out in the geometric theory that
hyperplanes for different values of $\beta$ are equivalent, and this 
is always the case if
$\exp(-\beta Q^2)$ is trace class for all $\beta>0$, see \S VII.6.  Thus in
order to simplify notation, we restrict our attention to the plane
$$
s_0+s_1+\cdots +s_n=1\;.
$$
Denote this value of the Radon transform by $\hx=\hx(\beta=1)$. 
When there is no chance of 
confusion, we simply write for the corresponding measure 
\be
d^{n}s=d^{n}s(1)\qquad\hensp{or}\qquad ds=d^{n}s(1)\;.
\ee

As a consequence of the 
trace class property of $e^{-\beta Q^2}$ for each $\beta>0$, the 
heat kernel regularization 
$X(s)$ and also $\hx$ is trace class.  The trace norm 
$\Vert\hx \Vert_1=\Tr\l((\hx^\ast \hx)^{1/2}\r)$ 
of $\hx$ satisfies 
\be
\Vert \hx\Vert_1 
\le {1\over n!} \Tr \l(e^{-Q^2}\r) \l(\prod^n_{j=0} \Vert b_j\Vert\r)\;.
\ee
We postpone the proof of (IV.14) to \S V, in conjunction with
the proof of other related bounds, see Corollary V.4.v.

\mbni{Symmetries of $\hx$: }

The group $\Z_2$, implemented by $\gamma$, and the group
${\fg }$, implemented by $U(g)$ both commute with $\exp(-\beta Q^2)$. 
Hence these groups act on $\hx$ by acting on the vertices of $X$. 
Let $\hx^\gamma$ denote the heat kernel regularization
$\gamma \hx\gamma^{-1}$ of $X^\gamma$ with vertices 
$$
X^\gamma = \{b^\gamma_0, b^\gamma_1\clips,b^\gamma_n\}\;.
$$
Similarly let $\hx^g$ denote the 
heat kernel regularization $U(g)\hx U(g)^*$ of $X^g$ with vertices
$$
X^g = \{b^g_0,b^g_1\clips,b^g_n\}\;.
$$

\subsection{Expectations and the Radon Transforms}

\mbni{Expectations  $\langle \hx;g\rangle$:}

The expectation $\langle \hx;g\rangle$ of a heat kernel regularization 
$\hx$ is defined by
\be
\langle \hx;g\rangle = \Tr (\gamma U(g)\hx)\; .
\ee
Since the expectation is a linear function of each vertex of $X$, we also 
use the notation which indicates this multilinearity, namely we denote
$\lra{\hx;g}$ by 
\be
\langle \hx ;g\rangle = \Tr (\gamma U(g)\hx)  = \langle
b_0,b_1\clips,b_n;g\rangle =
\langle b_0,b_1\clips,b_n;g\rangle_n\;.
\ee
Here we use the subscript $n$, when it may be helpful to clarify the number
of vertices.  The bound  (IV.14) ensures that  the expectation is continuous in
each vertex, and
\be
|\langle b_0,b_1\clips,b_n;g\rangle | \le {1\over n!} \Tr \l(e^{-Q^2}\r)\l(
\prod^n_{j=0} \Vert b_j\Vert\r) \;.
\ee

\mbni{Symmetries of Expectations:}

As a consequence of cyclicity of the trace, and the commutativity of
$\gamma$ and $U(g)$,
\be
\langle \hx ;g\rangle = 
\langle \hx ^\gamma;g\rangle = \langle \hx^g ;g\rangle =
\llangle \hx ^{g^{-1}};g\rrangle\;.
\ee
More generally, for $h\in {\fg }$, $\langle \hx ^h;g\rangle = \langle
\hx ;h^{-1} gh\rangle$.

Another symmetry of the expectation arises from cyclic permutation of the
vertices,
\be
\langle b_0,b_1\clips,b_n;g\rangle = \llangle
b_n^{g^{-1}\gamma},b_0\clips,b_{n-1};g\rrangle\;.
\ee
We also remark that the expectation is invariant under the infinitesimal
$d$.  This means that for all heat-kernel regularizations $\hx$,
\be
\langle d \hx ;g\rangle=0\;.
\ee
In particular, if $\hx^\gamma=\hx$, then $(d\hx)^\gamma=-d\hx$ and (IV.20)
vanishes by (IV.18).  On the other hand, if $\hx^\gamma=-\hx$, then 
$(d\hx)=Q\hx+\hx Q$, and (IV.20) vanishes on account of (IV.3) and cyclicity
of the trace.
Writing out (IV.20) in detail for $X$ with vertices $b_0\clips,b_n$, we
infer that
\be
\sum^n_{j=0} \llangle
b^\gamma_0,b^\gamma_1\clips,b^\gamma_{j-
1},db_j,b_{j+1}\clips,b_n;g\rrangle_n
=0 \; .
\ee

Another interesting identity for expectations is
\be
\langle b_0\clips,b_n;g\rangle_n = \sum^{n+1}_{j=1} \llangle
b_0\clips,b_{j-1},I,b_j\clips,b_n;g\rrangle_{n+1}\;.
\ee
This follows from a change of variables in the Radon transform (IV.10), for
$1\le j\le n+1$.  Let $\beta'_i = \beta_i$, $0\le i\le j-2$,
$\beta'_{j-1}=\beta_{j-1}+\beta_j$, $\beta'_i = \beta_{i-1}$, $j\le i\le
n$, and $\beta'_{n+1}=\beta_{j-1}$.  Then
\beq
&& \llangle  b_0\clips,b_{j-1},I\clips,b_n;g\rrangle_{n+1} \nn\\
&&\hskip.75truein  = 
\int_{\tilde\sigma_n }
\Tr
\l(\gamma U(g)b_0 e^{-\beta'_0Q^2}\cdots b_{j-1} e^{-\beta'_{j-1}Q^2} b_j
e^{-\beta'_j Q^2}\cdots b_n e^{-\beta'_nQ^2}\r)   
d\beta'_0\cdots d\beta'_n d\beta'_{n+1}\;, \nn\\
&&
\eeq
where $\tilde\sigma_n$
denotes the set
\be
0<\beta'_i\; , \quad 0<\beta'_{n+1}<\beta'_{j-1}\;,\quad \hbox{and}\quad
\beta'_0+\beta'_1 +\cdots + \beta'_n=1\; .
\ee
Note the integrand is 
independent of
$\beta'_{n+1}$.  Therefore the
$\beta'_{n+1}$ integration yields $\beta'_{j-1}$ times an integrand common
to every such term, $1\le j\le n+1$.  This latter integrand by itself would
integrate to $\llangle b_0\clips,b_n;g\rrangle_n$.  Summing over $j$
results in the integrand $\sum^n_{j=0}\beta'_j$ times the integrand
defining $\langle b_0\clips,b_n;g\rangle_n$.  But the constraint
$\sum^n_{j=0} \beta'=1$ in this integral reduces the sum to $\langle
b_0\clips,b_n;g\rangle_n$, so we infer (IV.22). 

\subsection{The  Cochain $\tau^{\jlo}$}
The JLO cochain \cite{JLO1} is the expectation whose $n^{\rm th}$
component is defined by  
\be
\tau^\jlo_n (a_0\clips,a_n;g)=\langle a_0,da_1\clips,da_n;g\rangle\;.
\ee
We assume that for $a\in A$ the norm is defined so that $\vv{\cdot}$
dominates the first Sobolev norm   defined by
\be
\vv{a}_1 = \Vert a\Vert +\Vert da\Vert\;.
\ee
Namely
\be
\vv{a}_1 \le \vv{a} \;.
\ee
In other words, every element of $\A$ has a bounded derivative, so each
element of such an $\A$ is the non-commutative generalization of a
continuous function.
As a consequence, we infer that  the expectation
$\tau^{\jlo} =
\{
\tau^{\jlo}_n\}$ is a cochain:

\mbni{Lemma IV.1.} {\it The expectation $\tau^\jlo$ is an element of the
space $\C(\A)$ of cochains, as defined at the start of \S II.  Furthermore
$\tau^\jlo$ extends by continuity and linearity from $\A$ to the subalgebra
$\B_1$ of operators
$b\in
\B(\CH)$ such that $\vv{b}_1<\infty$, and $\B_1$ is a Banach algebra.}

\mbni{Proof.} Clearly $\tau^\jlo_n$ is $(n+1)$-linear in $\A$.  We show
that $\tau^\jlo_n\in \C_n$.  This requires the symmetry (II.1) of cochains,
continuity in the norm of $\A$ and continuity in ${\fg }$.  In addition
$\tau^\jlo_n$ must vanish for $a_k=I$, $k=1,2\clips,n$.  But this latter fact 
follows from $dI=0$.

The required symmetry (II.1) for $\tau^\jlo_n$ is a consequence established
in (IV.18).  We combine this fact with the assumption (IV.8) which ensures
$d(a^{g^{-1}}) = (da)^{g^{-1}}$.  Thus 
\begin{eqnarray*}
\tau^\jlo_n
\!\l(a^{g^{-1}}_0\clips,a^{g^{-1}}_n;g\r)& = &\llangle
a^{g^{-1}}_0,d\!\l(a^{g^{-1}}_1\r)\clips, d\!\l(a^{g^{-1}}_n\r);g\rrangle \\
&=&
\llangle a^{g^{-1}}_0 , (da_1)^{g^{-1}}\clips, (da_n)^{g^{-1}} ;g\rrangle \\
&=&
\llangle a_0,da_1\clips, da_n;g\rrangle = \tau^\jlo_n (a_0\clips,a_n)
\end{eqnarray*} as
desired.

The continuity of $\tau^\jlo_n$ in $\A$ follows from (IV.17) and (IV.26).

\be
\l| \tau^\jlo_n (a_0\clips,a_n;g)\r|\le \frac1{n!} \Tr (e^{-Q^2}) \l(
\prod^n_{j=0} \vv{a_j}_1\r)\;.
\ee
Since $\vv{a}_1\le \vv{a}$, we have
\be
\vv{\tau^\jlo_n} \le {1\over n!} \Tr \l( e^{-Q^2}\r)\;.
\ee
Furthermore $U(g)$ is a continuous, unitary representation, so
$\tau^\jlo_n(a_0\clips$, $a_n;g)$ is continuous (pointwise) in ${\fg }$.  Thus
$\tau^\jlo_n\in \C_n$.  

We verify that the sequence $\tau^\jlo\in \C$.  The factor  ${1\over
n!}$ in (IV.29) ensures that
$$
n\vv{\tau_n}^{1/n} \le 0(1)\;,
$$
so the entire condition (II.3) is satisfied.

Thus $\tau^\jlo_n$ extends by continuity to $\B_1$.  Finally we verify that
$\B_1$ is a Banach algebra.  In fact for $a,b\in \B_1$, we infer from (IV.5)
that $d(ab)=(da)b+a^\gamma(db)$.  Hence 
$$
\vv{ab}_1 = \Vrt{ab} +\Vrt{d(ab)}\le \Vrt{a}\, \Vrt{b}+ \Vrt{da}\, \Vrt{b} +
\Vrt{a}\, \Vrt{db} \le \vv{a}_1 \vv{b}_1\;,
$$
as asserted. 
\mbni{Other Symmetries of $\tau^\jlo$:}

Remark that for $a_j\in \A$, the odd components of $\tau^\jlo$ vanish,
namely
\be
\tau^\jlo_{2n+1} (a_0\clips,a_n;g)=0\;.
\ee
This is a consequence of (IV.1,3,5) which ensures $(da)^\gamma=-da$.  Thus
using (IV.18),
\begin{eqnarray*}
\langle a_0,da_1\clips,da_n;g\rangle &=& \llangle
a^\gamma_0,(da_1)^\gamma\clips,(da_n)^\gamma;g\rrangle\\
&=& (-1)^n \langle a_0,da_1\clips,da_n;g\rangle\;,
\end{eqnarray*}
so $\tau^\jlo_n$ vanishes for odd $n$.

Another elementary identity, a consequence of (IV.21) and the  choice of
$a_j\in \A$ is that
\be
\sum^n_{j=0}(-1)^j \tau^\jlo_n
(a_0\clips,a_{j-1},da_j,a_{j+1}\clips,a_n;g)=0\;.
\ee

\subsection{The JLO Pairing and the Generating Functional}
\mbni{The Generating Functional $\J(z;a)$:}

We evaluate the generating functional $\J(z;a)$ of (III.15) for

$\tau=\tau^\jlo$.  Since $\tau^\jlo_{2n+1}=0$, we can also write for $a\in
\A$,
\be
\J^\jlo(z;a)=\sum^\infty_{n=0} (-z^2)^n \tr \tau_n\l(a,a\clips,a;g\r)\;.
\ee
Hence
\be
\J^\jlo (z;a) = \Tr\l(\gamma U(g)a e^{-Q^2+izda}\r)\;.
\ee

\mbni{A Formula for the Pairing:}
Using (IV.33), we infer that the pairing can be expressed simply.  
\mbni{Proposition IV.2.}{\it The pairing $\lra{\tau^\jlo,a}$ of the
cochain
$\tau^\jlo$ with $a\in \Mat_m(\A)$ satisfying $a^2=I$ can be written
\be 
\fz^Q(a;g)
=  \frac1{\sqrt\pi} \int^\infty_{-\infty} e^{-t^2} \Tr \l( \gamma U(g)
ae^{-Q^2 +itda}\r) dt\;. 
\ee 
Here Tr denotes both the trace on $\CH$ and the matrix trace in 
$\Mat_m(\A)$.   
}

\section{Fractionally Differentiable Structures}
We use the name {\it quantum harmonic analysis} for the study of 
fractional differentiability of operator valued functions.  
An interpolation space, in the quantum context, is a Banach algebra of 
operator-valued functions with fractional derivatives.  We distinguish 
quantum harmonic analysis from ``non-commutative harmonic analysis,'' 
a term used to denote the study of harmonic analysis on non-commutative 
groups.

\subsection{The Classical Picture}
Let us digress on a simple case --- we refer to it as the classical case ---
for purposes of motivation.  Take $\E = \oplus^n_{k=0} \E_k$ to be the
exterior algebra of smooth differential forms on the $n$-torus $\bbt^n$. 
Let $\E_k$ denote $k$-forms with the standard $L^2$ inner product.  We let
$\CH$ denote the Hilbert space of $L^2$ forms obtained by completing $\E$
as an inner product space with the inner product on $\E$ given by the sum
of the inner products on $\E_k$.  Thus $\CH=\oplus^n_{k=0}\CH^{(k)}$.  Also
define $\gamma$ as $(-1)^k$ on $\CH^{(k)}$, and let ${\fg }$ denote the group of
translations on $T^n$.  Thus ${\fg }$ acts unitarily on $\CH$, and for $f\in \CH$,
$(U(x)f)(x)=f(x-y)$.

Define $d$ to be the exterior derivative with domain $\E\subset \CH$.  Then
define $Q=d+d^\ast$.  Clearly $Q\gamma + \gamma Q=0$ on $\E$, and
$U(g)Q=QU(g)$.  The operator $Q$ with domain $\E$ is essentially 
self-adjoint.  Then $Q^2 = dd^\ast +d^\ast d = -\Delta$ is the Laplacian.  
Also $\exp(-\beta Q^2)$, $\beta\ge 0$, commutes with $\gamma$ and $U(a)$. 
Furthermore $\exp(-\beta Q^2)=e^{\beta \Delta}$ is trace class for every
$\beta>0$.

Alternatively, we can consider $\E$ as a subalgebra of $B(\CH)$, the
bounded, linear operators on $\CH$.  An element in $\E_k$ maps $\CH_{k'}$
to $\CH_{k+k'}$ by exterior multiplication.  We give this algebra the norm
\setcounter{equation}{0}
\be
\vv{b}_1 = \Vert b\Vert +\Vert db\Vert\;,\quad b\in \E\; ,
\ee
where $\Vert\cdot \Vert$ denotes the operator norm on $\CH$.  This agrees
with the $L^\infty$ norm defined on the coefficients of the form $b$. 
Define $\A_1$ as the completion of the smooth functions $\E_0$ (the
smooth zero forms) in the norm (V.1).  Hence $\A_1$ is the algebra of
Lipshitz continuous functions,
\be
\sup_x |a(x+y)-a(x)|\le M|y| \;.
\ee
Since (V.1) has the same form as the norm (IV.26), we can regard this
example as a special case of \S IV where we take $\vcv = \vcv_1$   and
$\A=\A_1$.  From this point of view, the material in \S IV belongs to the
study of the non-commutative Lipshitz class.

In order to distinguish the differentiable structure from the continuous
structure in the non-commutative case, one wants to study
the analogs of H\"older continuous classes, which in the classical case
would satisfy
\be
\sup_x |a(x+y)-a(x)|\le M|y|^\alpha\;, \quad 0<\alpha\le 1\;,
\ee
for  $\alpha$   the exponent of continuity.

Related to such classes are functions with fractional derivatives of order
$\alpha$.  The derivative $da$ of a H\"older continuous function is
unbounded.  However fractional derivatives may be bounded.  One way to
define an $L^p$ fractional derivative of order $\alpha$ of the function 
$a$ is to suppose that
\be
(-\Delta +I)^{\alpha/2}a(x)\in L^p(\bbt^n)\;,
\ee
for which an extensive theory exists in the classical case, see \cite{S}.  If
$a(x)$ is bounded, then  a natural norm on such functions is , $\Vert
a\Vert_{L^\infty}+\Vert (-\Delta+I)^{\alpha/2}a\Vert_{L^p}$.  If the norm
with $p=\infty$ exists, then one is ensured that the function $a(x)$ is
H\"older continuous for all continuity exponents $\alpha'<\alpha$.  For
$\alpha\le 1$ this norm
\be
\Vert a\Vert_{L^\infty} +\Vert (-\Delta +I)^{\alpha/2} a\Vert_{L^\infty}
\ee
is equivalent to
\be
\Vert a\Vert_{L^\infty} + \Vert
(-\Delta+I)^{-(1-\alpha)/2}da\Vert_{L^\infty}\;.
\ee
The norm (V.6) provides a natural measure of functions with derivatives of
order $\alpha$, or
of functions which are H\"older continuous with exponent $\alpha'< 
  \alpha$.  Other norms of classical analysis could also be studied.

In this section we pose these questions in the non-commutative case.  Thus
function space norms need to be replaced by operator norms, and there are
other corresponding adjustments.  We have the following dictionary
\medbreak
\begin{center}
\begin{tabular}{ll}
{\bf classical} & {\bf non-commutative} \\
\hline
function space & algebra $\A$ of linear transformations \\
exterior derivative $d$ & graded commutator with $Q$ \\
Laplacian & $Q^2$ \\
$(-1)^{\rm degree}$ & $\gamma$ \\
$L^\infty$-norm & operator norm\\
$L^p$-norm & $I_p$-Schatten norm \\
generalized function & operator-valued generalized function\\
Sobolev norms of a function & norm of maps between Sobolev spaces \\
tempered distribution  &  bounded map between Sobolev spaces\\
space of fractionally differential functions & interpolation space $\jba$\\
exterior derivative & graded commutator with $Q$\\
degree of regularity & local regularity exponent $\eloc$\\
regularity as a function of dimension & global regularity exponent $\eglo$\\
regularized current & heat kernel regularization\\
integral of   (current) & JLO-cochain $\tau^\jlo_n$

\end{tabular}
\end{center}
\medbreak

It is natural to define fractional derivatives in terms of the  scales
determined by $Q$.  Thus we say that a bounded operator  $a$ has a
derivative of order $\mu$ if $(Q^2 + I)^{\mu/2} a(Q^2 +I)^{-\mu/2}$ is
also bounded. We take the norm
\be
\Vert a\Vert +\Vert (Q^2+I)^{\mu/2} a (Q^2+I)^{-\mu/2}\Vert 
\ee
as the non-commutative version of (V.5).  Since $da$ is fundamental for the
theory of invariants, we prefer to pose our assumptions in terms like (V.6),
rather than (V.7).  We show in \S V.3 that (V.7) with $\mu>1$ leads us to
assume a bound on $\Vrt{(Q^2+I)^{-(1-\mu)/2}da}$ itself.  More generally,
it is possible to assume
\be
\Vrt{(Q^2+I)^{-\beta /2}da(Q^2 +I)^{- \alpha /2} }<\infty
\ee
for some $0\le \alpha,\beta$ and $0\le \alpha+\beta<1$. The order of
differentiability is $\mu=1-\beta$.  Thus we obtain not one space, but a
family of non-commutative spaces which generalize the space of bounded
functions with fractional derivatives in the sense of (V.6).  In fact, each of
these spaces gives rise to a theory of geometric invariants.  In \S V.5 we
introduce a family of {\it interpolation spaces} $\jba$ that provides a
natural framework in which to generalize the construction of 
\S IV.

\subsection{Sobolev Spaces in $\CH$}
We start with a Hilbert space $\CH$ and a fundamental Dirac operator
$Q=Q^\ast$ on $\CH$ with domain $\D$, as in \S IV.  We define Sobolev
spaces $\CH_p\subset \CH$ for $0\le p<\infty$ which are the domain of
$|Q|^p$ with a Hilbert space structure.  In order to simplify the discussion of
embeddings, let us consider the domain of the operator
$(Q^2+I)^{p/2}$, which we denote $\D((Q^2+I)^{p/2})$ or $\D_p= \D_p(Q)$
for short.  The Sobolev space $\CH_p=\CH_p(Q)$, $0\le p<\infty$, is the
domain 
$\D_p$, considered as a Hilbert space with inner product
\be
\lra{f,g}_{\CH_p} = \lra{(Q^2+I)^{p/2}f,(Q^2+I)^{p/2}g}\;.
\ee
The corresponding negative Sobolev space $\CH_{-p}$, for $p>0$, is the
completion of
$\CH$ in the norm determined by the inner product
\be
\lra{f,g}_{\CH_{-p}} = \lra{(Q^2+I)^{-p/2} f,(Q^2+I)^{-p/2}g}\;.
\ee
For $\alpha>\beta$ there is a natural embedding $\CH_\alpha\subset
\CH_\beta$.  With respect to this embedding the spaces $\CH_p$ and
$\CH_{-p}$ are dual, and for $p>0$ they define a Gelfand triple
\be
\CH_p \subset \CH \subset \CH_{-p}\;.
\ee
This is a standard device in the study of classical generalized functions or
distributions, see [Gelfand].  We also introduce the square root of  
the resolvent of the ``Laplacian''
$Q^2$,
\be
R=(Q^2+I)^{-1/2}\;,
\ee
so
\be
\lra{f,g}_{\CH_p} = \lra{R^{-p}f,R^{-p}g}_{\CH}\;.
\ee
In the classical case, the integral operator given by $R^{p/2}$, $p>0$, is
called the Bessel transform operator of order $p$.
We define
\be
\CH_\infty = \bigcap_p \CH_p\;,\quad\hensp{and}\quad \CH_{-\infty} =
\bigcup_p \CH_p\;.
\ee
Then for $p\ge 0$
\be
\CH_\infty \subset \CH_p\subset \CH_0 \subset \CH_{-p}\subset
\CH_{-\infty}\;,
\ee
and for $s>0$,
\be
e^{-sQ^2}:\CH_{-\infty}\to \CH_\infty\;.
\ee
\subsection{Some Facts}
 
\mbni{Schatten Classes $I_p$.}

The analog in the non-commutative case of $l_p$ spaces are the Schatten
ideals $I_p=I_p(\CH)$.  This is the subspace of compact operators on
$\CH$ for which the norm $\Vrt{\cdot}_p$ is finite.  Here
\be
\Vrt{b}_p = \Vrt{b}_{I_p} =  \l(\Tr\l((b^\ast b)^{p/2}\r)\r)^{1/p}\;.
\ee
When there may be a chance of confusion, we write $\Vrt{\cdot}_{I_p}$ for
$\Vrt{\cdot}_p$.  For $p=1$ this is the trace norm, and if $\Vrt{b}_p<\infty$
for some $p$, then $\Vrt{b}=\lim_{p\to\infty}\Vrt{b}_p$.  It is clear that
$\Vrt{b}_p\le
\Vrt{b}_{p'}$ if $p'\le p$.

The Schatten norms satisfy a H\"older inequality for $1\le r$, namely
\be
\Vrt{ab}_r \le \Vrt{a}_p\Vrt{b}_q\;,\quad {1\over p}+\frac1q =
\frac1r\;.
\ee
More generally, 
\be
\Vrt{a_0\cdots a_n}_r\le \prod^n_{j=0}\Vrt{a_j}_{p_j}\;,\quad
\sum^n_{j=0}\frac1{p_j} = \frac1r\;.
\ee

\mbni{The Beta Function $B_n$.}
Let $\eta_j>0$, $j=0,1\clips,n$.  Then define the beta function $B_n$ as
\be
B_n(\eta_0,\eta_1\clips,\eta_n) = {\prod^n_{j=0} \Ga (\eta_j)\over \Ga
\l(\sum^n_{j=0}\eta_j\r)}\;.
\ee
Here $\Ga(\cdot)$ denotes the gamma function.

We also define $\sigma_n \subset \R^{n+1}$ as the subset 
\be
\sigma_n = \l\{ s:s\in \R^{n+1}\;,\quad 0<s_j\;,\quad
\sum^n_{j=0}s_j=1\r\}\;.
\ee
A natural measure on $\sigma_n$ is $d^{n}s(1)$, as given 
in (IV.12), namely Lebesque measure 
restricted to the $n$-hyperplane $s_0+\cdots+s_n=1$.
Then we claim that 
\be
B_n(\eta_0\clips,\eta_n) 
= \int_{\sigma_n} \l(\prod^n_{j=0}s_j^{-1+\eta_j}\r)d^{n}s(1)\;,
\ee
namely $B_n$ is a Radon transform given by the hyperplane $\sigma_n$.
For $0<\beta$, let $\beta \sigma_n$ denote the set $\sigma_n$ scaled 
by $\beta$ and define the Radon transform 
\be
B_n(\eta_0\clips,\eta_n;\beta) = \int_{\beta
\sigma_n}\l(\prod^n_{j=0}s_j^{-1+\eta_j}\r)d^{n}s(\beta)\;.
\ee
Changing variables, we have
\be
B_n \l(\eta_0\clips,\eta_n;\beta\r) = \beta^{-1+\sum^n_{j=0} \eta_j} B_n
(\eta_0\clips,\eta_j;1)\;,
\ee
or
$$
\int_{\beta \sigma_n} \l(\prod^n_{j=0}s_j^{-1+\eta_j}\r) d^{n}s(\beta) =
\beta^{-1+\sum^n_{j=0} \eta_j} \int_{\sigma_n} \l(
\prod^n_{j=0}s_j^{-1+\eta_j}\r)d^{n}s(1)\;.
$$
Using the representation $\Ga(\eta) = \int^\infty_0 e^{-t} t^{-1+\eta}dt$,
we have
\begin{eqnarray*}
\Ga(\eta_0)\cdots \Ga(\eta_n) &=& \int_{s_n>0} \l(e^{-\sum^n_{j=0}s_j}\r)
\l(\prod^n_{j=0}s_j^{-1+\eta_j}\r)ds_0 ds_1 \cdots ds_n \\
&=&\int^\infty_0 d\beta 
       e^{-\beta} 
\int_{s_n>0}ds_0ds_1\cdots ds_n
\delta(s_0+s_1+\cdots +s_n - \beta)  
\l(\prod^n_{j=0}s_j^{-1+\eta_j}\r)\\
&=&\int^\infty_0 d\beta 
       e^{-\beta}\int_{\beta \sigma_n}
\l(\prod^n_{j=0}s_j^{-1+\eta_j}\r)d^{n}s(\beta) \\
&=& \int^\infty_0 d\beta e^{-\beta} 
\beta^{-1+\sum^n_{j=0}\eta_j} \int_{\sigma_n}
\l(\prod^n_{j=0} s_j^{-1+\eta_j}\r) d^{n}s(1) \\
&=& \Ga \l(\sum^n_{j=0} \eta_j\r) \int_{\sigma_n}
\l(\prod^n_{j=0}s_j^{-1+\eta_j}\r)d^{n}s(1)\;,
\end{eqnarray*}
where in the second to last equality we use (V.24).  Hence 
we have proved (V.22).

We remark that with $|\sigma_n|$ the measure of $\sigma_n$, and
$\beta>0$, we infer from (V.20,24) that
\beq
&&\hskip-.5in |\sigma_n|  =   B_n(1\clips,1) = \frac1{n!} \;,
\quad |\beta \sigma_n|=\frac{\beta^n}{n!}\;, 
\quad \nn\\
&&\hskip-.5in  B_n\l(\frac12,1\clips,1 \r) = {4^nn!\over
(2n)!}\;,\quad {\rm and}\quad B_n \l(\frac12,\frac12,1\clips,1\r) 
= {\pi\over (n-1)!}\;.
\eeq

\subsection{Operator-Valued Generalized Functions}
We formulate here the natural definition of a generalized function in the
non-commutative case, which reduces to a distribution in the classical
case.  A {\it generalized function}, or {\it operator valued distribution}, 
is a bounded, linear transformation between two spaces
$\CH_p=\CH_p(Q)$ for different values of $p$.
 We consider a bounded linear transformation $x$ with
domain
$\CH_{p_1}$ and range contained in $\CH_{p_2}$.  If $p_1=0$ and $p_2\ge 0$,
then $x$ is a bounded linear transformation on $\CH$.  If $p_1>0$ and
$p_2\ge 0$, then $x$ is an unbounded operator on $\CH$ with domain
$\D_{p_1}$.  If $p_1\ge 0$ and $p_2<0$, then $x$ is a sesquilinear form
with domain $\D_{p_2}\times \D_{p_1}$. 

The operator
$(Q^2+I)^{-p/2}$ defines a unitary isomorphism between $\CH_{p_1}$ and
$\CH_{p_1+p}$.  The norm of a bounded, linear transformation $x$ from
$\hpo$ to $\hpt$ is given on $\CH$ by
\be
\Vert x\Vert_{(p_2,p_1)} = \Vert x\Vert_{\hpo\to\hpt} = \Vert
R^{-p_2}x  R^{p_1}\Vert\;.
\ee
We often are interested in the case $p_1\ge 0$ and $p_2< 0$ of generalized
functions, for which
\be
\Vrt{x}_{(-p_2,p_1)} = \Vert x\Vert_{\hpo\to \CH_{-p_2}} = \Vert
R^{p_2 }x R^{p_1 }\Vert\;.
\ee

Let $\T(p_2,p_1)= \T(p_2,p_1;Q)$ denote the space of bounded, linear
transformations from
$\hpo(Q)$ to $\hpt(Q)$.  If $x\in \T(p_3,p_2)$ and $y\in \T(p_2,p_1)$   then
$xy\in
\T(p_3,p_1)$.  Clearly $Q$ is an element of $\T(p-1,p)$, with norm
$\Vert Q\Vert_{(p-1,p)}=1$.

For $\alpha,\beta>0$, any $x\in \T(-\beta,\alpha)$ defines a sesquilinear
form on $\D_\beta\times \D_\alpha$ in $\CH$.  For all $p$, the
space $\D_p\subset
\CH_\infty$, so $x$ is defined on $\CH_\infty\times \CH_\infty$.  Thus for
$0<s,t$, we infer from (IV.16) that
$$
e^{-sQ^2}x e^{-tQ^2}
$$
is bounded.  Thus   for $0<s\le 1$, $0<\alpha\le 1$,
\be
\Vrt{R^{-\alpha} e^{-sQ^2}} \le 2s^{-\alpha/2} \;,\hensp{and}
\Vrt{e^{-sQ^2}xe^{-tQ^2}} \le 4s^{-\beta/2} t^{-\alpha/2}
\Vrt{x}_{(-\beta,\alpha)} \;.
\ee
  We could on this account
define an alternative norm
\be
\vv{x}_{(-p_2,p_1)} = \sup_{0<s,t\le 1} \l(s^{\beta/2}t^{\alpha/2}
\Vrt{e^{-sQ^2}xe^{-tQ^2}}\r)\;,
\ee
and use (V.26) to define a slightly larger space of generalized functions
including $\T(-\beta,\alpha)$, namely the completion of
$\T(-\beta,\alpha)$ in the norm $\vv{x}_{(-\beta,\alpha)}\le
4\Vrt{x}_{(-\beta,\alpha)}$.

\mbni{Definition V.1.a.} {\it A non-commutative generalized function
$x$ is an element of $\T(-\beta,\alpha)$ for some $\alpha,\beta$.  If
$\alpha,\beta\ge 0$, we call $x$ a {\rm vertex} of type
$(\beta,\alpha)$, with respect to $Q$, (or {\rm vertex} for short).

{\bf b.} A regular set of $(n+1)$ ordered vertices $X=\{x_0\clips,x_n\}$,
with respect to $Q$, (for short, a {\it regular set of vertices}) is a set of
vertices $x_j$ of type $(\beta_j,\alpha_j)$, where $\alpha_j,\beta_j$
satisfy the following conditions:
\be
0< \eta_j = 1-\frac12 (\alpha_j+\beta_{j+1})\;,\quad j=0,1\clips,n\;. 
\ee
Here $\beta_{n+1}$ is defined by $\beta_{n+1}=\beta_0$.  

{\bf c.} The {\rm local
regularity exponent} $\eloc$ of the set $X$ is defined by the
\be
0<\eloc = \min_{1\le j\le n}\{\eta_j\}\;,
\ee
and the {\rm global regularity exponent} $\eglo$ of the set $X$ defined by
the mean exponent}
\be
\eglo = {1\over n+1} \sum^{n}_{j=0}\eta_j\;.
\ee
\medbreak

A given vertex $x_j$ is generally an element of several different 
spaces $\T(-\beta,\alpha)$; for instance, $Q$ is an element of
$\T(\mu-1,\mu)$ for every real $\mu$. We say that a set of vertices
$X=\{x_0\clips,x_n\}$
is regular if it satisfies Definition V.1.b for some given set
of $\{-\beta_j, \alpha_j\}$.
Note that
\be
0<\eloc \le \eglo \le 1\;.
\ee
Furthermore, if $\{x_0,x_1\clips,x_n\}$ is a regular set of vertices, then
so is any cyclic permutation
\be
\{ x_j,x_{j+1}\clips, x_n,x_0,x_1\clips,x_{j-1}\}\;.
\ee
The transformations $U(g)$ and $\gamma$ commute with $Q^2$ on $\CH$. 
Thus $U(g):\CH_p\to \CH_p$ and $\gamma:\CH_p\to \CH_p$.  We infer that
$x\in
\T (-\beta,\alpha)$ ensures
\be
x^\gamma=\gamma x\gamma\hensp{and} x^g = U(g)x U(g)^\ast \in
\T(-\beta,\alpha)\;.
\ee
It follows that if $\{x_0\clips,x_n\}$ is a regular set of vertices, then 
so is 
\be
\l\{ x^{g_0}_0,x^{g_1}_1\clips,x^{g_n}_n\r\} 
\ee
for $g_0,g_1\clips g_n\in {\fg }$.  Similarly any of the $x_j$'s may be replaced
by $x^\gamma_j$.

\mbni{Definition V.2.} {\it The {\rm heat kernel regularization} of a regular
set of vertices $X=\{x_0,x_1\clips,x_n\}$ with respect to $Q$ is defined for
$s\in \sigma_n$ by the following sesquilinear form on $\CH\times \CH$,
\be
X(s)=R^{\beta_0}x_0e^{-s_0Q^2} x_1 e^{-s_1Q^2}\cdots
x_ne^{-s_nQ^2}R^{-\beta_0}\;.
\ee
We take $X(s)=0$ for $s\not\in \sigma_n$.}
\medbreak

Note that $s\in \sigma_n$ ensures that each $s_j>0$.  Hence the form
(V.34) is bounded on $\CH\times \CH$, and $X(s)$ uniquely determines a
bounded, linear operator on $\CH$, which we also denote by $X(s)$. 
Furthermore, $\exp(-\beta Q^2)$ is trace class for all $\beta >0$,  so $X(s)$
is a trace class operator on $\CH$.

\mbni{Proposition V.3.} {\it Assume that $X(s)$ is the heat kernel
regularization {\rm (V.34)} of a regular set of vertices with respect to $Q$. 
Then for any $\mu$ in the interval $0<\mu<1$:
\begin{itemize}
\ritem{(i)} The trace norm of $X(s)$ is bounded for $s\in\sigma_n$, as 
defined in (V.21), by
\be
\Vrt{X(s)}_1 \le \l(2\mu^{-(1-\eglo)}\r)^{n+1}\Tr \l( e^{-(1-\mu)Q^2}\r) \l(
\prod^n_{j=0} s_j^{-(1-\eta_j)} \Vrt{x_j}_{(-\beta_j,\alpha_j)}\r)\;,
\ee
with $\eglo$ defined in {\rm (V.32)} and $\eta_j$ in {\rm (V.31)}.

\ritem{(ii)} The map $s\mapsto X(s)$ is continuous from $\sigma_n$ to $I_1$,
the Schatten ideal of trace class operators.  In fact the map is H\"older
continuous with  exponent $\eta'$ less than $\eloc\;$, up to the
boundary of $\sigma_n$.  For the Euclidean
distance
$|s-s'|$ sufficiently small, and with $m_1,m_2$ defined in {\rm (V.51)},
\be 
\Vrt{X(s)-X(s')}_1 \le m_1m_2^{n+1} |s-s'|^{\eta'} \l(\prod^n_{j=0}
s_j^{-1+\eta_j}\r)\l(\sum^n_{j=0} s_j^{-\eta'} \r)
\l(\prod_{j=0}^n||x_j||_{(-\beta_j, \alpha_j)} \r)\;.
\ee
Since $\eta'<\eloc$, the right hand side of {\rm (V.39)} is 
integrable over $s\in \sigma_n$.
\ritem{(iii)} If each $x_j\in \B(\CH)$, then for $s\in\sigma_n$,
\be
\Vrt{X(s)}_1\le \Tr (e^{-Q^2}) \l(\prod^n_{j=0}\Vrt{x_j}\r)\;.
\ee
\end{itemize}
}

\mbni{Proof.} Define the following operators $T_j,S_j$, $j=0,1\clips,n$:
\be
T_j = R^{\beta_j}x_jR^{\alpha_j}\;,\quad S_j = R^{-\alpha_j-\beta_{j+1}}
e^{-s_jQ^2}\;,
\ee
where $\beta_{n+1}:= \beta_0$.  Then $X(s)=T_0S_0T_1S_1\cdots
T_nS_n$.  Each $T_j$ is bounded, and
\be
\Vrt{T_j}=\Vrt{x_j}_{(-\beta_j,\alpha_j)}\;.
\ee
Each $S_j$ is in the Schatten class $I_{s_j^{-1}}$.  In fact, for $0<\mu<1$,
by the H\"older inequality for Schatten norms (V.18),
\beq
\Vrt{S_j}_{s^{-1}} 
&\le & \Vrt{R^{-\alpha_j-\beta_{j+1}}e^{-\mu s_jQ^2}}_{I_\infty} 
  \Vrt{e^{-(1-\mu)s_jQ^2}}_{I_{s^{-1}_j}} \nn\\
&\le & 2(\mu s_j)^{-(\alpha_j +\beta_{j+1})/2} 
  \Tr \l(e^{-(1-\mu)Q^2}\r)\;,
\eeq
where we use the bound (V.28) for the $\Vrt{\cdot}_{I_\infty}$
(operator) norm. Thus using H\"olders inequality (V.19) on $X(s)$ with the
exponent $\infty$ for
$T_j$ and the exponent $s^{-1}_j$ for $S_j$, and using
$\sum^n_{j=0}s_j=1$, we have (with the exponents $\eta_j$ and $\eglo$ 
defined in (V.31,32) )
$$
\Vrt{X(s)}_1\le 2^{n+1}\mu^{-(n+1)(1-\eglo)}\Tr \l( e^{-(1-\mu)Q^2}\r)
\l(\prod^n_{j=0} s_j^{-(1-\eta_j)} \Vrt{x_j}_{(-\beta_j,\alpha_j)}\r)
\;,
$$
which is  the bound (V.38).    Note that if $x_j\in \B(\CH)$,
$j=0,1\clips,n$, then
$\alpha_j=\beta_j = 0$, $j=0,1\clips,n$ and we can take $\mu=0$ in the
bound on $S_j$.  
In fact, we have $\Vrt{T_j} = \Vrt{x_j}$ and  
$\Vrt{S_j}_{I_{s_j^{-1}} }=\l(\Tr (e^{-Q^2})\r)^{s_j}$.  
Thus the factor
$2^{n+1}$ in (V.35) can be replaced by 1.  Also $\eta_j=1=\eglo$, 
for all $j$,
so in this case we have (V.36).  This completes the proof of (i) and (iii).

(ii).  In order to establish continuity of $s\to X(s)$ at $s$, we consider
$X(s)-X(s')$ where $s$, $s'\in \sigma_n$ and where $s'$ is sufficiently
close to $s$.  Let $\ts = \min_js_j$; note $s\in \sigma_n$ ensures $\ts>0$.
We suppose that $s'$ lies in the neighborhood of $s$ defined by
\be
\sup_j |s_j-s'_j|< \ep \ts\;.
\ee
We take $0<\ep<1$.  Thus
\be
|s_j - s'_j| s_j^{-1} \le \ep\;,\quad\hensp{and} s'_j\ge (1-\ep)s_j\;,\quad
j=0,1\clips,n\;.
\ee
The first inequality in (V.45) is a consequence of 
$$
|s_j-s'_j|\le \ep \ts \le \ep s_j\;,
$$
while the second inequality follows by
$$
s'_j = s_j +(s'_j-s_j)\ge s_j - |s_j-s'_j|\ge (1-\ep) s_j\;.
$$
We now show that on this set, and for any $\eta' <\eloc$,
\be
\Vrt{X(s)-X(s')}_1 \le 0(|s-s'|^{\eta'})\;. 
\ee
In other words, $s\mapsto X(s)$ is H\"older continuous with any 
exponent $\eta' <\eloc$. 

 We require a slightly different set of bounds from (V.43).
Let us denote $S_j(s)$ by $S_j$ and $S_j(s')$ by $S'_j$.  The bound (V.45)
ensures
\beq
\Vrt{S'_j}_{s^{-1}_j} &\le & \Vrt{R^{-\alpha_j - \beta_{j+1}} e^{-\mu
s'_jQ^2}} \, \Vrt{e^{-(1-\mu)s'_jQ^2}}_{I_{s^{-1}_j}} \nn\\
&\le & 2(\mu s'_j)^{-(\alpha_j+\beta_{j+1})/2} \l(\Tr
\l(e^{-(1-\mu)(s'_j/s_j)Q^2}\r)\r)^{s_j} \nn\\
&\le & 2\l( (1-\ep)\mu s_j\r)^{-(\alpha_j+\beta_{j+1})/2} \l(\Tr
\l(e^{-(1-\mu)(1-\ep)Q^2} \r)\r)^{s_j}\;.
\eeq
Furthermore we establish for $\eta' <\eloc$
\be
\Vrt{S_j -S'_j}_{s^{-1}_j} \le M s_j^{-1+(\eta_j-\eta')} |s_j-s'_j|^{\eta'}
\l(\Tr \l(e^{-(1-\ep)(1-\mu)Q^2}\r)\r)^{s_j}\;,
\ee
where
\be
M=2\mu^{-2+\eta_j} \ep^{1-\eta'} \; . 
\ee
Let $s_j(\alpha)=\alpha s_j + (1-\alpha) s'_j$ interpolate between $s_j$
and $s'_j$.  Then
\beq
S_j - S'_j &=& S_j(s_j(\alpha)) \big\vert^1_0 = \int^1_0
{dS_j(s_j(\alpha))\over d\alpha} d\alpha\nn\\
&=& \l(s'_j-s_j\r) \int^1_0 Q^2  S_j(s_j(\alpha)) d\alpha\nn\\
&=& (s'_j - s_j) \int^1_0 Q^2 R^{-(\alpha_j+\beta_{j+1})}
e^{-s_j(\alpha)Q^2} d\alpha\;.
\eeq
Thus
\beq
\Vrt{S_j-S'_j}_{s^{-1}_j} &\le & |s_j-s'_j| \int^1_0 \Vrt{Q^2
R^{-(\alpha_j+\beta_{j+1})} e^{-\mu s_j(\alpha)Q^2}} \,
\Vrt{e^{-(1-\mu)s_j(\alpha)Q^2}}_{s^{-1}_j} d\alpha\nn\\
&\le & |s_j-s'_j| \int^1_0 2(\mu
s_j(\alpha))^{-(2+\alpha_j+\beta_{j+1})/2} \l(\Tr \l(
e^{-(1-\mu)s_j(\alpha)Q^2}\r)\r)^{s_j} d\alpha\;.\nn\\
&&
\eeq
For $0\le \alpha\le 1$,
\be
s_j(\alpha)\ge (1-\ep)s_j\;.
\ee
Thus
\be
\Vrt{S_j-S'_j}_{s^{-1}_j} \le 2|s_j-s'_j| (\mu
s_j)^{-(2+\alpha_j+\beta_{j+1})/2} \l(\Tr
\l(e^{-(1-\ep)(1-\mu)Q^2}\r)\r)^{s_j}\;.
\ee
Write $s_j^{-(2+\alpha_j+\beta_{j+1})/2} =
s_j^{-(1-\eta')}s_j^{-(\alpha_j+\beta_{j+1}+2\eta')/2}$, and use (V.47). 
Thus
\be
\Vrt{S_j-S'_j}_{s^{-1}_j} \le 2|s_j-s'_j|^{\eta'} s_j^{-(\alpha_j
+\beta_{j+1} +2\eta')/2} \ep^{(1-\eta')} \mu^{-(2+\alpha_j+\beta_{j+1})/2} 
\l(\Tr \l( e^{-(1-\ep)(1-p)Q^2}\r)\r)^{s_j}\;.
\ee
With $\eta_j$ given in (V.27) and $M$ of (V.49) we have (V.48).

Now  write
\be
X(s) - X(s') = \sum^n_{j=0} T_0S_0T_1S_1\cdots T_j(S_j-S'_j)T_{j+1}
S'_{j+1}\cdots T_nS'_n \;.
\ee
Estimate $\Vrt{X(s)-X(s')}_1$ using H\"older's inequality in the $I_p$
norms, as in the derivation of the bound on $X(s)$.  Use the operator norm
on each $T_j$ and the $\Vrt{\cdot}_{s^{-1}_j}$-Schatten norm on $S_j$, on 
$S'_j$, or on $S_j-S'_j$.

We obtain from (V.38, 43, 47, 48) the following bound on (V.55):
\be
\Vrt{X(s)-X(s')}_1 \le m_1m_2^{n+1} |s-s'|^{\eta'} \l(\prod^n_{j=0}
s_j^{(-1+\eta_j)}\r) \l(\sum^n_{j=0} s_j^{-\eta'}\r)
\l(\prod_{j=0}^n||x_j||_{(-\beta_j, \alpha_j)} \r)\;.
\ee
Here
\be
m_1=\mu^{-1} \ep^{(1-\eta')} \Tr\l(e^{-(1-\ep)(1-\mu)Q^2}\r)\;,\hensp{and}
m_2=2\l((1-\ep)\mu\r)^{-(1-\eglo)}\;.
\ee
This completes the proof of the proposition.

\mbni{Corollary V.4.} {\it Assume $X(s)$ is the heat kernel regularization
{\rm (V.37)} for a regular set of vertices with respect to $Q$, and that
$\exp(-\beta Q^2)$ is trace class for all $\beta>0$.  Then with 
$d^{n}s=d^{n}s(1)$ defined in (IV.12),  
\begin{itemize}
\ritem{(i)} The Radon transform
\be
\hx =\int_{\sigma_n} X(s)d^{n}s 
\ee
exists and is a trace class operator on $\CH$.
\ritem{(ii)}  The 
 trace and integration of $\gamma U(g)X(s)$ commute, namely 
\be
\Tr (\gamma U(g)\hx ) = \int_{\sigma_n} \Tr (\gamma U(g) X(s)) d^{n}s\;.
\ee

\ritem{(iii)} For $s\in \sigma_n$, defined in (V.21), and for $0<\mu<1$, 
the quantity
\be
\Tr \l(\gamma U(g) X(s)\r) = \Tr \l(\gamma U(g) e^{-\mu s_n Q^2} x_0
e^{-s_0 Q^2} x_1 e^{-s_1 Q^2} \cdots 
e^{-s_{n-1} Q^2} x_n e^{-(1-\mu)s_nQ^2}\r)\;,
\ee
is independent of $\mu$.  Thus we define
\beq 
\Tr \l( \gamma U(g)x_0 e^{-s_0Q^2}\cdots x_ne^{-s_nQ^2}\r)
& =& \lim_{\mu\to 0+} \Tr \l(\gamma U(g) e^{-\mu s_nQ^2} x_0e^{-s_0Q^2}
    \cdots x_n e^{-(1-\mu)s_nQ^2}\r) \nn\\
&= &\Tr (\gamma U(g) X(s))\;.
\eeq  
In summary, we write $\Tr (\gamma U(g)X)$ as
\be
\lra{x_0,x_1\clips,x_n;g}_n = \int_{\sigma_n} \Tr \l(\gamma U(g)
x_0e^{-s_0Q^2}\cdots x_ne^{-s_nQ^2}\r) d^{n}s\;.
\ee
\ritem{(iv)} Given $0<\mu<1$, the expectation {\rm (V.62)} satisfies
\be
\l|\lra{x_0,x_1\clips,x_n;g}_n\r| \le m_1 m^{n+1}_2 \Ga \l(
(n+1)\eglo\r)^{-1}
\l(\prod^n_{j=0} \Vrt{x_j}_{(-\beta_j,\alpha_j)}\r)\;, 
\ee
for constants
\be
m_1 = \Tr \l(e^{-(1-\mu)Q^2}\r)\;,\quad m_2 = 2\Ga (\eloc)
\mu^{-(1-\eglo)} \;.
\ee
\ritem{(v)} If all $x_i\in \B(\CH)$, so $\alpha_i =\beta_i = 0$,
$i=1,2\clips,n$, then 
\be
\l|\lra{x_0\clips,x_n;g}_n\r| \le {1\over n!} \Tr \l(e^{-Q^2}\r)
\l(\prod_{j=0}^n \Vrt{x_j} \r)\;. 
\ee
\ritem{(vi)} The expectation $\lra{x_0,x_1\clips,x_n;g}_n$ satisfies  the
symmetries {\rm (IV.15--19)}.  Thus,
\beq
\hskip-.5in \lra{x_0 \clips, x_n ;g}_n 
&=& \lra{x^g_0 \clips, x^g_n;g}_n 
= \lra{x^\gamma_0 \clips,x_n^\gamma;g}_n
\;,
\\
\hskip-.5in \lra{x_0\clips,x_n;g}_n &=& \lra{  x_n^{g^{-1} \gamma}
,x_0,x_1\clips,x_{n-1}}_n \label{V.perm}\\
\hskip-.5in \lra{x_0\clips, x_n; g}_n &=& \sum^{n+1}_{j=1}
\lra{x_0\clips,x_{j-1} , I,x_j
\clips,x_n;g}_{n+1}\label{V.sum}
\eeq
and if both $\l\{ x_0,x_1\clips,  x_{j-1},Qx_j,x_{j+1}\clips,x_n\r\}$ 
and 
$\l\{ x_0,x_1\clips,  x_{j-1}, x_jQ,x_{j+1}\clips,x_n\r\}$ are also a
regular set of vertices for $j=0,1\clips,n$, then
\be
\lra{d\hx ;g}=\sum^n_{j=0} \lra{x_0^\gamma,x^\gamma_1\clips, x^\gamma_{j-1} ,
dx_j,x_{j+1}\clips, x_n;g}_n=0\;. \label{V.leib}
\ee
\end{itemize}
}

\mbni{Proof.} (i--ii) We showed in Propostion V.3 i--ii, that 
$s\mapsto X(s)$ is H\"older continuous from $\sigma_n$ to the Schatten 
class $I_1$. Using this bound, we infer that the Radon transform
$\int_{\sigma_n}X(s)ds$ exists on the unit hyperplane, and the integral
and trace commute
$$
\int_{\sigma_n}\Tr\l(X(s)\r) d^{n}s = \Tr \l(\int_{\sigma_n}X(s)d^{n}s\r)\;.
$$
This is also the case with $X(s)$ replaced by $TX (s)$, for $T\in \B(\CH)$. 
In particular (V.59) holds.

(iii) We evaluate $\Tr\l(\gamma U(g) X(s)\r)$ for $s\in \sigma_n$.  With the
notation (V.41), 
$\Tr\l(\gamma U(g) X(s)\r)=\Tr\l(\gamma U(g)T_0S_0\cdots T_nS_n\r)$.  
Each $S_j$ is trace class, and for $0<\mu<1$, both $e^{-\mu Q^2}$ 
and $e^{-(1-\mu)Q^2}$ are trace class.  Thus  
\begin{eqnarray*}
\Tr \l(\gamma U(g)X(s)\r) &=& \Tr \l( \gamma U(g)T_0 S_0\cdots
T_nS_n e^{\mu s_nQ^2}e^{-\mu s_nQ^2}\r)\\
 &=& \Tr \l( e^{-\mu s_nQ^2}\gamma U(g)T_0 S_0\cdots
T_nS_n e^{\mu s_nQ^2}\r)\\
&=& \Tr \l( \gamma U(g)e^{-\mu s_nQ^2}R^{\beta_0}x_0R^{\alpha_0}S_0
     \cdots T_nR^{-\alpha_n}R^{-\beta_0}e^{-(1-\mu)s_nQ^2}\r) \;.
\end{eqnarray*}
Here we use the fact that $Q^2$ commutes with $\gamma$ and with $U(g)$. 
Also, $R^{\beta_0}$ commutes with $Q^2$, with $\gamma$, and with $U(g)$.
Therefore we can also cyclically permute $R^{\beta_0}$ in the trace to yield 
\beq
\Tr \l(\gamma U(g) X(s)\r) &=& 
\Tr \l( \gamma U(g)e^{-\mu s_n Q^2}R^{\beta_0}x_0R^{\alpha_0}S_0
     \cdots T_nR^{-\alpha_n}R^{-\beta_0}e^{-(1-\mu)s_nQ^2}\r)\nn\\
&=& \Tr \l( \gamma U(g)e^{-\mu s_n Q^2}x_0R^{\alpha_0}S_0
     \cdots T_nR^{-\alpha_n}e^{-(1-\mu)s_n Q^2}\r)\nn\\
&=& \Tr \l( \gamma U(g) e^{-\mu s_n Q^2} x_0 e^{-s_0Q^2}x_1e^{-s_1Q^2}
     \cdots e^{-s_{n-1}Q^2}x_n e^{-(1-\mu)s_nQ^2}\r) \;.\nn
\eeq
Since this is true for any $\mu$ in the range,  $\Tr (\gamma
U(g)X(s))$ is independent of $\mu$, and we have established (V.61).  This  
completes the proof of (iii).

(iv--v) Note that
\beq
\l| \lra{x_0,x_1\clips,x_n;g}\r| &=& |\Tr (\gamma U(g)\hx)| 
\le \Vrt{\gamma U(g)} \, \Vrt{\hx}_1\nn\\
&\le & \int_{\sigma_n} \Vrt{X(s)}_1 d^{n}s\;.
\eeq
Thus the bounds (V.63--65) are established by integrating (V.38) over
$\sigma_n$, and using the definition of $B_n$, see (V.20).  Note that
$\eta_j<1$, and $\Ga(\eta_j)$ is monotonic decreasing on (0,1). Thus
$\Ga(\eta_j)\le \Ga(\eloc)$.

(vi) The symmetries (V.66, 67, 69) can be established as the corresponding
symmetries for $\Tr (\gamma U(g)X(s))$, expressed as (V.60).  Then we
integrate over
$\sigma_n$.  In the case of (V.68), we follow the argument in \S IV leading
to (IV.22).

\subsection{Interpolation Spaces}
In this section we define certain Banach algebras $\jba$ consisting of
operators
$b\in
\bh$ with a bounded fractional derivative.  We call these spaces {\it
interpolation spaces}.  These spaces are a natural framework for the study
of the JLO cochain, and in \S VI we introduce algebras $\A\subset \jba$ to
study $\tau^\jlo$ on $\C(\A)$.

As in previous sections let
$Q=Q^\ast$ with domain
$\D=\CH_1\subset \CH$, and let $R=(Q^2+I)^{-1/2}$.  We say that $b$ has a
bounded derivative of order $\alpha>0$, if $b$ is a bounded linear
transformation on $\CH_\alpha$.  In other words, the form
$R^{-\alpha}bR^\alpha$ defines a bounded element of $\bh$, which we denote
\be
R^{-\alpha}bR^\alpha\in \bh\;.
\ee
In the notation of \S V.2, $b\in \T(\alpha,\alpha)$.  Let us define
$\fj_\alpha$ as $\bh\cap \T(\alpha,\alpha)$ with the norm
\be
\Vrt{b}_{\fj_\alpha} = \l(\Vrt{b}+\Vrt{R^{-\alpha}b R^\alpha}\r)\;.
\ee
If $b\in \fj_\alpha$,
then  as a blinear form on $\D_\alpha\times \D_\alpha$, 
\be
\lra{R^{-\alpha} g,bR^\alpha f} = \lra{g,R^{-\alpha}bR^\alpha f} 
\ee
and
\be
\l|\lra{R^{-\alpha}g,bR^\alpha f}\r| = \Vrt{g}\, \Vrt{f}\,
\Vrt{b}_{\fj_\alpha}\;.
\ee
Since $Q$ is self-adjoint, $bR^\alpha\in \D\l( (R^{-\alpha})^\ast
\r)=\D(R^{-\alpha})$.  It follows that if $a,b\in \fj_\alpha$, then $ab\in
\fj_\alpha $ and
\beq
\Vrt{ab}_{\fj_\alpha} &=& \Vrt{ab} + \Vrt{R^{-\alpha}a R^\alpha R^{-\alpha}
bR^\alpha} \nn\\
&\le & \Vrt{a}\, \Vrt{b} + \Vrt{R^{-\alpha} aR^\alpha}\, \Vrt{
R^{-\alpha}bR^\alpha} \le \Vrt{a}_{\fj_\alpha} \Vrt{b}_{\fj_\alpha}\;.
\eeq
Thus $\fj_\alpha$ is a Banach algebra.

It is useful to characterize fractional differentiability not by the property
(V.71), but rather by some properties of $db=Qb-b^\gamma Q$.  The reason
is that the expression $db$ arises in the  geometric
interpretation of   $\A$, as studied in \S IV.

We now define such a family   of subalgebras of $\fj_{1-\beta}$, which we
denote by 
$\jba\subset \fj_{1-\beta}$.  Let
$0\le \alpha,\beta$ and $0\le \alpha+\beta<1$.  If $b\in \bh$, then $b\in
\jba$ if $db \in \T(-\beta,\alpha)$.  In other words, $\jba$ consists of
elements $b$ of $\bh$ such that (in the notation of \S V.3) $db$ is a vertex
of type $(\beta,\alpha)$.  We give $\jba$ the norm
\beq
\Vrt{b}_{\jba} &=& \Vrt{b}+c_{\alpha+\beta} \Vrt{db}_{(-\beta,\alpha)}
\nn\\ &=& \Vrt{b} +c_{\alpha+\beta} \Vrt{R^\beta dbR^\alpha}\;.
\eeq
Here we define $c_\mu$ for $0\le \mu<1$ by
\be
c_\mu = \sup_{0\le \delta\le 1} 2\delta \int^\infty_0 (1+t^{-1})^{\delta/2}
(1+t)^{-1-(1-\mu)/2}dt\;. 
\ee
Note that $c_\mu$ is greater than the $\delta=1$ value in (V.77),
which in turn is monotonic in $\mu$.  Thus
\be
c_\mu\ge 2\int^\infty_0 t^{-1/2}(1+t)^{-1} dt = 2\pi >1\;.
\ee
Also $c_\mu$
diverges logarithmically as
$\mu
\nearrow 1$.

\goodbreak
\mbni{Proposition V.5.} {\it Let $0\le \alpha,\beta$ and $0\le
\alpha+\beta<1$.  Then
\begin{itemize}
\ritem{i)}  
$\jba \subset \fj_\delta$ for all \be
0\le \delta \le 1-\beta\;.
\ee
In this case
$$
\Vrt{b}_{\fj_\delta} \le 2\Vrt{b}_\jba\;.
$$
\ritem{ii)} Let $\delta$ satisfy $-(1-\alpha)\le \delta\le 1-\beta$.  Then
\be
\jba \subset \T(\delta,\delta)\;.
\ee
Furthermore,
\be 
\Vrt{b}_{(\delta,\delta)}  = \Vrt{R^{-\delta} bR^\delta}  \le  \Vrt{b}
+\frac12 c_{\alpha+\beta}\Vrt{db}_{(-\beta,\alpha)}  
\le   \Vrt{b}_{\jba} \;. 
\ee  
\ritem{iii)} If $a\in \jba$, then $a\in \T(-\beta,-\beta)\cap
\T(\alpha,\alpha)$.  Also if $a,b\in \jba$, then both $(da)b$ and $a(db)$ are
elements of $\T(-\beta,\alpha)$.  Also
\be
\Vrt{(da)b}_{(-\beta,\alpha)} \le \Vrt{da}_{(-\beta,\alpha)}
\Vrt{b}_{(\alpha,\alpha)} 
\ee
and
\be 
\Vrt{a(db)}_{-\beta,\alpha)} \le   \Vrt{a}_{(-\beta,-\beta)}
\Vrt{db}_{(-\beta,\alpha)} \;.
\ee 
 \end{itemize} }
 
\mbni{Corollary V.6.} {\it Let $a,b\in \jba$, with $0\le
\alpha,\beta$, and  $\alpha+\beta<1$.  Then
\begin{itemize}
\ritem{(i)} Leibniz Rule: The relation 
\be
d(ab) = (da)b+a^\gamma(db)\;, 
\ee
is an identity of elements in $\T(-\beta,\alpha)$, namely between vertices
$d(ab),(da)b$, and $a^\gamma (db)$ of type $(\beta,\alpha)$.

\ritem{ii)} The space $\jba$ is a Banach algebra, so  
\be
\Vrt{ab}_{\jba} \le \Vrt{a}_{\jba} \Vrt{b}_{\jba} \;.
\ee
\end{itemize} }

\medbreak
For $t\ge 0$, introduce the operators $R(t) = (Q^2 +(1+t)I)^{-1/2}$ and
$R=R(0)$.  Then the spectral theorem ensures that for $\mu \ge 0$,
\be
\Vrt{R(t)^\mu} \le (1+t)^{-\mu/2}\;,\quad \Vrt{R^{-\mu}R(t)^\mu}\le
1\;,\quad \hensp{and} \Vrt{QR(t)}\le 1\;.
\ee
We use a standard representation for $R^\mu$, $0<\mu<2$, which is a
consequence of the Cauchy integral theorem applied to the function
$z^{-\mu/2}$, namely
\be
R^\mu = {\sin (\pi \mu/2)\over \pi} \int^\infty_0 t^{-\mu/2} R(t)^2 dt\;.
\ee

\mbni{Proof of Proposition V.5 and Corollary V.6.} (i) We estimate the
$\fj_\delta$ norm of
$b\in \jba$. Let 
$\D_2=\D(Q^2)= \hbox{Range} (R(t)^2)$. On the form domain $\D_2\times
\D_2$,  and for $0\le \delta\le 1-\beta$ we write
\be
\Vrt{b}_\jd = \Vrt{b}+\Vrt{R^{-\delta} bR^\delta} \le
2\Vrt{b}+\Vrt{R^{-\delta}[b,R^\delta]}\;.
\ee
We now study $R^{-\delta}[b,R^\delta]$.  Using (V.87),
\be
[b,R^\delta] = {\sin (\pi \delta/2)\over \pi} \int^\infty_0 t^{-\delta/2}
[b,R(t)^2] dt\;.
\ee
On $\D_2\times \D_2$, 
\beq
[b,R(t)^2 ] &=& R(t)^2 [R(t)^{-2} ,b] R(t)^2\nn\\
&=& R(t)^2 (Qdb+db^\gamma Q)R(t)^2\;.
\eeq
Here $b^\gamma = \gamma b\gamma$, and $(db)^\gamma = -db^\gamma$. 
Hence
\be
R^{-\delta}[b,R^\delta] = {\sin(\pi \delta/2)\over \pi} \int^\infty_0
t^{-\delta/2} R^{-\delta} R(t)^2 (Qdb+db^\gamma Q) R(t)^2 dt\;.
\ee
We can bound (V.91) using (V.86). We use the following to estimate the first
term on the right of (V.91),
\beq
 \Vrt{R^{-\delta} R(t)^2 QdbR(t)^2} &\le & \Vrt{R^\delta R(t)^2
QR^{-\beta}}
\,
\Vrt{db}_{(-\beta,\alpha)} \Vrt{R^{-\alpha}R(t)^2} \nn\\
  &\le & (1+t)^{-(1-\beta-\delta)/2}(1+t)^{-(2-\alpha)/2}
\Vrt{db}_{(-\beta,\alpha)}\;,\nn\\
&&
\eeq
provided $\delta\le 1-\beta$, and $\alpha\le 2$, as assumed in (V.79). 
Furthermore
$db^\gamma = -(db)^\gamma$, and the unitarity of $\gamma$, along with
$\gamma R=R\gamma$ ensures that for all $(p,q)$,
\be
\Vrt{db^\gamma}_{(p,q)} = \Vrt{db}_{(p,q)} \;.
\ee
We therefore estimate in a similar fashion the second term on the right of
(V.91) namely
\be
\Vrt{R^{-\delta}R(t)^2 db^\gamma QR(t)^2} \le (1+t)^{-(2-\delta-\beta)/2}
(1+t)^{-(1-\alpha)/2} \Vrt{db}_{(-\beta,\alpha)} \;, 
\ee
provided $\beta+\delta\le 2$ and $\alpha\le 1$.  But $\beta+\delta\le 1$ by
(V.79); also $0\le \alpha$, $\beta$ and $\alpha+\beta<1$ ensures
$\alpha<1$.  Hence (V.94) does hold.  

Using (V.92, 94), we bound (V.91). 
There are two similar bounds for the two terms in (V.91).  We also use
$\sin x\le x$ for $0\le x\le \pi/2$.  Thus
\beq
\Vrt{R^{-\delta}[b,R^\delta]} &\le & \delta \int^\infty_0 t^{-\delta/2}
(1+t)^{-1-(1-\alpha-\beta)/2+\delta/2} dt \Vrt{db}_{(-\beta,\alpha)} \nn\\
&\le &\frac12 c_{\alpha+\beta} \Vrt{db}_{(-\beta,\alpha)}\;.
\eeq
Here $c_{\alpha+\beta}$ is defined in (V.77), and is relevant since both
$0\le \alpha+\beta<1$ and $0\le \delta\le 1-\beta\le 1$.  Hence we
conclude that $b\in \jd$ for $0\le \delta\le 1-\beta$.

To estimate the norm $\Vrt{b}_\jd$, using (V.88) we have 
\beq
\Vrt{b}_\jd &  \le &
2\Vrt{b}+\Vrt{R^{-\delta}[b,R^\delta]} \nn\\
&\le & 2\Vrt{b} + \frac12 c_{\alpha+\beta}\Vrt{db}_{(-\beta,\alpha)} \nn\\
&\le & 2\Vrt{b}_\jba\;.
\eeq
This completes the proof of that $\jba\subset \fj_\delta$, and hence of part
(i) of the proposition.  

(ii)  We have also proved part (ii) in case $0\le \delta\le
1-\beta$.  In fact
\beq
\Vrt{b}_{(\delta,\delta)} &=& \Vrt{R^{-\delta} bR^\delta} \le
\Vrt{b}+\Vrt{R^{-\delta}[b,R^\delta]} \nn\\
&\le & \Vrt{b} +\frac12 c_{\alpha+\beta} \Vrt{db}_{(-\beta,\alpha)} \le
\Vrt{b}_\jba\;.
\eeq
Thus to complete the proof of (ii), we need to verify the case $\alpha -1\le
\delta\le 0$.  In that case,  we show equivalently that $b\in 
\T(-\delta,-\delta)$ for
$0\le
\delta\le 1-\alpha$.  Thus we need to verify that $R^\delta bR^{-\delta}$ is
bounded.  Write
\beq
R^\delta bR^{-\delta } &=& b+[R^\delta,b]R^{-\delta}\nn\\
&=& b-{\sin (\pi \delta/2)\over \pi} \int^\infty_0 t^{-\delta/2} R(t)^2
(Qdb+db^\gamma Q) R(t)^2R^{-\delta} dt\;.\nn\\
&&
\eeq
Now we use the estimates (V.86), which yield 
\be
\Vrt{R(t)^2 QdbR(t)^2R^{-\delta}} \le
(1+t)^{-1-(1-\delta-\beta)/2+\delta/2} 
  \Vrt{db}_{(-\beta,\alpha)} \;,
\ee
 as long as both $\beta\le 1$ and $\alpha+\delta\le 2$.  Since  we assume
$0\le
\alpha,\beta$ and $\alpha+\beta<1$, it follows that $\beta<1$.  Also
$\alpha<1$, and since 
$0\le \delta\le 1-\alpha$, we infer that  $\alpha+\delta\le 1$.  Thus both
conditions are met and  (V.99) holds.  Likewise 
\be
\Vrt{R(t)^2 db^\gamma QR(t)^2R^{-\delta}} \le
(1+t)^{-1-\alpha/2+\gamma/2} \Vrt{db}_{(-\beta,\alpha)}\; , 
\ee
if both $\beta\le 2$  and $\alpha+\delta\le 1$.  Both these conditions also
  hold.  Thus from (V.98--100) we infer that 
\beq
\Vrt{R^\delta  bR^{-\delta}} = \Vrt{b}_{(-\delta,-\delta)} &\le & \Vrt{b}
+ \delta \int^\infty_0 t^{-\delta/2}
(1+t)^{-1-(1-(\alpha+\beta))/2+\delta/2} dt \Vrt{db}_{(-\beta,\alpha)}
\nn\\
&\le & \Vrt{b} +\frac12 c_{\alpha+\beta} \Vrt{db}_{(-\beta,\alpha)} \;.
\eeq
Hence $b\in \T(-\delta,-\delta)$ and (V.80--81) hold  as claimed.

(iii) Let us assume $a\in \jba$.  Then from (ii), and the restrictions $0\le
\alpha,\beta$ and $\alpha+\beta<1$, we infer
\be
a \in \T(-\beta,-\beta) \cap \T(\alpha,\alpha)\;. 
\ee
 Thus we can estimate
$(da)b$ as a map from $\CH_\alpha$ to  $\CH_{-\beta}$ as 
$$
\Vrt{(da)b}_{(-\beta,\alpha)} \le \Vrt{da}_{(-\beta,\alpha)}
\Vrt{b}_{(\alpha,\alpha)}\;,
$$
showing (V.82).  Likewise, since $\gamma$ commutes with $Q$,
\be
\Vrt{\alpha^\gamma}_{(-\beta,-\beta)} = \Vrt{a}_{(-\beta,\beta)}\;,
\ee
and
\be
\Vrt{\alpha^\gamma(db)}_{(-\beta,\alpha)} \le
\Vrt{a^\gamma}_{(-\beta,-\beta)} \Vrt{db}_{(-\beta,\alpha)} \le
\Vrt{a}_{(-\beta,\beta)} \Vrt{db}_{(-\beta,\alpha)}\;,
\ee
which is (V.83).  We therefore conclude that $(da)b$ and $a^\gamma (db)$
are both elements of $\T(-\beta,\alpha)$.  This completes the proof of the
proposition.

To establish the corollary, note $b\in \T(\alpha,\alpha)$, $Q\in
\T(\alpha-1,\alpha)$ and $a^\gamma\in \T(\alpha-1,\alpha-1)$, according
to (V.80).  Therefore $a^\gamma Qb\in \T(\alpha-1,\alpha)$.  The important
conclusion here  is that $a^\gamma Qb$ is defined as a sesquilinear form on
$\CH\times \CH$ with some domain; in fact the domain is
$\D_{1-\alpha}\times \D_\alpha$.  But $\beta<1-\alpha$, so
$\D_{1-\alpha}\subset \D_\beta$, and  $\D_{1-\alpha}\times
\D_\alpha\subset \D_\beta\times \D_\alpha$, which is contained in  the
domain of
$(da)b$ and $a^\gamma(db)$.  Furthermore $Qab$ and $(ab)^\gamma Q$ are
both forms on the  domain $\D_1\times \D_1\subset
\D_{1-\alpha}\times \D_\alpha$.  Thus on $\D_1\times \D_1$,  we have
the identity
\beq
d(ab) &=& Qab-(ab)^\gamma Q\nn\\
&=& Qab-a^\gamma Qb+a^\gamma Qb-a^\gamma b^\gamma Q\nn\\
&=& (da)b+a^\gamma (db)\;. 
\eeq
However, by Proposition V.5.iii, each term on  the right side of (V.105)
extends by continuity to
$\D_\beta\times \D_\alpha$.  Thus $d(ab)$ also extends by continuity  to
this domain, and the identity (V.84) holds in $\T(-\beta,\alpha)$.  We have
therefore demonstrated the Leibniz rule (V.84) as an identity on
$\T(-\beta,\alpha)$.

Finally we estimate $\Vrt{ab}_\jba$ for $a,b\in \jba$.  Using (V.81--84),
and the definition (V.76) of the norm on $\jba$, we conclude
\beq
\Vrt{ab}_{\jba} &=& \Vrt{ab}+c_{\alpha+\beta}
\Vrt{d(ab)}_{(-\beta,\alpha)} \le \Vrt{a}\,\Vrt{b} +c_{\alpha+\beta} \l(
\Vrt{(da)b}_{(-\beta,\alpha)} + \Vrt{a(db)}_{(-\beta,\alpha)} \r)\nn\\
&\le & \Vrt{a}\, \Vrt{b}+c_{\alpha+\beta} \Vrt{da}_{(-\beta,\alpha) } \l(
\Vrt{b}+\frac12 c_{\alpha+\beta} \Vrt{db}_{(-\beta,\alpha)} \r)\nn\\
&&  +
c_{\alpha+\beta} \Vrt{db}_{(-\beta,\alpha)} \l( \Vrt{a}+\frac12
c_{\alpha+\beta} \Vrt{da}_{(-\beta,\alpha)}\r)\nn\\ &=&
\l(\Vrt{a}+c_{\alpha+\beta}\Vrt{da}_{(-\beta,\alpha)} \r) \l(
\Vrt{b}+c_{\alpha+\beta} \Vrt{db}_{(-\beta,\alpha)} \r)  = \Vrt{a}_{\jba}
\Vrt{b}_\jba\;.
\eeq 
Thus $\jba$ is a Banach algebra, and the proof of the
corollary is complete.

\subsection{Generalized Schatten  Classes}
In \S V.4 we introduced the spaces $\T(p_2,p_2)$ of generalized functions
as bounded, linear transformations from $\hpo$ to $\hpt$.  It is convenient
to introduce subspaces of $\T(p_2,p_1)$ which are Schatten  $I_p$ classes,
with
$\T(p_2,p_1)$ being the $I_\infty$ case.   We measure $I_p$ size in terms
of the Schatten norm (V.17).    We say $R^{-p_2}x
R^{p_1}\in I_p$, if the bilinear form $R^{-p_2}x R^{p_1}$ uniquely
determines an operator in $\bh$ which belongs to the Schatten ideal
$I_p$.  Thus for $1\le p$, define the generalized Schatten class 
\be
\T(p_2,p_1;p) = \l\{ x:x\in \T(p_2,p_1)\;,\quad R^{-p_2}x R^{p_1}\in
I_p\r\}\;.
\ee
Let $\T(p_2,p_1;p)$ be a normed space with norm
\be
\Vrt{x}_{\T(p_2,p_1;p)} = \Vrt{R^{-p_2} x R^{p_1}}_{I_p}\;.
\ee

The norms $\T(p_2,p_1;p)$ satisfy a H\"older inequality, as a consequence
of the inequality (V.18) for Schatten class $I_p$ norms.

\mbni{H\"older Inequality:}  Let $x_j\in \T(\alpha_j,\alpha_{j+1};p_j)$,
$j=0,1\clips, n$, where $1\le p_j$,  $\sum^n_{j=0} p_j^{-1} = p^{-1} \le 1$.
Then
$x_0 x_1\cdots x_n\in \T(\alpha_0,\alpha_{n+1};p)$ and
\be
\Vrt{x_0x_1\cdots x_n}_{\T(\alpha_0,\alpha_{n+1};p)} \le \prod^n_{j=0}
\Vrt{x_j}_{\T(\alpha_j,\alpha_{j+1}; p_j)} \label{V.vvv} 
\ee
Using the results of \S V.5, we  arrive at certain relations between
$\jba$ and $\T(-\beta,\alpha;p)$.

\mbni{Proposition V.7.} {\it Let $Q=Q^\ast$ and $e^{-sQ^2} \in I_1$, for all
$s>0$.  Let $b\in \jba$ for $0\le \alpha,\beta$, and $0\le \alpha+\beta<1$. 
Then
\begin{itemize}
\ritem{(i)} 
\be
Qdb\in \T(-\beta-1,\alpha)\; , \qquad db^\gamma Q\in
\T(-\beta,\alpha+1)\;, 
\ee
and
\be
d^2 b = Qdb +(db^\gamma) Q = [Q^2,b]\in \T(-\beta-1,\alpha+1)\; .
\ee
 
\ritem{(ii)} For $0\le s$, as a form on $\CH_\infty\times \CH_\infty$,
\be
[b,e^{-sQ^2}] = \int^s_0 e^{-tQ^2}d^2 be^{-(s-t)Q^2} dt\;.
\ee
Both sides of {\rm (V.112)} also define operators in $\bh$.
\ritem{(iii)} For  $0<\ep<s$, define 
\be
H_\ep := \int^{s-\ep}_\ep e^{-tQ^2} d^2 be^{-(s-t)Q^2} dt\in
\T(\alpha,-\beta;s^{-1})\;.
\ee
Furthermore, as $\ep,\ep'\to 0+$,
\be
\Vrt{H_\ep-H_{\ep'}}_{\T(\alpha,-\beta;s^{-1})}\to 0\;.
\ee
The corresponding limit $H_0=\lim_{\ep\to 0+}H_\ep$ is {\rm (V.112)}. 
Thus
\be
[b,e^{-sQ^2}] \in \T(\alpha,-\beta;s^{-1}) \;.
\ee
\ritem{(iv)} Let $0<\mu<1$, and let
\beq
M&=&M(\alpha,\beta,\mu,s)\nn\\
&=&4\mu^{-\alpha-\beta-\frac12}B_1
((1-\alpha-\beta)/2,(2-\alpha-\beta)/2))
\l(\Tr \l(e^{-(1-\mu)Q^2}\r)\r)^s  \;.
\nn 
\eeq
Then
\be
\Vrt{[b,e^{-sQ^2}]}_{\T(\alpha,-\beta;s^{-1})} \le
Ms^{\frac12-(\alpha+\beta)}\Vrt{b}_{\jba}\;. 
\ee
\end{itemize}}

\mbni{Proof.} (i) Let   $\D_2=D(Q^2)$.  The identity (V.111)  can be
established on the domain
$\D_2\times \D_2$, where 
$$
Q^2 b-bQ^2 = Q(Qb-b^\gamma Q) +(Qb^\gamma - bQ)Q = d(db)=d^2b\;.
$$
This can be written
$d^2b= Qdb+(db^\gamma)Q$, which is the algebraic relation (V.111).  Since
$b\in \jba$, in particualr $db\in \T(-\beta,\alpha)$.  Hence $Qdb\in
\T(-\beta-1,\alpha)\subset \T(-\beta-1,\alpha+1)$.  Also $db^\gamma =
-(db)^\gamma \in \T(-\beta,\alpha)$.  Thus $db^\gamma Q\in
\T(-\beta,\alpha+1)\subset \T(-\beta-1,\alpha+1)$.  Hence the domain
inclusions (V.110--111) hold.

(ii) As a form on $\CH_\infty\times \CH_\infty$, and using (V.110--111),
we infer that  
\be
[b,e^{-sQ^2}] = -e^{-tQ^2} be^{-(s-t)Q^2} \Big|^{t=s}_{t=0} = \int^s_0 e^{-
tQ^2 }
d^2 be^{-(s-t)Q^2} dt\;.
\ee
Thus (V.112) is an identity for sesquilinear forms.  The left side is an
element of
$\bh$, and therefore so is the right side.

(iii)  As
$e^{-tQ^2/2}$ is trace class and $d^2b\in \T(-\beta-1,\alpha +1)$, clearly
$e^{-tQ^2}d^2 be^{-(s-t)Q^2}$ is trace class for $0<t<s$.  Furthermore, using
H\"olders inequality on
$$
\l(e^{-(1-\mu)t Q^2}\r) \l(e^{-\mu tQ^2} R^{-\alpha}
(d^2b)R^{-\beta}e^{-\mu(s-t)Q^2}\r)\l(e^{-(1-\mu)(s-t)Q^2}\r)\;,
$$
with exponents $t^{-1},\infty, (s-t)^{-1}$ respectively, we see that 
\beq
   &&\Vrt{e^{-tQ^2} (d^2b)e^{-(s-t)Q^2}}_{\T(\alpha,-\beta;s^{-1})}
= 
\Vrt{e^{-tQ^2} R^{-\alpha}(d^2 b)R^{-\beta}e^{-(s-t)Q^2}}_{s^{-1}}
  \nn\\
&&\qquad \le 4\mu^{-\frac12 - (\alpha+\beta)} \Tr\l(e^{-(1-\mu)Q^2}\r)^s 
\l( t^{-(\alpha+\beta+1)/2}(s-t)^{-(\alpha+\beta)/2}
\Vrt{R^{-\alpha}QdbR^{-\beta}}_{(-\alpha-\beta-1,\alpha+\beta)}\r.\nn\\
&&\qquad\qquad \l. +t^{-(\alpha+\beta)/2}(s-t)^{-(\alpha+\beta+1)/2}
\Vrt{R^{-\alpha}(db^\gamma)
QR^{-\beta}}_{(-\alpha-\beta,1+\alpha+\beta)}\r)\;.
\eeq
Here we have used (V.110--111) as well as (V.28).  Note
\be
\Vrt{R^{-\alpha}QdbR^{-\beta}}_{(-\alpha-\beta-1,\alpha+\beta)} \le
\Vrt{db}_{(-\beta,\alpha)}\;,
\ee
and
\be
\Vrt{R^{-\alpha}(db^\gamma)QR^{-\beta}}_{(-\alpha-\beta,1+\alpha+\beta)}
\le \Vrt{db}_{(-\beta,\alpha)}\;.
\ee
Thus integrating (V.118) and using (V.20,24) we obtain
\be 
 \Vrt{H_\ep}_{\T(\alpha,-\beta;s^{-1})}  \le 4\mu^{-\frac12
-(\alpha+\beta) } s^{\frac12-(\alpha+\beta)} B_1((1-\alpha-\beta)/2,
(2-\alpha-\beta)/2)
\l(\Tr\l(e^{-(1-\mu)Q^2} \r)\r)^s \Vrt{db}_{(-\beta,\alpha)} \;.  
\ee 
This shows that $H_\ep \in \T(\alpha,-\beta;s^{-1})$ and the 
bound on $\Vrt{H_\ep}_{\T(\alpha,-\beta;s^{-1})}$ is of the form (V.116),
uniformly in $\ep$.  We now establish convergence of $H_\ep$ in this norm. 
In fact for
$\ep' >\ep$, the expression $H_\ep-H_{\ep'}$ is just the integral (V.112)
restricted to the intervals $t\in [\ep,\ep']$ and $t\in [s-\ep',s-\ep]$.  We
therefore obtain  
  the bound
\be
\Vrt{H_\ep - H_{\ep'}}_{\T(\alpha,-\beta;s^{-1})}
\le o(1)\mu^{-\frac12-(\alpha+\beta)} s^{-(\alpha+\beta+1)/2} \;,
\ee
as $\ep,\ep'\to 0$.  
Thus we have established the convergence as $\ep\to 0+$ of $H_\ep$ in
$\T(\alpha,-\beta;s^{-1})$.  Since $H_0$ is equal to $[b,e^{-sQ^2}]$, as we
saw in (V.112), we have the bound (V.116) also for the limit.  This
completes the proof of the proposition.  

We end this section with a useful corollary.

\mbni{Corollary V.8.} i. {\it Consider the set $X=\l\{
x_0,x_1\clips,x_n\r\}$. Let $y_j$ and $z_j$ be elements of an interpolation
space $\jba$, where $0\le \alpha,\beta$ and $0\le
\alpha+\beta<1$.   Suppose  each $x_j$ is one of the forms
\be
 y_j,\enspace dy_j,\enspace d^2y_j, \enspace y_j(dz_j), \hensp{or} (dy_j)
z_j \;, 
\ee 
but where no two adjacent $x_j$'s are of the form $d^2y_j$. 
(Here we consider $x_j$ and $x_{j+1}$ adjacent, as well as 
$x_0$ and
$x_n$ adjacent.)  Then $X$   is a regular
set of vertices with respect to $Q$. 

{\rm (ii)} Let $X^\jlo_n=\{a_0,da_1\clips,da_n\}$, where $a_j\in \jba$. 
Then there exist constants $m_1,m_2<\infty$ such that the trace norm of
the Radon transform $\hx^\jlo_n$, defined in {\rm (V.58)}, of  $X^\jlo_n(s) $
satisfies the bound
\be
\Vrt{\hx^\jlo_n}_1 \le
m_1m^{n+1}_2\mu^{-(\alpha+\beta)(n+1)/2}\l({1\over
n!}\r)^{\frac12+\l({1-\alpha-\beta\over 2}\r)} \Tr \l(e^{-(1-\mu)Q^2 }\r) 
\l( \prod^n_{j=0} \Vrt{a_j}_\jba\r)\; .
\ee

{\rm (iii)} For $\mu$ fixed,
\be
n^{1/2} \Vrt{\hx^\jlo_n}^{1/n}_1 \le 0 (n^{-(1-\alpha-\beta)/2} ) \;,
\ee
where $ (1-\alpha-\beta)>0$.

{\rm (iv)} Let $a_j\in \jba$, and define 
\beq
X_1&=& \{ a_0,da_1\clips,(da_{j-1})a_j,da_{j+1},\clips, da_n\} \;,\\
X_2&=& \{a_0,da_1 \clips, da_{j-1}, a_j da_{j+1}\clips, da_n\} \;, 
\eeq
and
\be
X_3 = \{a_0,da_1\clips, da_{j-1} , d^2 a_j , da_{j+1} \clips, da_n\} \; . 
\ee
Then
\be
X_1(s)-X_{2}(s) = \int^{s_{j-1}}_0 X_3 (s_0\clips,s_{j-2} , t,s_{j-1}-t,
s_j\clips , s_{n-1}) dt\;.
\ee

{\rm (v)} After integration over $s\in \sigma_n$,
\be
\hx _1-\hx _2=\hx _3\;,
\ee
or in terms of the expectations {\rm (V.58)}
\beq
&&  \lra{a_0,da_1\clips,(da_{j-1}) a_j,da_{j+1}\clips,da_n;g}_{n-1}   - 
\lra{a_0,da_1\clips,da_{j-1},a_jda_{j+1}\clips,da_n;g}_{n-1}\nn\\
&&\hskip2in  =  \lra{a_0,da_1,da_2\clips, da_{j-1} ,d^2
a_j,da_{j+1}\clips,da_n;g}_n
\;.
\eeq

{\rm (vi)} There are constants $m_1,m_2<\infty$ such that for  $0<\mu<1$,
\beq 
&&\l| \lra{a_0,da_1\clips,da_j,d^2a_j,da_{j+1}\clips,da_n;g}_n\r| \nn\\
&&\hskip1in \le
m_1m_2^{n+1}\mu^{-(\alpha+\beta) n/2-1} \Tr \l( e^{-(1-\mu)Q^2 }\r) \l(
{1\over n!}\r)^{1-(\alpha+\beta)/2} \l( \prod^n_{j=0}
\Vrt{a_j}_{\jba} \r)\;.\nn\\
\eeq
}

\mbni{Proof.}  (i) For $y,z\in \jba$,  $y\in \T(0,0)$ and  $dy\in
\T(-\beta,\alpha)$.  Furthermore  
$$d^2 y \in \T(-\beta-1,\alpha+1),\quad ydz\in
\T(-\beta,\alpha) \hensp{and} (dy)z\in \T(-\beta,\alpha)\;.
$$
This is a
consequence of the definition of $\jba$ and Proposition V.6.  The most
singular case occurs with $[(n+1)/2]$ vertices $x_j=d^2 y_j$, interspersed
between vertices in $\T(-\beta,\alpha)$.  (Here $[\;\cdot\;]$ denotes the
integer part.)  Thus with an even number of vertices (odd $n$)
\be
0<\eloc = \eglo = (1- \alpha-\beta)/2 \;.
\ee
In the case of an odd number of vertices,
\be
\eloc = (1- \alpha-\beta)/2< \eglo\;.
\ee
In either case $X$ is a regular set.

(ii--iii) In the case that $X=X^\jlo_n$, we have
\be
\eta_0 = 1-\beta/2\;,\quad \eta_j=\frac12 +\l( {1-\alpha-\beta\over 2}\r)
\;,\quad j=1,2\clips,n\;,
\ee
so
\be
\eloc = \frac12 + \l( {1-\alpha-\beta\over 2}\r) >\frac12 \hensp{and}   
(n+1)\eglo = (n+1)\eloc + \alpha/2\;.
\ee
The bounds (V.124--125) then follow from the bound (V.63) and the
asymptotics of the $\Gamma$ function.  

(iv--v) For $s\in \sigma_n$, $X_1(s)$ and $X_2(s)$ are trace class.  Also
\beq
X_1(s) -X_2(s)&=&a_0e^{-s_0Q^2} da_1\cdots e^{-s_{j-2}Q^2} da_{j-1}
[a_j,e^{-s_{j-1}Q^2}] da_{j+1}e^{-s_jQ^2}  \nn\\
&&\hskip1.5in \times \cdots da_n
e^{-s_{n-1}Q^2}\;.
\eeq 
Using Proposition V.7.iii--iv, the commutator in
(V.137) is an element of
$\T(\alpha,-\beta,s^{-1}_{j-1})$, with norm bounded  by
$Ms^{\frac12-(\alpha+\beta)}_{j-1} \Vrt{a_j}_{\jba}$.  Therefore
$$
da_{j-1} [a_j,e^{-s_{j-1}Q^2}] da_{j+1}\in \T(-\beta,\alpha;s^{-1}_{j-1})\;,
$$
and
\beq
 \Vrt{da_{j-1}[a_j,e^{-s_{j-1}Q^2}]da_{j+1}}_{\T
(-\beta,\alpha;s^{-1}_{j-1})}
 &\le& \Vrt{da_{j-1}}_{\T(-\beta,\alpha;\infty)}
\Vrt{[a_j,e^{-s_{j-1}Q^2}]}_{\T(\alpha,-\beta;s^{-1}_{j-1})}  
\Vrt{da_{j+1}}_{\T(-\beta,\alpha;\infty)} \nn\\ &    
 \le &
Ms^{\frac12-(\alpha+\beta)}_{j-1} \prod^{j+1}_{k=j-1}
\Vrt{a_j}_\jba\;. 
\eeq
On the other hand, Proposition V.7.ii shows that 
\beq
X_1 (s)-X_2(s)&=& \int^{s_{j-1}}_0 a_0e^{-s_0Q^2}da_1 e^{-s_1Q^2}\cdots
da_{j-1}e^{-tQ^2} d^2a_je^{-(s_{j-1}-t)Q^2}  \nn\\&&\hskip1in
 \times  da_{j+1}e^{-s_jQ^2} \cdots da_ne^{-s_{n-1}Q^2}
\cdots da_n e^{-s_{n-1}Q^2} dt\nn\\
&=& \int^{s_{j-1}}_0
X_3(s_0,s_1\clips,s_{j-2},t,s_{j-1}-t,s_j\clips,s_{n-1})dt\;, 
\eeq
which is (V.129).  Integrating over $\sigma_{n-1}$ yields (X.130).  The
bound (V.132) then follows by an analysis of $\Vrt{X_3(s)}_1$ similar to
the proof of (V.124).

\section{Cocycles}
\setcounter{equation}{0}
Throughout this section   take  $\A$ to be a subalgebra of an interpolation
space $\jba$ introduced in \S V.5.     We begin this section by showing the
$\tau^\jlo$ extends in this case to be an element of $\C(\A)$, including
the formula for the pairing with a root of $I$.  We define
define a fractionally-differentiable structure.  Finally we show that
$\tau^\jlo$ is a cocycle.  These facts are
preliminary to the next section where we show that the pairing is actually a
homotopy invariant.   

\subsection{The JLO-Cochain Extends to Interpolation Spaces}
In this sub-section we extend the JLO-cochain from the framework in \S IV
where the space of cochains $\C(\A)$ live    over an algebra $\A$ of
differentiable functions (i.e., $da\in \bh)$, to the case that $\A$ is
contained in one of the interpolation spaces $\jba$.  Thus $a\in \A$ will
have a fractional derivative of order $1-\beta$ and $da\in
\T(-\beta,\alpha)$ will be a generalized function. We require 
\be
\A\subset \jba\hensp{for some} 0\le \alpha,\beta,\hensp{and} 0\le
\alpha+\beta<1\;.
\ee  
We also require that the
norm
$\vcv$ on $\A$ satisfy
\be
\vv{a}_\jba \le \vv{a}\;.
\ee
Otherwise, we retain the basic hypotheses of \S IV.  The Hilbert space $\CH$
is
$\Z_2$ graded by $\gamma$ and carries a continuous unitary representation
$U(g)$.  We assume that $Q=Q^\ast$ commutes with $U(g)$ and $Q\gamma
+\gamma Q=0$.  We assume that $\exp(-\beta Q^2)$ is trace class for all
$\beta>0$.  
We assume that $\A$ is pointwise invariant under the action of $\gamma$,
and invariant under the action of $U(g)$, $U(g) \A U(g)^\ast \subset \A$.

\mbni{Definition VI.1.} We call the quintuple 
$$
\{ \CH,Q,\gamma,U(g),\A\} \label{V.vv}
$$
satisfying the above hypotheses
a $\Theta$-summable, fractionally-differentiable structure.   
\mbni{Proposition VI.2.} {\it Let $\{\CH,Q,\gamma,U(g),\A\}$ be a
$\Theta$-summable, fractionally-differentiable structure. Then
$\tau^\jlo\in \C(\A)$.  There exist $m <\infty$ such that 
\be
\vv{\tau_n^\jlo} \le m^{n+1} \l({1\over
n!}\r)^{\frac12+\l({1-\alpha-\beta\over 2}\r)} \Tr\l(e^{-Q^2/2}\r)
\;.\label{V.v}
\ee 
}

\mbni{Proof.} For $a_j\in \jba$, we have already established in Corollary
V.8.ii, with $\mu=\frac12$, that $X^\jlo_n$ is trace class with trace norm
bounded by
$$
m^{n+1}\Tr\l(e^{-Q^2/2}\r) 
\l({1\over n!}\r)^{\frac12+\l({1-\alpha-\beta\over 2}\r)}
\l(\prod^n_{j=0} \Vrt{a_j}_\jba\r)\;.
$$
 Since $\Vrt{a_j}_\jba\le \vv{a_j}$, we
conclude that $\tau^\jlo_n$ is defined on $\A^{n+1}$ and that
$\vv{\tau^\jlo_n}$ satisfies (\ref{V.v}).  Since $\alpha+\beta<1$, this
entails  the ``entire'' condition $n^{1/2}\vv{\tau^\jlo_n}^{1/n}\to 0$.  Thus 
$\tau^\jlo\in
\C(\A)$.

Having extended the notion of $\tau^\jlo$ to a fractionally-differentiable
structure, we now observe that the pairing
$\lra{\tau^\jlo,a}$ has the same representation as in the differentiable 
case.

\mbni{Corollary VI.3.} {\it Let $\{\CH,Q,\gamma,U(g),\A\}$ be a
$\Theta$-summable, fractionally-differentiable structure.  Let $a\in
\Mat_m(\A^{{\fg }})$ satisfy $a^2=I$.  Then
\be 
\fz^Q(a;g) = \lra{\tau^\jlo,a} 
=  {1\over \sqrt{\pi}} 
\int^\infty_{-\infty} e^{-t^2}\Tr \l(\gamma U(g)ae^{-Q^2 +itda}\r)dt\;.
\ee
Here $\Tr$ denotes both the trace on $\CH$
and the matrix trace in
$\Mat_m(\A)$, in case $m>1$. }

\subsection{The JLO-Cochain is a Cocycle}
\mbni{Proposition VI.4.} {\it Let $\{\CH,Q,\gamma , U(g),\A\}$ be a
$\Theta$-summable, fractionally-differentiable structure.  Then the cochain
$\tau^\jlo\in
\C(\A)$ is a cocycle for
$\part$, namely
\be
\part \tau^\jlo=0\;.
\ee
}

\medbreak

\mbni{Remarks.} 1. The cochain $\tau^\jlo$ was originally defined in 
\cite{JLO1}, for the differentiable $da\in \B(\CH)$, where the cocycle 
condition was also established.  This cochain has been investigated again 
in several different contexts, see \cite{GS,EFJL,K1,Q,KL} for example.  
Our presentation  is self contained.

2. The only known  cocycles for $\C(\A)$ are elements
$[\tau^\jlo]$, where $\tau^\jlo$ is defined by some $Q$. (Here $Q$ gives
rise either to the $\Theta$-summable case considered above, or 
to the class of cochains satisfying the KMS-condition. 
See \cite{K1,JLO2,JLW} for this extension, applicable in the differentiable 
case.)  Taking the larger space of cochains $\D(\A)\supset \C(\A)$, and 
the corresponding coboundary operator $\partial$,  Connes earlier gave 
another cocycle $\tau^C$, see \cite{C2}.  This cocycle is convenient 
because it satisfies a ``normalization'' 
condition, central to Connes' analysis of pairings of cocycles in $\D$,
and also used by him in other studies.  
Furthermore, Connes showed that any cocycle $\tau\in\D(\A)$ is cohomologous 
to a normalized cocycle in $\D$. 
The cocycle $\tau^C$ is determined by an operator $F$ satisfying $F^2=I$ and 
$F\gamma+\gamma F=0$.  
With $\tau^\jlo$ the cocycle determined by $Q$,
and with $\tau^C$ the cocycle determined by an appropriate $F=F(Q)$, 
Connes has shown \cite{C3} (for the differentiable case)
that $\tau^C$ and $\tau^\jlo$ are cohomologous. 
In other words, there is a cochain $G\in \D$ such that
$\tau^\jlo=\tau^C+\part G$.  

On the other hand, cocycles in $\C(\A)$ are not normalized in Connes' sense.  
As discussed in \S III, by working with the cochains $\C(\A)$ we avoid 
the need to consider this normalization.  Furthermore, a pairing can be 
defined for all cochains in $\C(\A)$, rather than just for cocycles.
The importance of pairing a cocycle then rests on the pairing 
yielding an invariant, as dicussed in \S VII.

\mbni{Proof.}  It was shown in (IV.25) that if $\alpha=\beta=0$, then
evaluated on $\A$, 
$\tau^\jlo_{2n+1}=0$.  By the symmetry (V.66), this extends to
$\A\subset \jba$.  Hence to establish the cocycle condition in
$\C(\A)$, it is sufficient to show that for all odd $n$,
\be
B\tau^\jlo_{n+1} = -b\tau^\jlo_{n-1} = \langle
da_0\clips,da_n;g\rangle \;.
\ee 
By Corollary V.8.i, the right side of (VI.6) is a well-defined expectation. 
We prove below that for $n$ odd,
\be
\l( B\tau^\jlo_{n+1}\r)(a_0\clips,a_n;g)=\tau^\jlo_n
(da_0,a_1\clips,a_n;g)\;,
\ee
and
\be
\l(b\tau^\jlo_{n-1}\r)(a_0\clips,a_n;g)=\sum^n_{j=1} (-1)^{j-1}
\tau_n^\jlo (a_0,a_1\clips,da_j\clips,a_n;g)\;.
\ee
Starting from the definition (II.19) of $B$, and the symmetry (V.67),
$B\tau^\jlo_{n+1}$ equals 
\beq
\l(B\tau^\jlo_{n+1}\r)(a_0\clips,a_n;g) &=& \sum^n_{j=0}(-1)^j \llangle
I,da^{g^{-1}}_{n-j+1}\clips, da^{g^{-1}}_n,
da_0\clips,da_{n-j};g\rrangle_{n+1}\nn\\ &=&\sum^n_{j=0} \llangle
da_0\clips, da_{n-j} , I,da_{n-j+1}\clips,da_n;g\rrangle_{n+1}\nn\\
&=& \sum^{n+1}_{j=1}
\lra{da_0\clips,da_{j-1},I,da_j\clips,da_n;g}_{n+1}\nn\\
 &=& \langle
da_0\clips,da_n;g\rangle_n\;.
\eeq 
In the last step we use (V.68). This establishes the first part of (VI.9).

On the other hand, recall (II.9),
$$
\l(V(0)\tau^\jlo_{n-1}\r) (a_0\clips,a_n;g) = \langle
a_0a_1,da_2\clips,da_n;g\rangle_{n-1}\;.
$$
Likewise for $1\le r\le n-1$,
\beq 
  \l( V(r)\tau^\jlo_{n-1}\r) (a_0\clips,a_n;g)  
&= &(-1)^r
\lra{a_0,da_1\clips,da_{r-1},d(a_ra_{r+1}),da_{r+2}\clips,da_{n+1};g}_n\nn
\\
& =& (-1)^r
\langle a_0\clips,a_r da_{r+1},da_{r+2}\clips,da_n;g\rangle_{n-1} \nn\\
&& +
(-1)^r
\langle a_0,da_1\clips, (da_r) a_{r+1},da_{r+2}\clips,da_n;g\rangle_{n-1}\;.
\eeq
Note that here we have expanded the vertex $d(a_ra_{r+1})\in
\T(-\beta,\alpha)$ into the sum of two vertices $(da_r)a_{r+1}$ and
$a_r(da_{r+1})$, each of them in $\T(-\beta,\alpha)$.  This is justified in
Corollary V.6.i.
Also for $r=n$, using (V.67),
\beq
  \l(V(n)\tau^\jlo_{n-1}\r) (a_0\clips,a_n;g)  
&=& (-1)^n \llangle
a^{g^{-1}}_n a_0,da_1\clips,da_{n-1};g\rrangle_{n-1}\nn \\
&=& (-1)^n \langle a_0,da_1\clips,(da_{n-1}) a_n;g\rangle_{n-1} 
\;.
\eeq
Thus
\beq
 \l( b\tau^\jlo_{n-1} \r) (a_0\clips,a_n;g) 
& = &\llangle a_0 a_1,da_2\clips, da_n;g\rrangle_{n-1} - \llangle
a_0,a_1da_2\clips,da_n;g\rrangle_{n-1}\nn\\
&&+ \sum^{n-1}_{r=1} (-1)^r \l( \llangle a_0\clips,(da_r)
a_{r+1},da_{r+2}\clips,da_j;g\rrangle_{n-1}  \r.\nn\\
&&\l.- \llangle
a_0\clips,da_r,a_{r+1}da_{r+2}\clips,da_n;g\rrangle_{n-1}\r)\nn\\
& =&\sum^{n-1}_{r=0} (-1)^r \llangle a_0 ,da_1\clips,d^2 a_{r+1}\clips,
da_n;g\rrangle_n\;.
\eeq
The final identity in (VI.12) involves combining terms using Corollary V.8.v. 
Hence using (V.69) we have
\be
\l(b\tau^\jlo_{n-1} \r) \l(a_0\clips,a_n;g\r) = -\langle
da_0,da_1\clips,da_n;g\rangle_n\;,
\ee
completing the proof of (VI.8), and hence the proof of (VI.5).

\section{\ \ Homotopy Invariants} %
\subsection{\ The Main Result: JLO Pairing is Invariant}
\setcounter{equation}{0}
In this section we consider the pairing of a family $\tau^\jlo(\la)$ of
cocycles with a square root $a$ of $I$. These cocycles arise from a family of
non-commutative, fractionally-differentiable structures on $\A$.  The
advantage to pairing
$\tau^\jlo(\la)$ with $a\in {\rm Mat}_m(\A^{\fg })$ rather than to pairing an
arbitrary family of cochains $\tau(\la)$, is the fact that the pairing 
function (given in various forms in III.11, 28, 29, and  31)  namely
\be
\fz^Q(a;g) = \llangle \tau^\jlo (\la),a\rrangle\; , 
\ee
is a constant function of $\la$.  Hence we
obtain an {\it invariant}, and we call the continuous variation in 
$\la$ a {\it homotopy}.  In other words, each $\lra{\tau^\jlo_{(\la)},p}$ 
is a homotopy invariant.  

Our basic result is to give conditions that are sufficient to prove that
$\tau^\jlo(\la)$ is continuously differentiable in $\la$.  Under these
hypotheses, the pairing function $\lra{\tau^\jlo(\la,a}$ is actually
constant.  This invariant is in general not integer valued, but may be in
certain special cases. 

In particular we assume that the $\la$-dependence of $\tau^\jlo(\la)$, and
hence that the pairing of $\lra{\tau^\jlo(\la),p}$, arises from the
$\la$-dependence of
$Q(\la)$. Here $Q(\la)$ generates $\tau^\jlo(\la)$ as described in \S IV, and 
the parameter $\la$ lies in an open interval $\La = (\la_1,\la_2)\subset
\R$.  Our operator $Q(\la)$ is a self-adjoint operator on $\CH$, and we
suppose that
$Q(\la)$ has the general form
\be
Q(\la) = Q+q(\la)\;.
\ee
We regard $Q=Q^\ast$ as defining a basic $\tau^\jlo$, and $q(\la)$ as
providing a deformation of $Q$ and a perturbation $\tau^\jlo(\la)$  of
$\tau^\jlo$.  

Let us state the main result of this section.
\mbni{Theorem VII.1.i} {\it Let $\{\CH, Q(\la),\gamma,U(g)$, $\A\}$ be a
regular family of $\Theta$-summable, fractionally differentiable, 
structures as defined in
\S VII.3.  Then the corresponding family of JLO cocycles
$\{\tau^\jlo(\la)\}$ is continusouly differentiable as a function $\La\to
\C(\A)$, and there is a continuous family
$\{h(\la)\}\subset
\C(\A)$ such that for all $\la\in \La$,
\be
{d\over d\la}\tau^\jlo(\la)=\part h(\la)\;.
\ee

{\rm (ii)} The function $g \rightarrow \tau^{\jlo}(\lambda)$ is continuous
function of 
$g\in {\fg }$ from ${\fg }$ to $\C(\A)$, uniformly for $\lambda$ in compact subsets
of $\Lambda$.
}

An immediate consequence of
the fact that
$d\tau^\jlo(\la)/d\la =\part h$ is the invariance of
$\fz^{Q}(a;g)$.  In particular,   continuous differentiability of
$\tau^\jlo(\la)$ ensures, c.f.\ Proposition~III.6, that the pairing
$\lra{\tau^\jlo(\la),a}$ is also a continuously differentiable function.  
Hence 
\be
{d\over d\la} \lra{\tau^\jlo(\la),a} = 
\langle \part h(\la),a\rangle = 0\; .
\ee
We established in Proposition III.1 the vanishing of the pairing function
defined in (VI.4) on coboundaries. 
Thus we have
\mbni{Corollary VII.2.i} {\it For a regular family of non-commutative
structures, the JLO-pairing $\lra{\tau^\jlo(\la),p}$ is constant.  For
$\la_1,\la_2\in \La$, 
\be
\fz^Q(a;g) = \lra{\tau^\jlo(\la_1),a} = \lra{\tau^\jlo(\la_2),a}\; . 
\ee
Furthermore
\be
\tau^\jlo(\la_2) = \tau^\jlo(\la_1) + \part H 
\ee
where $H=H(\la_1,\la_2) = \int^{\la_2}_{\la_1} h(\la)d\la$. 

{\rm (ii)} In addition (VII.5) is a continuous function of $g$. 
}\medbreak

The considerations in this section are both algebraic and analytic. While the
former are universal,   the latter  are crucial.  Hence the relevant
analytic groundwork has already been prepared in \S V.  The invariants
$\llangle
\tau^\jlo,a\rrangle$ do take different values for a given $\A$ and a given
$a$.  
In fact, even a bounded, differentiable family $q(\la)$ requires work to
establish the continuous differentiability of $\tau^\jlo(\la)$.  The work is
not that much greater for the class of regular  perturbations which we
consider here and which include  a wide variety of interesting examples.
Sections VII.2--5 are devoted to a precise formulation of a regular
deformation and to the proof of this theorem.  In the literature  there are
special cases of this result, for example \cite{EFJL}, but there is no general
and easily verifiable conditions on $q(\la)$, like the one we give here, which
result in a homotopy.  In fact, the differentiability of $\tau^\jlo(\la)$ was
never completely analyzed, even for bounded perturbations $q(\la)$.
\subsection{\ Regular Linear Deformations}
In this subsection we outline a class of {\it regular linear deformations}
$\{\CH,Q(\la)$ , $\gamma,U(g),\A\}$  of
the JLO cochain $\tau^\jlo$ defined for a particular $\{\CH,
Q,\gamma,U(g)$, and
$\A\}$.  We denote the family of $Q$'s by $\{Q(\la)\}$ and the
family of cochains by $\tau^\jlo(\la)$.  These families depend on a fixed
$\CH,\gamma,U(g)$, and $\A$.

We first compile a list of assumptions.

\mbni{a.} {\it The Starting Point.} The undeformed problem is
given by the structure introduced earlier in \S V.7.  It is defined by a self
adjoint $Q=Q^\ast$ with domain $\D$, acting on a Hilbert space $\CH$.  The 
heat kernel $\exp(-\beta Q^2)$ is assumed to be trace class for every 
$\beta>0$.  There is a $\Z_2$ grading $\gamma$ on $\CH$ for which $Q\gamma +
\gamma Q=0$.  There is a continuous, unitary representation $U(g)$ on
$\CH$ of a compact Lie group ${\fg }$, and $U(g)Q=QU(g)$.  There is a
Banach algebra of observables $\A$, with 
\be
\A\subset  \jba\;,
\ee
and where $\jba$ is an interpolation space introduced in \S V.5.  We require
that $0\le \alpha,\beta$ and $0\le \alpha+\beta<1$.  We also require that
the norm $\vcv$ of $\A$ satisfy
\be
\vv{a}_\jba\le \vv{a}
\ee
for all $a\in \A$.  Thus elements of $\A$ have $0<1-\beta$ fractional
derivatives with respect to $Q$.  We assume that the algebra
$\A$ is pointwise invariant under the action of $\gamma$, namely
$\alpha^\gamma = \gamma
\alpha\gamma=a$ for $a\in \A$.  Furthermore $\A$ is invariant under the
action of $U(g)$, namely $a^g = U(g)aU(g)^\ast \in \A$ for $a\in \A$.
This structure defines a JLO cocycle
$\tau^\jlo$ and a non-commutative, fractionally-differentiable, structure
$\{\CH,Q,\gamma,U(g),\A\}$.

\mbni{b.} {\it A Family of Regular Linear Perturbations.} A family of regular
deformations of $\{\CH,Q,\gamma,U(g),\A\}$ is defined by a family $q(\la)$
of regular perturbations of $Q$ on the space $\CH$.  Let $q$ denote a
symmetric operator on
$\CH$ with domain $\D$.  We suppose there are constants $0\le a $,
$M<\infty$    such that on the domain $\D\times \D$,
\be
q^2\le a^2 Q^2+M^2\;.
\ee
In other words, $q$ is a bounded map from the Sobolev space $\CH_1$, 
defined in \S V.2, to $\CH_{-1}$.  We define
the family $\{q(\la)\}$ of regular perturbations\footnote{The
condition (VII.9) ensures $\Vrt{qf} \le a\Vrt{Qf} + M\Vrt{f}$, a condition
introduced by T. Kato to study $Q+q$, see \cite{K2}.  The relevant case $a<1$
corresponds in our present case to the bound $\mu a<1$ of (VII.9).  If $0<a$
may be chosen arbitrarily small (which may require $M(a)$ large), then $q$
is said to be infinitesimally small compared with $Q$.  In that case $\mu$
may be chosen arbitrarily large.
}
parameterized by real $\la$, and the family $\{Q(\la)\}$ of perturbed
operators, by
\be
q(\la) = \la q\;,\hensp{and} Q(\la)=Q+q(\la)\;.
\ee
The linearity of $q(\la)$ in $\la$ is the linearity in the sub-section title. 
Here $\la$ belongs to a bounded open interval $\La$,
$$
\la\in \La =(-\mu,\mu),\hensp{and} 0<\mu<a^{-1}\;.$$

In addition to the bound (VII.9), we will need another bound: for some
$0<\ep<1$,
\be
\Vrt{R^{1-\ep}qR^\ep} + \Vrt{R^\ep q(\la)R^{1-\ep}} \le O(1)
\ee
where $R=(Q^2+I)^{-1/2}$.  If $q(\la)$ is essentially self-adjoint on $\D$,
then (VII.11) follows automatically from (VII.9), in fact for all 
$0\le \ep\le 1$.  Howerver, if $q(\la)$ is not essentially self-adjoint on 
$\D$, we also assume (VII.11) for some $0<\ep$.

\mbni{c.} {\it Symmetries.} We assume that $\gamma$ and $U(g)$ of
Assumption (a), also are symmetries of $\{Q(\la)\}$ in the sense that
\be
Q(\la)\gamma+\gamma Q(\la)=0\;,\quad U(g) Q(\la)=Q(\la)U(g)
\ee
for all $\la\in \La$ and for all $g\in {\fg }$.  Of course this is ensured by
$\gamma q+q\gamma=0$ and $U(g)q=qU(g)$.

\mbni{d.} {\it The Algebra $\A$.} We assume that the algebra $\A\subset
\jba$ is  independent of $\la$.  It is necessary that for $a\in \A$, the
differential $d_\la a$, as $\la$ varies, remains in $\jba$.  Thus we require
that the norm $\vcv$ on $\A$ satisfies 
$$
\vv{a}\ge \Vrt{a} +\sup_{\la\in \La}\Vrt{d_\la a}_{(-\beta,\alpha)}
$$
for some $\alpha,\beta$ where
$$
0\le \alpha,\beta\;,\quad \alpha+\beta<1\;.
$$
Since $\La$ is an interval $[-\mu,\mu]$, this is ensured by
$$
da\in \T(-\beta,\alpha) \hensp{and} [q,a]\in \T(-\beta,\alpha)\;.
$$

\mbni{Definition VII.3.} {\it A family $\{\CH,Q(\la)\}$ of operators
satisfying Assumptions (a--b) are a regular (linear) $Q$-family.  A family
$\{\CH, Q(\la),\gamma,U(g),\A\}$ which satisfies Assumptions (a--d) is a
regular linear  family of $\Theta$-summable, fractionally-differentiable,
non-commutative  structures.}

\mbni{Remark.}  We generalize the notion of a regular family in \S VII.3,
replacing linearity by an additional assumed estimate.  
\mbni{Proposition VII.4.} {\it Let $\{Q(\la)\}$ denote a regular (linear)
$Q$-family.  Then
\begin{itemize}
\ritem{a)} For each $\la\in \La$, $Q(\la)$ is self-adjoint on the domain $\D$.
\ritem{b)} There are constants $\tilde M_1,\tilde M_2<\infty$ such that 
\be
Q^2\le \tilde M_1^2 (Q(\la)^2 +I)
\ee
for all $|\la|\le \mu$.  Here $\tilde M_1=\max \{2,2\mu M, (1-\mu
a)^{-1},(1-\mu a)^{-1} \mu M\}$.  Also for all $\la,\la'\in \La$,
\be
q(\la)^2 \le \tilde M^2_2 (Q(\la')^2+I)\;,
\ee
where $\tilde M^2_2 = \tilde M^2_1+(\mu M)^2$.
\ritem{c)} For all $\beta>0$, and all $|\la|\le \mu$, $\exp(-\beta Q(\la)^2)$
is trace class and
\be
Tr \l( e^{-\beta Q(\la)^2}\r) \le e^\beta \Tr \l(e^{-\beta Q^2\tilde
M^{-2}_1}\r)\;.
\ee
For given $\beta$, this bound is uniform for $\la$ in a compact subset of
$\La$.
\end{itemize}
}

\mbni{Corollary VII.5.} {\it For $\la$ in any compact subset of $\La$, the
regular linear   family
$\{\CH,Q(\la),\gamma,U(g),\A\}$ determines a bounded family
$\{\tau^\jlo(\la)\}$ of JLO-cocycles on $\A$.}

\mbni{Proof.} (a) A symmetric operator $Q(\la)$ on the domain $\D$ is self
adjoint if and only if for some $\alpha>0$, $(Q(\la)\pm i\alpha)\D=\CH$. 
This is the statement that the resolvents $(Q(\la)\pm i\alpha)^{-1}$ exist
and are bounded.  By the spectral theorem for $Q=Q^\ast$, we infer
$\Vrt{(Q\pm i\alpha)^{-1}}\le \alpha^{-1}$ and $\Vrt{(Q(Q\pm
i\alpha)^{-1}}\le 1$.  Thus (VI.4) ensures that for all $\la\in \La$,
$$
\Vrt{q(\la) (Q\pm i\alpha)^{-1}} \le (\mu a)^2 +(\mu M/\alpha)^2\;. $$
Assumption (b) ensures $|\la|a<\mu a <1$. Thus $\Vrt{q(\la)(Q\pm
i\alpha)^{-1}}<1$, for $\alpha$ sufficiently large, uniformly in $\la\in \la$. 
It follows that the series
\be
(Q\pm i\alpha)^{-1} \sum^\infty_{n=0} (-q(Q\pm i\alpha)^{-1})^n 
\ee
converges in norm.  This is $(Q+q\pm i\alpha)^{-1}$, as can be verified from
the series expansion.  Since the domain of (VII.13) is $\CH$, the range of
$(Q+q\pm i\alpha)$ is $\CH$.  The range of $(Q\pm i\alpha)^{-1}$ is $\D$, so
the domain of $Q+q\pm i\alpha$ is contained in $\D$.  However $Q+q\pm
i\alpha$ is originally defined on all of $\D$, so that is its domain.

(b) Remark that (VII.14) follows from (VII.13), using (VII.9).  In fact
\beq
q(\la)^2 &\le & (\la a)^2 Q^2 +(\la M)^2 \le (\la a)^2 \tm^2_1 (Q(\la')^2+I) 
+ (\la M)^2 \nn\\
&\le & \l( (\mu a)^2 \tm^2_1+ (\mu M)^2\r) \l(Q(\la ')^2+I\r) \nn\\
&\le & \l(\tilde M_1^2+ (\mu M)^2\r)\l(Q(\la')^2+I\r)\;. \nn
\eeq

So now we establish (VII.13).  On the domain $\D\times \D$ for
sesquilinear forms, it follows from the Schwarz inequality that for any
$\ep >0$,
\be
\pm (q(\la)Q+Qq(\la)) \le \ep Q^2 + {1\over \ep} q(\la)^2\;.
\ee
Thus on $\D\times \D$, we infer from  (VII.17) and (VII.9) that 
\beq
\hskip-36pt Q^2 &=& (Q(\la)-q(\la))^2 =
Q(\la)^2-q(\la)^2-(q(\la)Q+Qq(\la))\nn\\
\hskip-36pt &\le & Q(\la)^2 + \ep Q^2 +
\l({1\over \ep}-1\r) q(\la)^2 \nn\\ 
\hskip-36pt &\le & Q(\la)^2 + \l( \ep + (\la a)^2
\l({1\over \ep}-1\r)\r)Q^2 + (\la M)^2
\l({1\over \ep}-1\r)\; . 
\eeq

\mbni{Case 1.} $|\la|a\le 1/2$: In this case choose $\ep=1/2$ in (VI.18). 
Then $\ep+(\la a)^2\l({1\over \ep}-1\r)\le {3\over 4}$, so collecting the
$Q^2$ terms in (VII.18) gives 
\be
\frac14 Q^2 \le Q(\la)^2 +(\la M)^2\le Q(\la)^2 + (\mu M)^2\;.
\ee 
Hence (VII.13) holds with $\tm_1=2\max \{1,\mu M \}$.

\mbni{Case 2.} $\frac12 \le |\la|a\le \mu a<1$: In this case choose $\ep =
|\la|a$ in (VII.18).  
Then the coefficient of $Q^2$ in (VII.18) is 
$1-\ep-(\la a)^2 \l({1\over
\ep}-1\r) =(1-|\la|a)^2\ge (1-\mu a)^2$, and $\l({1\over \ep} -1\r)\le 1$.
Thus (VII.18) ensures that 
\be
(1-\mu a)^2 Q^2 \le Q(\la)^2 +\la^2M^2\le Q(\la)^2+\mu^2 M^2\;.
\ee
Thus (VII.13) holds with $\tm_1 =(1-\mu a)^{-1}\max \{1,\mu  M \}$. 
Thus in both cases, (VII.13) holds with
\be
\tm_1 = \max \{ 2,2\mu M,(1-\mu a)^{-1},(1-\mu a)^{-1}\mu M\}\; .
\ee
This completes the proof of (b).

(c) Using (VII.13),
$$
\tm^{-2}_1 Q^2-I\le Q(\la)^2\;.
$$
We infer that if $E_i(Q^2)$ is the $i$th eigenvalue of $Q^2$, counting in
increasing order, then  by the minimax principle, $\beta
\tm^{-2}_1 E_i(Q^2)-\beta\le \beta E_i(Q(\la)^2)$.  Assumption (a) includes
the assertion that $\exp(-\beta Q^2)$ is trace class for all $\beta$.  Thus
(VII.15) follows.
\medbreak
This completes the proof of the proposition.  The corollary follows.  In fact
Assumptions (a--d) plus the fact that $Q(\la)=Q(\la)^\ast$ and $\Tr
\l(e^{-\beta Q(\la)^2}\r)<\infty$, with a uniform bound for $\la$ in a
compact subset of $\La$, ensure the existence of $\{\tau^\jlo (\la)\}$.  The
fact that this family is bounded   then follows as a
consequence of (IV.6), along with   (VII.15).

\mbni{Proposition VII.6.} {\it If $\{Q(\la\}$ denotes a regular, linear 
$Q$-family, then the Sobolev spaces $\CH_p(Q(\la))$, with $p\in [-1,1]$ and
$\la$ in a compact subset of $\La$, are independent of $\la$.
}

\mbni{Proof.} We require that if $f\in \CH_p(Q(\la))$ then $f\in
\CH_p(Q(\la')$, and if $f_n\to f\in \CH_p(Q(\la))$, then $f_n\to f$ in
$\CH_p(Q(\la'))$.  It is sufficient to establish this for $p\ge 0$, from which
the result for $p\le 0$ follows from the duality of $\CH_p$ with
$\CH_{-p}$.  Furthermore for $p=1$, we have verified (Proposition VII.4(b))
that $\D(Q(\la))=\D(Q)$ for all $\la\in \La$, and hence that
$\CH_p(Q(\la))=\D((Q(\la)^2+I)^{1/2}) = \D(Q(\la))=\CH_p(Q)$ is independent
of $Q$.

The statement about convergence for $p=1$ is equivalent to the
existence of constants $\tm_1,\tm_2$ such that for all $\la\in \La$, 
\be
(Q^2+I) \le \tm^2_1(Q(\la)^2+I)\;,\hensp{and} Q(\la)^2+I\le
\tm^2_2(Q^2+I)\;.
\ee
The first inequality was proved in Proposition VII.2(b), while the second
follows from 
$$
Q(\la)^2 = (Q+q(\la))^2 =Q^2 + q(\la)^2 + q(\la)Q+Q q(\la)\le 2(Q^2+q(\la)^2)\;,
$$
along with assumption (VII.9).

For $0\le p\le 1$, the desired results for $\CH_p$ follow from the
inequalities
$$
(Q^2+I)^p\le \tm^p_1 (Q(\la)^2 +I)^p\;, \quad (Q(\la)^2+I)^p \le \tm^p_2
(Q^2+I)^p\;.$$
But suppose $0\le A^2\le B^2$ is a monotonicity relation for invertible
operators on a domain $\D\times \D$, where $A$ and $B$ are essentially self
adjoint on
$\D$.  Then automatically \be
A^{2p}\le B^{2p}
\ee
for all $0\le p\le 1$.  This completes the proof.

\subsection{\ Regular Deformations}
In \S VII.2 we studied regular linear deformations $Q(\la)=Q+\la q$ of $Q$,
and the resulting family $\{\CH,Q(\la),\gamma,U(g),\A\}$ of
$\Theta$-summable, fractionally differentiable structures.  In this section
we replace $\la q$ by a family $q(\la)$.  We require that $q(\la)$ satisfies
the assumptions of \S VII.2, and in addition we make as assumption  on
the derivative of $q(\la)$ with respect to $\la$.  Replacing Assumption (b) in
\S VII.2, we formulate the following:
\mbni{b$'$.} {\it A Family of Regular Perturbations.} We assume that for
each $\la\in \La$, the operator $q(\la)$ is a symmetric operator on the
domain $\D=\D(Q)$.  We assume that there are constants $0\le a<1$ and
$0\le M<\infty$ such that for all $\la$ in a compact subset of $\La$, the
inequality
\be
q(\la)^2\le a^2Q^2+M^2
\ee
holds on $\D\times \D$.  We define
\be
Q(\la)=Q+q(\la)\;.
\ee

Note that if $q$ is an operator with domain $\D$ which satisfies (VII.24) on
$\D\times \D$, then $q$ is an element of $\T(0,1)$.  Furthermore, if $q$ is
symmetric, then $q$ determines uniquely an element of $\T(-1,0)$ given by
the adjoint sesquilinear form.  Conversely, if $q$ is a symmetric
sesquilinear form on $\D_\infty\times \D_\infty$, and if furthermore
$q\in \T(0,1)\cap \T(-1,0)$, then $q$ uniquely determines a symmetric
operator on the domain $\D$.  Thus we may consider $q(\la)$ as an operator
with domain $\D$ or as an element of $\T(0,1)\cap \T(-1,0)$.

According to (VII.24),
$$
q(\la)\in \T(0,1)\cap \T(-1,0)
$$
and $q(\la)$ varies over a bounded set for $\la$ in a compact subset of
$\La$.  As in Assumption (b), we also require that $q(\la)\in
\T(-\ep,1-\ep)\cap \T(-1+\ep,\ep)$ for some interval
$0<\ep<\ep_0$.  We combine these requirements by assuming that
\be
\Vrt{q(\la)}_{(-\ep,1-\ep)} \le O(1)
\ee
for all $\ep\in [0,\ep_0]\cup [1-\ep_0,1]$ with some $\ep_0>0$. 
Furthermore the bound (VII.26) is uniform for $\la$ in a compact subset of
$\La$.  The assumption (VII.26) for all $0\le \ep\le 1$ follows
automatically from (VII.24) in case $q(\la)$ is essentially self-adjoint on
$\D$.

 It is in the latter
sense that we make an assumption about the differentiability of
$q(\la)$.  We assume that for
$\la,\la'\in \La$, the difference quotient
$$
\delta(\la,\la')={q(\la) -q(\la')\over \la-\la'}\;,$$
which is an element of $\T(0,1)\cap \T(-1,0)$, converges in both these
spaces as $\la'\to\la$.

Thus we assume that there exists a symmetric operator $\dq(\la)$ with
domain $\D$, which is the derivative of $q(\la)$ in the sense that for
$\la,\la'$ in a compact set of $\La$,
\be
\lim_{\la'\to\la} \Vrt{\delta(\la,\la')-\dq(\la)}_{(0,1)} +
\lim_{\la'\to\la} \Vrt{\delta(\la,\la')-\dq(\la)}_{(-1,0)}=0\;.
\ee
We also want $\dq(\la)$ to be continuous in $\la$, in the space $\T(0,1)$. 
Thus we suppose that
\be
\lim_{\la'\to\la} \Vrt{\dq(\la)-\dq(\la')}_{(0,1)} +\lim_{\la'\to\la}
\Vrt{\dq(\la)-\dq(\la')}_{(-1,0)} =0\;.
\ee
\medbreak

We let (b$'$) replace the assumption (b) of the previous subsection. 
We retain assumptions (a, c, d).  In the linear case of \S VII.2, $q(\la)=\la
q$, so $\delta(\la,\la')=q=\dq(\la)$.  As $q\in \T(0,1)$, the limits
(VII.27--28) hold trivially, and $(b')$ is an automatic consequence of $(b)$.

\mbni{Definition VII.7.} {\it A family $Q(\la)$ satisfying the Assmptions
{\rm (a)} of {\rm \S VII.2} and {\rm (b$'$)} above (including {\rm
(VII.24--28)}) is a regular $Q$-family.  The family
$\{\CH,Q(\la),\gamma,U(g),\A\}$ which satisfies Assumptions {\rm (a, b$'$,
c, d)} is a regular family of $\Theta$-summable,
fractionally-differentiable, non-comutative structures.}

\mbni{Proposition VII.8.} {\it Let $\{Q(\la)\}$ denote a regular
$Q$-family.  Then for $\la$ in any compact subset of $\La$, the
conclusions of Proposition VII.4 and Corollary VII.5 hold with $a<1$
replacing $\mu a<1$, and with $M$ replacing $\mu M$ in all estimates.}   
\medbreak
The proof of this proposition parallels that of the Proposition VII.4 and
Corollary VII.5.  We just replaced the bound $\la^2 q^2 \le \mu^2 a^2 Q^2
+\mu^2 M^2$ by (VII.24).  Self-adjointness of $Q(\la)$ follows as before. 
Furthermore the three inequalities all flow from the inequality
$q(\la)^2\le a^2 Q^2 +M^2$.  Thus we end up with the modified form of
(VII.13--15), where $a$ replaces $\mu a$ and where $M$ replaces $\mu
M$. Similarly we can derive the inequalities (VII.22).  Thus we also have
proved
\mbni{Proposition VII.9.} {\it Let $Q(\la)$ denote a regular $Q$-family. 
Then the Sobolev spaces $\CH_p(Q(\la))$, with $p\in [-1,1]$, and $\la$ in
any compact subset of $\La$, are independent of $\la$.}
\medbreak
We now proceed to study the   differentiability of $\exp(-sQ(\la)^2)$.  Let
us define the difference quotient of heat kernels by 
\be
\Delta(\la,\la')=(\la-\la')^{-1} \l(e^{-sQ(\la)^2} - e^{-sQ(\la')^2}\r)\;.
\ee

\mbni{Proposition VII.10.} 
{\it Let
$Q(\la)$ be a regular
$Q$-family and let
$0<s\le 1$.  Let $\la$ belong to a compact subset of $\La$.  Let 
\be
Y(\la)=\int^s_0 e^{-tQ(\la)^2} d_\la \dq(\la)e^{-(s-t)Q(\la)^2} ds\;,
\label{VII.10}
\ee
where $d_\la \dq(\la)=Q(\la) \dq(\la) + \dq(\la) Q(\la)$.  Also 
 let $X=\{I,d_\la \dq(\la)\}$ be a two-vertex set.  Then
\begin{itemize}
\ritem{(i)} $X$ is a regular set with respect to $Q$.
\ritem{(ii)} The operator $Y(\la)$ of {\rm (VII.30)} is related to the heat
kernel regularization of $X$ by
\be
Y(\la) = \int^s_0X(t,s-t)dt \label{VII.11}
\ee
\ritem{(iii)} Let $0\le \alpha,\beta$ and $0\le \alpha+\beta<1$ and
$0<s<1$.  Consider  $Y(\la)$ and
$\Delta(\la,\la')$ for
$\la,\la'$ in a compact subset of $\La$.  They  are bounded uniformly in
$\T(\alpha,-\beta;s^{-1})$, as defined in
\S {\rm V.6}.  There exists $M<\infty$ such that 
\be 
\Vrt{Y}_{\T(\alpha,-\beta;s^{-1})}
 + \Vrt{\Delta}_{\T(\alpha,-\beta;s^{-1})}
 \le    Ms^{-(\alpha+\beta)/2}   \;.\label{VII.12}
\ee 
\ritem{(iv)} The derivative of $e^{-sQ(\la)^2}$ in
$\T(\alpha,-\beta;s^{-1})$ is
$-Y$.  In fact for any $\ep>0$, 
\be
  \Vrt{\Delta+Y}_{\T(\alpha,-\beta;s^{-1})}
\le o(1) s^{-(\alpha+\beta+\ep)/2} \label{VII.13}
\ee
where $o(1)\to 0$ as $|\la-\la'|\to 0$. 
\end{itemize}
We write
\be
{d\over d\la} e^{-sQ(\la)^2} = -\int^s_0 e^{-tQ(\la)^2}d_\la \dq(\la)
e^{-(s-t)Q(\la)^2}ds\;.
\ee
 }  

\mbni{Remark.}  The fact that the derivative (VII.34) exists not just as a
limit of difference quotients in $\bh$, but also as a limit
in the space $\T(\alpha,-\beta;s^{-1})$ is crucial.  It is this fact which will
allow us to differentiate the expression for $\tau^\jlo_n(\la)$ in terms of
the expectations which define $\tau^\jlo_n(\la)$.  In other words, it
establishes the commutativity of differentiation with respect to $\la$ and
the trace and integration over $s$.
\mbni{Proof.} (i)  We assume in (b$'$) that $\dq(\la)\in \T(0,1)\cap
\T(-1,0)$.  Furthermore as explained in (b), $q(\la)\in \T(0,1)\cap
\T(-1,0)$, so also $Q(\la)\in \T(0,1)\cap \T(-1,0)$.  As a consequence, both
$\dq(\la) Q(\la)$ and $Q(\la)\dq (\la)$ and therefore $d_\la \dq(\la)$ are
elements of $\T(-1,1)$.  Hence the two vertex set has $\alpha_0 = \beta_0
=0$ and $\alpha_1 = \beta_1 = 1$, giving $\eta_0 = \eta_1=\frac12$.  Thus
according to Definition V.1,
$X$ is a regular set with respect to $Q$.  

(ii)  Note $X(t,s-t)\in I_1$ for $t,s-t>0$, and $\Vrt{X(t,s-t)}_1$ satisfies
the bound (V.38), which is integrable over $s$.  This defines $Y(\la)$ in
(VII.31) and also proves (ii).

(iii) We next estimate $\Vrt{Y(\la)}_{\T(\alpha,-\beta;s^{-1})}$.  In fact
$$
\Vrt{Y(\la)}_{\T(\alpha,-\beta;s^{-1})} \le \int^s_0
\Vrt{X(t,s-t)}_{\T(\alpha,-\beta, s^{-1})} dt\;,
$$
so it is sufficient to estimate the integrand for $t,s-t>0$.  We have,
\beq
\Vrt{X(t,s-t)}_{\T(\alpha,-\beta;s^{-1})} &=&
\Vrt{R^{-\alpha}X(t,s-t)R^{-\beta} }_{I_{s^{-1}}}\nn\\
&=& \Vrt{R^{-\alpha}e^{-tQ(\la)^2}d_\la \dq(\la)e^{-(s-t)Q(\la)^2}
R^{-\beta}}_{I_{s^{-1}}}\;.\nn
\eeq
Here $R=(Q^2+I)^{-1/2}$.  By Proposition VII.9, there is a constant $\tilde
M_3<\infty$, independent of $\la$, in a compact subset of $\La$, such that
for
$R(\la)=(Q(\la)^2+I)^{-1/2}$,
$$
\Vrt{R^{-1}R(\la)}\le \tm_3\;,
$$
and hence
\be
\Vrt{R^{-\alpha}R(\la)^\alpha} \le \tm^\alpha_3\;,\quad 0\le \alpha\le
1\;.
\ee
 Thus by H\"older's
inequality and (V.28)
\beq
  \Vrt{X(t,s-t)}_{\T(\alpha,-\beta;s^{-1})}  &\le& \tm_3^{\alpha+\beta}
\Vrt{R(\la)^{-\alpha-1}e^{-tQ(\la)^2/2}} \,
\Vrt{R(\la)^{-\beta-1}e^{-(s-t)Q(\la)^2/2}} \nn\\
&  &\times \Vrt{R(\la)d_\la \dq(\la) R(\la)} \,
\Vrt{e^{-tQ(\la)^2/2}}_{I_{t^{-1}}}
\Vrt{e^{-(s-t)Q(\la)^2/2}}_{I_{(s-t)^{-1}}}\nn\\
&  \le &4\tm_3^{\alpha+\beta} \l(\Tr \l(e^{-Q(\la)^2/2}\r)\r)^s
(t/2)^{-(\alpha+1)/2} 
  \Vrt{R(\la)d_\la \dq(\la)R(\la)}\;.\nn\\
&&
\eeq
Since $d_\la \dq(\la)\in \T(-1,1)$, it follows that
$$ 
\Vrt{R(\la)d_\la \dq(\la) R(\la)} \le \tm^2_3 \Vrt{d_\la
\dq(\la)}_{(-1,1)}\;.
$$
Furthermore, $\alpha+\beta<1$, and as we may take $\tm_3>1$,  
\beq
\hskip-.5in  \Vrt{X(t,s-t)}_{\T(\alpha,-\beta;s^{-1})}  &\le& 16M^3_3
t^{-(\alpha+1)/2} (s-t)^{-(\beta+1)/2} \nn\\
\hskip-.5in &&\enspace  \times \l( \Tr\l(e^{-Q(\la)^2/2}\r)\r)^s \Vrt{d_\la
\dq(\la)}_{(-1,1)}\;.
\eeq
Integrating over $s$ we therefore obtain
\be \Vrt{Y(\la)}_{\T(\alpha,-\beta;s^{-1})} \le \tm_4 s^{-(\alpha+\beta)/2}
\l(\Tr\l( e^{-Q(\la)^2/2}\r)\r)^s \Vrt{d_\la \dq(\la)}_{(-1,1)} 
\ee
with $M_4 = 16 M^3_3 B_1\l( {1-\alpha\over 2}, {1-\beta\over 2} \r)$.

Next we derive a similar bound on
$\Vrt{\Delta(\la,\la')}_{\T(\alpha,-\beta;s^{-1})}$.  In this case we recall
from Proposition VII.9 with $p=1$, that the domain $\D(Q(\la))$ of $Q(\la)$
is $\la$ independent for $\la$ in a compact subset of $\La$.  Thus
$Q(\la)^2$ is a sesquilinear from on $\D(Q(\la'))\times \D(Q(\la'))$.  Hence
\beq
\Delta (\la,\la') &=& (\la-\la')^{-1} e^{-tQ(\la)^2}
e^{-(s-t)Q(\la')^2}\Big\vert^{t=s}_{t=0} \nn\\
&=& (\la-\la')^{-1}\int^s_0 e^{-t Q(\la)^2} (Q(\la')^2 -Q(\la)^2)
e^{-(s-t)Q(\la')^2} dt\;.\nn\\
&&
\eeq
Note Range$(e^{-tQ(\la)^2})\subset \D(Q(\la))=\D(Q)=\D$.  We have on
$\D\times \D$ the form identity
\be
\Delta(\la,\la')=-\int^s_0 e^{-tQ(\la)^2} \l(Q(\la)\delta(\la,\la')
+\delta(\la,\la') Q(\la')\r)e^{-(s-t)Q(\la')^2}dt\;.
\ee
Since $\delta(\la,\la')\in \T(0,1)\cap \T(-1,0)$, we can repeat the proof of
the bound on $Y(\la)$ to obtain
\be  \Vrt{\Delta(\la,\la')}_{\T(\alpha,-\beta;s^{-1})}  \le  \tm_5
s^{-(\alpha+\beta)/2} \l(\Tr\l(e^{-\tm^{-2}_1 Q^2/2}\r)\r)^s   \l(
\Vrt{\delta(\la,\la')}_{\T(0,1)}+\Vrt{\delta(\la,\la')}_{\T(-1,0)}\r)\;,
\ee 
where $\tm_5=16 e\tm^4_3 B_1\l({1-\alpha\over 2},{1-\beta\over 2}\r)$. 
Since $s$ is bounded, and $\exp(-\beta Q^2)$ is trace class for all
$\beta>0$, this proves (VII.32).

(iv) Up to now we have only used the uniform bound on $\delta(\la,\la')$. 
However, in order to prove differentiability we need to establish
convergence to zero of the sum $\Delta+Y$.   
We express $\Delta+Y$ as a sum of five terms, and we then show that each
term converges to zero in $\T(\alpha,-\beta;s^{-1}$).  With the notation
\be
S(t)=e^{-tQ(\la)^2} \;,\quad S'(t)=e^{-tQ(\la')^2}\;,
\ee
use (VII.41, 44) to write
\be 
\Delta+ Y  =  -\int^s_0 S(t) Q(\la) \l(\delta S'(s-t)-\dq S(s-t)\r) dt
-\int^s_0 S(t) \l(\delta Q(\la')S'(s-t)-\dq Q(\la)S(s-t)\r) dt\;.
\ee 
We
expand this as a sum of differences
\be
\Delta +Y = -\sum^5_{j=1} \int^s_0 Z_j (t,s-t)dt\;,
\ee
with
\beq
Z_1 (t,s-t) &=& S(t)Q(\la)(\delta-\dq) S'(s-t)\;, \\
Z_2(t,s-t)&=&S(t)Q(\la)\dq (S'(s-t)-S(s-t))\;,\\
Z_3(t,s-t) &=& S(t)(\delta-\dq) Q(\la') S'(s-t)\;,\\
Z_4(t,s-t) &=& S(t)\dq \delta S'(s-t)(\la'-\la)\;,
\eeq
and
\be
Z_5 (t,s-t) = S(t)\dq Q(\la)(S'(s-t)-S(s-t))\;.
\ee
We now show that for any $\ep>0$, each of these terms satisfies
\be
\Vrt{Z_j(t,s-t)}_{\T(\alpha_1\beta;s^{-1})}\le
o(1)t^{-(1+\alpha)/2}(s-t)^{-(1+\beta +\ep)/2}\;,
\ee
where $o(1)\to 0$ as $|\la-\la'|\to 0$.  We then integrate (VII.50) over
$0<t<s$.
Since $0\le \alpha,\beta$ and $\alpha+\beta<1$, it follows that $\beta<1$. 
Therefore we can choose $\ep$ so that $(1+\beta+\ep)/2<1$.  Then the
integral converges, and we obtain
\be
\Vrt{\Delta+Y}_{\T(\alpha,-\beta;s^{-1})}\le
o(1)s^{-(\alpha+\beta+\ep)/2}\;,
\ee
as claimed.

Let us begin by proving the bound on $Z_1(t,s-t)$.  Observe that for $t>0$,
$s-t>0$, by H\"older's inequality
\beq
\Vrt{Z_1(t,s-t)}_{\T(\alpha,-\beta;s^{-1})} &=& 
\Vrt{R^{-\alpha}Z_1(t,s-t)R^{-\beta}}_{I_{s^{-1}}} \nn\\
&\le & \Vrt{R^{-\alpha}S(t)Q(\la)}_{I_{t^{-1}}} \Vrt{(\delta-\dq)R} \,
\Vrt{R^{-1}S'(s-t)R^{-\beta}}_{I_{(s-t)^{-1}}} \nn\\
&\le & \Vrt{R^{-\alpha}S(t/3)} \, \Vrt{S(t/3)}_{I_{t^{-1}}}
\Vrt{S(t/3)Q(\la)} \, \Vrt{(\delta-\dq)R} \nn\\
&&\qquad \Vrt{R^{-1}S'(s-t)/3)}\, \Vrt{S'( (s-t)/3)}_{I_{(s-t)^{-1}}}
\Vrt{S'(s-t)R^{-\beta}} \;.\nn\\
&&
\eeq
Using bounds (VII.35) and (V.28), we bound the first term on the right
\be
\Vrt{R^{-\alpha}S(t/3)} \le \tm_3^\alpha \Vrt{R(\la)^{-\alpha} S(t/3)} \le
2\tm^\alpha_3(t/3)^{-\alpha/2}\;.
\ee
Similarly the third, fifth, and seventh terms on the right are bounded.  Thus
\beq
\Vrt{Z_1(t,s-t)}_{\T(\alpha,-\beta;s^{-1})} &\le &
3^2 2^3 (3\tm_3)^{(\alpha+\beta+1)/2} t^{-(1+\alpha_/2}
(s-t)^{-(1+\beta)/2} \nn\\
&&\times \l(\ Tr\l(e^{-Q(\la)^2/3}\r)\r)^t \l(
\Tr\l(e^{-Q(\la')^2/3}\r)\r)^{s-t}\Vrt{(\delta -\dq)}_{(0,1)}\;.\nn\\
&&
\eeq
By Proposition VII.8, we have the bound (VII.15) for $\la$ and for $\la'$. 
Thus
$$
\l(\Tr\l(e^{-Q(\la)^2/3}\r)\r)^t \l(\Tr\l(e^{-Q(\la')^2/3}\r)\r)^{s-t}\le
\tm^s_4
$$
where $\tm_4$ is bounded uniformly in $\la,\la'$ in the compact subset of
$\La$.  Hence we include all constants together in one constant $\tm_5$ to
give
\be
\Vrt{Z_1(t,s-t)}_{\T(\alpha-\beta;s^{-1})} \le \tm_5
t^{-(1+\alpha)/2}(s-t)^{-(1+\beta)/2} \Vrt{\delta-\dq}_{(0,1)}\;.
\ee
The hypothesis (VII.27) ensures that $\Vrt{\delta-\dq}_{(0,1)}$ is $o(1)$ as
$|\la-\la'|\to 0$.  Thus (VII.55) is bounded by (VII.50) with $\ep=0$, and the
bound on $Z_1$ has been proved.  The bound on $Z_3$ is similar, except that
we use (VII.28) to conclude $\Vrt{\delta-\dq}_{(-1,0)}=o(1)$.

Next we consider the bound on $Z_4$.  Here we use the uniform bounds
$$
\Vrt{\dq}_{(-1,0)} \le O(1)\;,\quad \Vrt{\delta}_{(0,1)} \le O(1)
$$
to proceed as above to establish
\be
\Vrt{Z_4(t,s-t)}_{\T(\alpha-\beta;s^{-1})} \le O(|\la-\la'|)t^{-(1+\alpha)/2}
(s-t)^{-(1+\beta)/2}\;,
\ee
so the explicit coefficient $\la'-\la$ provides Lipshitz continuity of this
term.  In particular, we have established (VII.50) for $Z_1,Z_3,Z_4$, and
the bound holds for all three with $\ep=0$.

Let us now inspect $Z_2$  which requires a different method.  Let
$P_n$ denote the orthogonal projection in $\CH$ onto the subspace for
which $Q^2\le n$. Decompose $\CH= P_n\CH\oplus (I-P_n)\CH$ into two
subspaces on which $Q^2 \le n$ and $Q^2 >n$ respectively.  Also   write
\be
Z_2 (t,s-t)=Z_2(t,s-t)P_n+Z_2(t,s-t)(I-P_n)\;.
\ee
We claim that given $\ep_1>0$, we can chose $n_0$ sufficiently large so
that for $n>n_0$, we have
\be
\Vrt{Z_2(t,s-t)(I-P_n)}_{\T(\alpha-\beta;s^{-1})} \le \ep_1
t^{-(1+\alpha)/2} (s-t)^{-(1+\beta+\ep)/2}
\ee
for all $\la,\la'$ in the compact subset of $\La$.  In fact, choose $0<\ep$
sufficiently small so that $\beta+\ep<1$.  We  repeat the type of
estimate  in (VII.52)  above to obtain the bound
\beq
 \Vrt{Z_2(t,s-t)}_{\T(\alpha,-\beta-\ep;s^{-1})}  
&  \le &O(t^{-(1+\alpha)/2})
\Vrt{S'(s-t)-S(s-t)}_{\T(1,-\beta-\ep;(s-t)^{-1})}
\nn\\
& \le& O(1) t^{-(1+\alpha)/2} (s-t)^{-(1+\beta+\ep)/2}\;. 
\eeq
We use here 
$\Vrt{R^{-1}S'(s-t)R^{-\beta-\ep}}_{I_{(s-t)^{-1}}}\le O(1)$,  and
similarly 
\hbox{$\Vrt{R^{-1}S(s-t)R^{-\beta-\ep}}_{I_{(s-t)^{-1}}} \le O(1)$.}

On the other hand
\be
\Vrt{R^\ep (I-P_n)}= \Vrt{(Q^2+I)^{-\ep/2}(I-P_n)} \le (n+1)^{-\ep/2}\;.
\ee
Thus we have
\beq
\Vrt{Z_2(t,s-t)(I-P_n)}_{\T(\alpha,-\beta;s^{-1})}  &\le &
\Vrt{Z_2(t,s-t)R^{-\ep}R^\ep (I-P_n)}_{\T(\alpha,-\beta;s^{-1})} \nn\\
&\le & \Vrt{Z_2(t,s-t)}_{\T(\alpha,-\beta-\ep;s^{-1})} \Vrt{R^\ep(I-P_n)}
\nn\\
&\le & O\l( (n+1)^{-\ep/2}\r) t^{-(1+\alpha)/2}
(s-t)^{-(1+\beta+\ep)/2}\;.
\eeq
Hence for $n_0$ sufficiently large and $n\ge n_0$, we
have $O((n+1)^{-\ep/2})\le
\ep_1$, and (VII.58) holds.

We also claim that if we choose a fixed $n>n_0$, then
\be
\Vrt{Z_2(t,s-t)P_n}_{\T(\alpha,-\beta;s^{-1})} \le
o(1)t^{-(1+\alpha)/2}\;.
\ee
In fact $\exp(-Q^2)$ is trace class so $P_n\CH$ is a finite-dimensional
subspace of $\CH$.  The dimension of $P_n\CH$ is fixed once $n$ is
fixed.   Furthermore, the operator $T$ defined by
$$
T=R^{-\alpha} Z_2(t,s-t)R^{-\beta}P_n\;,
$$
yields
\be
T^\ast T=P_nR^{-\beta}Z_2(t,s-t)^\ast
R^{-2\alpha}Z_2(t,s-t)R^{-\beta}R_n\;, 
\ee
which acts on the finite-dimensional subspace $P_n\CH$.  Thus  the
absolute value
$|T| =(T^\ast T)^{1/2}$ of $T$ also acts on $P_n\CH$.  We therefore can write
\be   
\Vrt{T}_{I_{s^{-1}}} = \l(\Tr_{P_n\CH} \l(|T|^{s^{-1}}\r)\r)^s\;,
\ee
 with the trace restricted to $P_n\CH$.  But
\be
\Vrt{Z_2(t,s-t)P_n}_{\T(\alpha,-\beta;s^{-1})} = \Vrt{T}_{I_{s^{-1}}}\;,
\ee
so we can evaluate the $\T(\alpha,\beta;s^{-1})$ norm of $Z_2P_n$ on the
subspace $P_n\CH$.  

Let $f_j$, for $j=1,2\clips ,N$  be an orthonormal basis for $P_n\CH$.  We
claim that for each $j$,
\be
\lra{f,|T|^{-s^{-1}} f_j}^s \le o(1)t^{-(\alpha+1)/2}\;,
\ee
where $o(1)\to 0$ as $|\la-\la'|\to 0$.  As a consequence of (VII.66) and
of the fixed, finite dimensionality of $P_n\CH$, we infer (VII.62).

To prove (VII.66),   note that
\be
\Vrt{R^{-\beta}P_n} \le \Vrt{(Q^2+I)^{\beta/2} P_n} \le (n+1)^{\beta/2}
\;.
\ee
Also
\be
R^{-\alpha} Z_2(t,s-t) = R^{-\alpha} S(t) Q(\la)\dq (\la)
\l(S'(s-t)-S(s-t)\r)\;.
\ee
As a consequence of Proposition VII.9, we have the bounds (IV.22) for all
$\la$ in a compact subset of $\La$.  In particular, we have
$$
Q(\la)^2 \le \tm^2_2 (Q^2+I)
$$
on the subspace $P_n\CH$.  We therefore conclude that $\Vrt{Q(\la)P_n}$ is
bounded by $\tm_2(n+1)$.  In other words, if $P_m(\la)$ is the orthogonal
projection in $\CH$ onto the subspace on which $Q(\la)^2\le m$, we have
$$
P_n\CH \subset P_{\tm_2(n+1)} (\la)\CH\;,
$$
for each $\la$ in the compact subset of $\La$.  On the other hand the other
inequality (VII.22) ensures 
$$
Q^2 \le \tm^2_1(Q(\la)^2+I)\;,
$$
so 

$$
P_{\tm_2(n+1)} (\la)\CH\subset P_{\tm_1(\tm_2(n+1)+1)}\CH\;.
$$
We conclude that $(S'(s-t)-S(s-t))P_n $ has a range in $P_{n_1}\CH$, where
$n_1 = \tm_1 (\tm_2 (n+1)+1)$.  Thus
\beq
\Vrt{\dq (\la) (S'(s-t)-S(s-t))P_n } 
&=& \Vrt{\dq(\la)R
R^{-1}P_{n_1}(S'(s-t)-S(s-t) ) P_n} \nn\\
&\le & \Vrt{\dq}_{(0,1)} (n_1+1) \Vrt{(S'(s-t)-S(s-t)) P_n}\;.\nn\\
&& 
\eeq
Combining (VII.68) with the bound
\be
\Vrt{R^{-\alpha}S(t) Q(\la)}\le O(t^{-(1+\alpha)/2})\;, 
\ee
uniformly on a compact subset of $\La$, we can bound $|T|$ in norm by
\beq 
\Vrt{\, |T|\,} &=&  \Vrt{(T^\ast T)^{1/2}} \le \Vrt{T^\ast T}^{1/2} \nn\\
&\le&
O\l( t^{-(1+\alpha)/2}\r)(n_1+1)(n+1)^{\beta/2} \Vrt{\dq}^{1/2}_{(0,1)}
\Vrt{\dq}^{1/2}_{(-1,0)}
\Vrt{(S'(s-t)-S(s-t))P_n}\;. 
\eeq 
For $n\ge n_0$ fixed, the constants in (VII.71) are uniform in $\la$. 
However, $\Vrt{(S'(s-t)-S(s-t))P_n} = o(1)$ as $|\la-\la'|\to 0$, for  this
norm is calculated on a given, finite dimensional subspace of $\CH$.  Thus
$$
\Vrt{\, |T|\, } =  \l( o(1)t^{-(1+\alpha)/2}\r)\;, $$
with $o(1)\to 0$ as $|\la -\la'|\to 0$.  Likewise
$$
\Vrt{\, |T|^{s^{-1}}}\le\l( o(1) \l(t^{-(1+\alpha)/2}\r)\r)^{s^{-1}}
$$
and
$$
\lra{f,|T|^{s^{-1}}f_j}^s \le o(1)t^{-(1+\alpha)/2}\;.
$$
Hence we have proved (VII.66) and (VII.62).

We now combine (VII.58) with (VII.62) to give
\beq
\Vrt{Z_2(t,s-t)}_{\T(\alpha,-\beta;s^{-1})} &\le & 
\Vrt{Z_2(t,s-t)(I-P_n)}_{\T(\alpha,-\beta;s^{-1})} + \Vrt{Z_2(t,s-t)
P_n}_{\T(\alpha,-\beta;s^{-1})}\nn\\ &\le & o(1)t^{-(1+\alpha)/2}
(s-t)^{-(1+\beta+\ep)/2}\;.
\eeq 
Here we use $1\le (s-t)^{(1+\beta+\ep)/2}$
in the bound on $Z_2 P_n$.  Thus we have established (VII.50) in the case of
$Z_2$.

The proof of the bound (VII.50) for $Z_5$ is a minor modification on the
proof for  $Z_2$, and it also results in the bound
\be
\Vrt{Z_5(t,s-t)}_{\T(\alpha,-\beta;s^{-1})} \le o(1) t^{-(1+\alpha)/2}
(s-t)^{-(1+\beta+\ep)/2}\;.
\ee
Hence we have completed the proof of (VII.33) and of the proposition.

In  the course of establishing the proposition, we have used a method which
gives a useful bound on $\Delta$.  We state this separately,

\mbni{Proposition VII.11.} {\it Let $Q(\la)$ be a regular $Q$-family and let
$0<s\le 1$.  Let $\la,\la'$ belong to a compact subset of $\La$.  Let $0\le
\beta<1$.  Then for any $\ep>0$, 
\be 
\Vrt{e^{-sQ(\la)^2}-e^{-sQ(\la')^2}}_{\T(\beta,-1;s^{-1})}  + 
\Vrt{e^{-sQ(\la)^2} -e^{-sQ(\la')^2} }_{\T(1,-\beta;s^{-1})} \le  
o(1)s^{-(1+\beta+\ep)/2}\;,
\ee  
where $o(1)\to 0$ as $|\la-\la'|\to 0$.}

\mbni{Proof.}  We inspect the second term in (VII.74).  The bound on this
difference in the norm
$\T(1,-\beta;s^{-1})$ was proved in the course of our proof of the bound
(VII.72) on 
$Z_2(t,s-t)$.  To be explicit,
\beq
\Vrt{e^{-sQ(\la)^2} -e^{-sQ(\la')^2} }_{\T(1,-\beta;s^{-1})} &=&
\Vrt{R^{-1} \l(e^{-sQ(\la)^2} - e^{-sQ(\la')^2} \r)R^{-\beta}}_{I_{s^{-1}}}\nn\\
&\le & \Vrt{R^{-1}\l(e^{-sQ(\la)^2}-e^{-sQ(\la')^2}\r)
R^{-\beta}(I-P_n)}_{I_{s^{-1}}} \nn\\
&&+ \Vrt{R^{-1}\l(e^{-sQ(\la)^2} 
-e^{-sqQ(\la')^2}\r) R^{-\beta}P_n}_{I_{s^{-1}}}\;,
\eeq
where $P_n$ denotes the orthogonal projection onto the subspace of $\CH$
on which $Q^2\le n$.  As in the proof of (VII.72), and with the notation used
there,
\beq
\Vrt{R^{-1}(S(s) -S'(s)) R^{-\beta}(I-P_n)}_{I_{s^{-1}}} &\le &
\Vrt{R^{-1}(S(s)-S'(s)) R^{-\beta-\ep}}_{I_{s^{-1}}} \Vrt{R^\ep(I-P_n)}\nn\\
&\le & \Vrt{R^{-1}(S(s)-S'(s)) R^{-\beta-\ep}}_{I_{s^{-1}}} (n+1)^{-\ep}\nn\\
&\le & O(n+1)^{-\ep} s^{-(1+\beta+\ep)/2} \le
\ep_1s^{-(1+\beta+\ep)/2}\;.
\eeq
Here $\ep_1>0$ is given and $n$ is chosen sufficiently large, depending on
$\ep_1$.  Likewise for fixed $n$, 
\be
\Vrt{R^{-1}(S(s)-S'(s)) R^{-\beta} P_n}_{I_{s^{-1}}} \le o(1)\;,\quad
\hensp{as} |\la-\la'|\to 0\;.
\ee
This is a consequence of an analysis of $T^\ast T$, where
\be
T=R^{-1}(S(s)-S'(s))R^{-\beta}P_n\;.
\ee
We infer that 
\be
T^\ast T=P_nR^{-\beta}(S(s) -S'(s))R^{-2} (S(s)-S'(s))R^{-\beta}P_n
\ee
is bounded using (VII.67) and the argument following, for $n$ fixed, by
\be
\Vrt{\,|T|\,} \le \Vrt{T^\ast T}^{\frac12} \le O(1) \Vrt{(S(s)-S'(s))P_n}\le
o(1)\;.
\ee
Hence
\be
\l( \Tr_{P_n\CH} \l(|T|^{s^{-1}}\r)\r)^s\le o(1)
\ee
and (VII.77) holds.  But (VII.76--77) show that
$$
\Vrt{S(s)-S'(s)}_{\T(1,-\beta;s^{-1})} \le o(1) s^{-(1+\beta+\ep)/2}\;,
$$
which bounds the second term on the left of (VII.74).  The bound on the first
term of (VII.74) is just the corresponding dual bound on the adjoint of
$S(s)-S'(s)$.  Hence it follows and  the proposition is proved.

\subsection{\ The Basic Cochains $L(\la)$ and $h(\la)$}
We define the two families of cochains $L(\la)$ and $h(\la)$ as follows.  Let
\beq
 L_n(\la) (a_0\clips,a_n;g)  & = & \sum^n_{j=1} \langle a_0,d_\la
a_1\clips,d_\la a_{j-1}, [\dq(\la),a_j] , d_\la a_{j+1}\clips, d_\la
a_n;g\rangle_n \nn\\
&&   -  \sum^n_{j=0} \langle a_0,d_\la a_1\clips,d_\la a_j,d_\la
\dq(\la),d_\la a_{j+1}\clips, d_\la a_n;g\rangle_{n+1}\;,\nn\\
 & = & \sum^n_{j=1} \lra{a_0,d_\la a_1\clips, [\dq (\la),a_j]\clips,d_\la
a_n;g}_n \nn\\
&&\ - \sum^n_{j=0} \tau^\jlo_{n+1} (a_0,a_1\clips,a_j,\dq
(\la)\clips,a_n;g)\;,  
\eeq 
and let
\be h_n(\la) (a_0\clips,a_n;g) = (-1)^{[n/2]} \sum^n_{k=0} (-1)^k
\lra{a_0,d_\la a_1\clips,d_\la a_k,\dq(\la),d_\la a_{k+1}\clips,d_\la
a_n;g}_{n+1}\; . 
\ee 
The analytic properties of these cochains are a consequence of the
groundwork we have laid.

\mbni{Proposition VII.12.} {\it Let $\{\CH,Q(\la),\gamma,U(g),\A\}$ be a
regular family of $\Theta$-summable, fractionally-differentiable
structures.  Then

{\rm (i)}  The families $\{L(\la)\}$ and
$\{h(\la)\}$ are bounded families of cochains in $\C(\A)$.

{\rm (ii)} 
Furthermore $L_{2n+1}(\la)=0$ and $h_{2n}(\la)=0$.

{\rm (iii)} For each $j$, $0\le j\le n$,  define the functional 
$$
\tilde h^{(j)}_n(\la) (a_0\clips,a_n;g) = h_n(\la)
(a_0,a_1\clips,da_j\clips,a_n; g)\;.
$$
Then $\tilde h^{(j)}_n$ is  an element of $\C_n(\A)$.  

{\rm (iv)} The functional $\tilde h(\la)$ defined by 
\be
\tilde h(\la) = \l\{ \tilde h_n(\la)\r\} = \l\{ \sum^n_{j=0} (-1)^{j+1}
h^{(j)}_n(\la)\r\}   \label{VII.tilde h}
\ee
}
is an element of $\C(\A)$.

{\rm (v)} The function $g \rightarrow h(\lambda)$ is continuous in $g$ as a 
map from ${\fg }$ to $\C(\A)$, uniformly for $\lambda$ in compact subsets of 
$\Lambda$.
 
\mbni{Proof.} It is clear that for any $a_k=I$, $k\ne 0$, both $L_n(\la)$ and
$h_n(\la)$ vanish, as they must for cochains in $\C$.  Furthermore, the
expressions (VII.22) have the invariance property (II.1)  The symmetry
property under $\gamma$ ensures that $L_{2n+1}=0$ and $h_{2n}=0$, when
evaluated on $\A$.

Thus in order to ensure that $L(\la)$ and $h(\la)$ are in $\C$ we must
establish a uniform bound of the form of (II.31), as well as continuity in
$\la$ of the form (II.33).    To prove the uniform bound (II.31), we must make
sure that the expectation in each case is for a regular set of vertices.
This is true for each possibility. 

We remark that for $a\in \A$, $\dq(\la)a$ and $a\dq(\la)$ are vertices of
type (1,0) and (0,1) respectively.  Thus the sets
$$
X_n = \{a_0,d_\la a_1\clips,d_\la a_{j-1},\dq(\la) a_j , d_\la
a_{j+1}\clips,d_\la a_n\}
$$
and
$$
Y_n = \{ a_0,d_\la a_1\clips,d_\la a_{j-1},a_j\dq {(\la)},d_\la
a_{j+1}\clips , d_\la a_n\}
$$
have regularity exponents (V.31, 32) given by
$$
\eloc^{X_n} =\frac12 \min \{ 1-\alpha,1-\beta\}>0 \;,\quad \eglo^{X_n}
>1-\l( {\alpha+\beta\over 2} \r)   -{1\over 2(n+1)}
+{\alpha+\beta\over 2(n+1)} >\frac12\;,
$$
for $n$ sufficiently large, and
$$
\eloc^{Y_n}=\frac12 \min \{ 1-\alpha,1-\beta\}>0  \;,\quad
\eglo^{Y_n}>1-\l( {\alpha+\beta\over 2} \r)   -{1\over 2(n+1)}
+{\alpha+\beta\over 2(n+1)} >\frac12 \;,
$$
for $n$ sufficiently large.  Thus both $X_n$ and $Y_n$ have expectations in
$\C(\A)$, according to the estimate of Corollary V.4.iv.  Here we use
Proposition VII.6 which ensures that norms defined with $Q(\la)$ or with
$Q$ are equivalent.  The sum of $n$ such $X_n$ and $Y_n$ also defines a
cochain in $\C(\A)$, as $n^{1/n}\to 1$.

On the other hand, the second sum of (VII.82) for  $L_n(\la)$ has one vertex
$Q(\la)\dq(\la) + \dq(\la)Q(\la)=d_\la \dq(\la)$.  This vertex is of type
$(-1,1)$.  Thus again we may apply  Proposition VII.6 if we verify that
$\eloc>0$ and $\eglo>\frac12$ for the set of vertices
$$
Z=\{a_0,d_\la a_1\clips,d_\la a_j,d_\la \dq(\la),d_\la a_{j+1}\clips,d_\la
a_n\}.
$$
In this case
$$
\eloc^{Z_n}=\min \l\{ \l({1-\alpha\over 2}\r),\l({1-\beta\over 2}\r)\r\}>0
$$
and as before $\eglo^{Z_n}>\frac12$ for $n$ sufficiently large.  Thus
$L(\la)\in \C(\A)$.  For the cochain $h(\la)$ we proceed similarly.  However
$\dq(\la)\in \T(0,1)$, so proceeding as above we conclude $h(\la)\in
\C(\A)$.

Next we consider the expectation $\tilde h^{(j)}_n$.  Here one
vertex of type $d_\la^2a_j = Q(\la)d_\la a_j+d_\la a_jQ(\la)$ may
occur.  The vertex adjacent to this vertex will be either a vertex
$a_, d_\la a_k$, or $\dq(\la)$.  Such a pair of vertices will give
rise to $\eta j$'s in the range $0<(1-\alpha-\beta)/2 \le \eta j\le
\max((1-\alpha)/2, 1-\beta/2)$.  This does not affect the above
discussion, so each $\{\tilde h^{(j)}\} \subset \C(\A)$.  Likewise
$\tilde h\in \C(\A)$.  Both these bounds are uniform for $\la\in
\La$ in a compact set.

In order to establish the continuity of $h(\la)$ in $g$, we need to
analyze the difference $h(\la;g)-h(\la;g')$, for $g$, $g'$ nearby
elements of ${\fg }$.  We let $P_r$ denote the orthogonal projection
in $\CH$ onto the subspace $Q^2\le r$.  Let $\mu =
(1-\alpha-\beta)<4$, so $0<\mu<\eloc/2$ and $\mu$ is
independent of $n$.  Also let $R=(Q^2 +I)^{-1/2}$.  Then
\begin{eqnarray*}
U(g) -U(g') &=& P_r (U(g)-U(g'))\\
&&\quad + R^{-\mu} R^\mu(I-P_r)(U(g)-U(g'))\;.\end{eqnarray*}

We insert this relation in place of $U(g)-U(g')$ in each term in the
difference
$$
h_n(\la)(a_0\clips,a_n;g)-h_n(\la) (a_0\clips,a_n; g') \;.
$$
Expand into $2(n+1)$ terms using definition (VII.83).
 We will show that for $g$ sufficiently close to $g'$, the norm of
each of these terms is small, with a coefficient $o(1)$ uniform for
$\la$ in a compact and with the large-$n$ behavior
$o(1)M^{n+1}(n!)^{-\frac12 - \l( {1-\alpha-\beta\over 2}\r)}$. 
Hence $h(\la)$ is continuous in $g\in {\fg }$, uniformly on compact
sets of $\La$.

To complete the proof we argue why each term is small.  First
choose $r$ sufficiently large, which will make all the $(I-P_r)$
contributions small.  In fact we use cyclicity of the trace and the
fact that $R^\mu$ commutes with $\gamma$ to move $R^\mu$ to
the last factor in the trace, adjacent to $\exp(-s_{n+1}Q(\la)^2)$. 
Furthermore, we have established in Proposition VII.6, 9 that for
$\la$ in a compact subset of $\La$, $\Vrt{(Q(\la)^2+I)^{+\mu/2}
R^{-r}}$ uniformly bounded.  Thus the factor $\exp(-s_{n+1}
Q(\la)^2)R^{-\mu}$ can be replaced by
$\exp(-s_{\mu+1}Q(\la)^2)(Q(\la)^2+I)^\mu$ times a uniformly
bounded operator.  Since $|mu\le \eloc/2$, the factor
$(Q(\la)^2+I)^\mu$ can be absorbed into $\alpha_{n+1}$, with the
only effect being to change the combinatorial constant from one
vertex.  We are thus left with the factor
$$
\Vrt{R^\mu(I-P_r)}\le (r^2+1)^{-\mu/2}
\;,$$
which is $o(1)$ uniformly in all constants if $r$ is sufficiently
large.

Now fix $r$ and consider the first term $P_r(U(g)-U(g'))$.  Note
$Q^2$ and $U(g)$ commute, so $U(t)-U(g')$ acts on $P_r\CH$. 
While $U(g)$ is only strongly continuous on $\CH$, strong
continuity and norm continuity agrees on the finite dimensional
subspace $P_r\CH$.  Thus
$$
\Vrt{P_r(U(g)-U(g'))} \le o(1)\;,
$$
where $o(1)$ is bounded uniformly for fixed $r$, and where
$o(1)\to 0$ as $g^{-1} g'\to e$, the identity in ${\fg }$.  Thus both
terms in the expansion of $U(g)-U(g')$ give a small coeffient
times a uniform bound in $\C(\A)$, and the proof of Proposition
VII.12 is complete.

\mbni{Proposition VII.13.} {\it Assume that $Q(\la)$ is a
regular family of perturbations and $\A\in \jba$   with $0\le
\alpha,\beta$ and $0\le \alpha+\beta<1$.  Then 
\begin{itemize}
\ritem{(i)} The families of cochains $L(\la)$, $h(\la)\in \C(\A)$ are
continuous as maps from $\la\in \La\to \C(\A)$.

\ritem{(ii)} The family $\tau^\jlo (\la)\in \C(\A)$ is differentiable in
$\la$, and $d\tau^\jlo(\la)/d\la=L(\la)$.  Hence $\tau^\jlo$ is continuosly
differentiable.
\end{itemize} 
}
\medbreak
 
\mbni{Proof.}  (i) The continuity of $L(\la)$ and of $h(\la)$ in $\la$ flows
from an analysis of the differences $L(\la)-L(\la')$ and $h(\la)-h(\la')$. 
Taking $\la,\la'$ in a compact subset of $\La$, we obtain uniform estimates
\be
\vv{L_n(\la)-L_n(\la')} +\vv{h_n(\la)-h_n(\la')}\le o(1) m^{n+1}\l(
{1\over
n!}\r)^{\frac12+\ep}
\ee
for some $\ep>0$.  Here $o(1)$ is independent of $n$ and $o(1)\to 0$ as
$|\la-\la'|\to 0$.  To obtain this bound, write out the differences of $L_n$
or of $h_n$ using the definitions (VII.82--83).  

Each term in these sums has the form 
$\int \Tr\l(\gamma U(g)(X(s;\la)-X(s;\la'))\r)ds$, 
where $X(s,\la)$ is a product of $n$ or $n+1$
vertices (of the form $a_0$, $d_\la a_j$, $\dq(\la), d_\la \dq(\la)$, or
$[\dq(\la),a_j]$) and an equal number of heat kernels.  Thus each difference
$X(s;\la)-X(s;\la')$ can be expanded further as a sum of $2n$ or $w(n+1)$
terms, with exactly one difference in each.  Here this difference is either
the difference of a vertex at two values of $\la$, or else a difference of
heat kernels at two values of $\la$.

For each term, we repeat the uniform bounds as in the proof of Proposition
VII.12.  These bounds, however, can be improved through the presence of the
difference, which will ultimately give a coefficient $o(1)$.  If the
difference is a difference of heat kernels, use the bound of Proposition
IVV.11.  Suppose the difference occurs in the $j^{\rm th}$ heat kernel,
counting from zero on the left.  This yields a factor $o(1)$, as well as the
uniform bound, with the loss of an arbitrarily small power $s^{-\ep}_j$
from this one vertex.

Next let us consider terms for which the difference is a difference of a
vertex.  We need to consider each generic possibility.  If the vertex is
$d_\la a_j$ the difference has the form
\be
d_\la a_j-d_{\la'} a_j = [q(\la)-q(\la'),a_j] = (\la-\la')[\delta,a_j]\;,
\ee
where $\delta$ denotes the difference quotient for $q(\la)$.  The only
possible problem arises if $\delta$ is adjacent to a vertex containing
$\dq(\la)$.  Each term has at most two vertices with $q$ or $\dq$'s.  In this
case we use (VII.26) for $0<\ep$ or $\ep<1$.  For example, in the term
\be
\cdots (\la-\la') e^{-s_{j-1} Q(\la)^2} a_j \delta e^{-s_jQ(\la')^2}\dq (\la')
a_{j+1} e^{-s_{j+1}Q(\la')^2} \cdots\;,
\ee
we use the identity
\be
a_j\delta R^{1-\ep} = R^{-\ep}(R^\ep a_j R^{-\ep}) (R^\ep \delta R^{1-\ep})
\ee
to transfer $R^{-\ep}$ through $a_j$ and away from the heat kernel
$\exp(-s;Q(\la')^2)$ which is sandwiched between two factors of $q$ or
$\dq$.  By Proposition V.5.ii, $\Vrt{R^\ep a_jR^{-\ep}}=\Vrt{a_j}\le
\Vrt{a_j}_{\jba}$ for $0<\ep\le 1-\alpha$.  Also $\Vrt{R^\ep \delta
R^{1-\ep}}= \Vrt{\delta}_{(-\ep,1-\ep)} \le O(1)$ by (VII.26).  Thus (VII.87)
can be bounded by $O(|\la-\la'|)$, times the usual bound.  The special
treatment of the two vertices with factors of $q$ or $\dq$ does not change
$\eglo$ for $n$ large.  The $O(n^2)$ terms in $L_n$ and $h_n$ are bounded by
$m^n$ and do not affect the power of $1/n!$ in the overall estimate.  Thus
(VII.85) holds for any $\ep\le (1-\alpha-\beta)/2$, and part (i) is proved.

(ii) To establish differentiability of $\tau^\jlo(\la)$,  we write out the
difference quotient and subtract $L(\la)$.  Then
\be
{\tau^\jlo_n (\la)-\tau^\jlo_n(\la')\over \la-\la'} -L_n(\la)
\ee
can be expressed as a sum of differences, as in the proof of (i).  Two types
of terms occur in the difference quotient: the difference quotient for heat
kernel factors and the difference quotient for vertices.  These terms are in
1-1 correspondence with the terms in $L_n$; the difference quotients for
heat kernels correspond to the $d_\la \dq(\la)$ vertices in $L_n(\la)$ while
the difference quotients for vertices correspond to the $[\dq(\la),a_j]$
vertices in $L_n$.  After combining these terms, the estimates of the terms
with vertex differences proceed as in the proof of part (i).  The terms iwth
heat kernel difference quotients can be treated using the bounds of
Propositions VII.10--11.  The bounds are similar to the bounds proved
earlier, so we do not give the details.

\subsection{\ Deformations of $\tau^\jlo(\la)$ Yield  a Coboundary}
We establish that ${d\tau^\jlo(\la)\over d\la} = L(\la)=\part h(\la)$, with
$h(\la)$ defined in (VII.28).  In other words, we establish the constancy of
$\lra{\tau^\jlo(\la),a}$ under a homotopy as explained in \S VII.1.  This
completes the proof of Theorem VII.1 and of Corollary VII.2.

\mbni{Proposition VII.14.} {\it If $Q(\la)$ is a regular family of
perturbations and $\A\subset \jba$, then $\tau^\jlo(\la)\in \C(\A)$
satisfies
$$
{d\over d\la} \tau^\jlo(\la) = \part h(\la)\;. 
$$
}

\mbni{Proof.} In Proposition VII.13.ii we established that
$d\tau^\jlo(\la)/d\la$ exists and equals $L(\la)$, and furthermore that
$L(\la)$ is a continuous function from $\La$ to $\C(\A)$.  We also have
shown the existence and continuity of $h(\la)$.  So now we need only show
that $L=\part h$.

In Proposition VII.12.ii, we have shown that (evaluated on $\A$) both
$h_{2n}=0$ and $L_{2n+1}=0$.  Thus we need only verify that
\be
L_{2n}(\la) = (Bh_{2n+1})(\la)+(bh_{2n-1})(\la)\; . \label{VII.key}
\ee
We begin by proving that 
\be
(Bh_{2n+1})(\la)(a_0\clips,a_n;g)=-h_{2n} (da_0,a_1\clips,a_n;g)\;,
\label{VII.Bh}
\ee
where the right side is $\tilde h^{(0)}_{2n}$ defined in Proposition VII.12.iii,
and shown there to be an element of $\C_n(\A)$.  In fact, we establish
(\ref{VII.Bh}) starting from the definition 
$$ 
(Bh_{2n+1}) (a_0\clips,a_n;g)  
  = \sum^{2n}_{j=0} h_{2n+1}(I,a^{g^{-1}}_{2n-j+1}\clips,
a^{g^{-1}}_{2n}, a_0\clips,a_{2n-j};g)\;.
$$
We expand the right side according to the definition (VII.82) of $h$.  Note
that
$[(2n+1)/2]=n$.  Also, we use the cyclic permutation symmetry
(\ref{V.perm}) of expectations to permute $da_0$ into the zero$^{\rm th}$
position.  We obtain 
\goodbreak
\beq
&&  (Bh_{2n+1}) (a_0\clips,a_{2n};g) \nn\\
&&\hskip.5in   =  (-1)^n \sum^{2n}_{j=0} \l(
\sum^j_{k=0} (-1)^k
\l\langle I,da^{g^{-1}}_{2n-j+1}\clips,da^{g^{-1}}_{2n-j+k} ,
\dq(\la),da^{g^{-1}}_{2n-j+k+1}\clips,\r. \r.  \nn\\
&&\hskip 2.5in  \l.
da^{g^{-1}}_{2n},da_0\clips,da_{2n-j};g\r\rangle_{2n+2} \nn\\
&& \hskip.75in   + \l. \sum^{2n-j}_{k=0} (-1)^{k+j+1}
\lra{I,da^{g^{-1}}_{2n-j+1}\clips,
da^{g^{-1}}_{2n},da_0\clips,da_k,\dq(\la)\clips,da_{2n-j};g}_{2n+2} \r)
\nn\\
& &\hskip.5in  =  (-1)^n \sum^{2n}_{j=0} \l( \sum^{2n}_{k=2n-j}(-1)^{k+1}
\l\langle da_0,da_1\clips,da_{2n-j},I\clips,  da_k,\dq(\la)\clips,
da_{2n};g\r\rangle_{2n+2} \r.\nn\\
 &&
\hskip.75in \l. +
\sum^{2n-j}_{k=0}(-1)^{k+1}
\lra{da_0,da_1\clips,da_k,\dq(\la),da_{k+1}\clips,da_{2n-j},I\clips,
da_{2n};g}_{2n+2}\r)\; .\nn\\
&& \label{V.expanded}
\eeq
We now combine the terms in (\ref{V.expanded}) which have a particular
value of $k$.  For these terms, the factor $I$ occurs in the position $1,
2\clips,2n$, while the factor $\dq(\la)$ always follws $da_k$.  These
terms all have the same sign$(-1)^{n+k+1}$.  Thus using (\ref{V.sum}) we
have
\beq
  (Bh_{2n+1}) (a_0\clips,a_{2n};g) & = &
(-1)^{n+1}\sum^{2n+1}_{k=0}(-1)^k 
\langle da_0,da_1\clips,da_k,\dq(\la),  
da_{k+1}\clips,da_{2n};g\rangle_{2n+1}\nn
\\
& = & -h_{2n}(da_0,a_1\clips,a_{2n};g)\;,\nn
\eeq
as claimed in (\ref{VII.Bh}).

Next we evaluate $bh_{2n-1}$.  We claim that
\beq
  &&(bh_{2n-1}) (a_0\clips,a_{2n};g)   \nn\\
 &&\hskip.5in =  -\sum^n_{j=1} (-1)^j
h_{2n}(a_0,a_1\clips,d_\la a_j\clips,a_{2n};g)\nn\\
&& \hskip.75in  + \sum^{2n}_{j=1} \lra{a_0,d_\la a_1\clips,d_\la
a_{j-1},[\dq(\la),a_j]\clips,d_\la a_{2n};g}_{2n}\; . \label{VII.bh}
\eeq
In fact the definition of $b$ yields
\beq
 (bh_{2n-1})(a_0\clips,a_{2n};g)& = &\sum^{2n-1}_{j=0} (-1)^j h_{2n-1}
(a_0\clips,a_j a_{j+1}\clips,a_{2n};g)\nn\\
&&  + (-1)^{2n}h_{2n-1} \l(a^{g^{-1}}_{2n}
a_0,a_1\clips,a_{2n-1};g\r)\;. \label{VII.exp1}
\eeq
We expand using the definition of $h$, giving a sum of $n(2n+1)$ terms. 
We then expand further each of the terms in (\ref{VII.exp1}) arising from
$1\le j\le 2n-1$.  Each such term is a sum of expectations with one vertex
of the form $d(a_ja_{j+1})$, which we expand as
$$
d(a_ja_{j+1})=(da_j)a_{j+1}+a_j(da_{j+1})\;.
$$
We end up expressing $bh_{2n-1}$ as a sum of $8n^2$ expectations.  We
want to combine the terms in pairs, as in the proof of (VI.8).  However, the
expectations also contain one $\dq(\la)$ vertex; thus we are missing the
terms of the form
$$
\sum^{2n}_{j=1}\lra{a_0,d_\la a_1\clips,a_j\dq(\la)-\dq(\la)a_j\clips,
d_\la a_{2n} ;g}_{2n}\;.
$$
Thus we add and subtract this sum.  Hence (using $(-1)^n=-(-1)^{[(2n-1)/2]}$)
\beq
(bh_{2n-1}) (a_0\clips,a_{2n};g) &=& \sum^{2n}_{j=1} \lra{a_0,d_\la
a_1\clips, [\dq(\la),a_j]\clips,d_\la a_{2n};g}_{2n} \nn\\
&&- \sum^{2n}_{j=1} (-1)^j h_{2n} (a_0,a_1\clips,a_{j-1},d_\la
a_j,a_{j+1}\clips, a_{2n};g)\;, \nn
\eeq
as claimed.

We now add (VII.91) and (VII.93) to yield 
\beq
  (\part h)_{2n}(a_0\clips,a_{2n};g)& =& -\sum^{2n}_{j=0} (-1)^j
h_{2n}(a_0\clips,d_\la a_j\clips,a_{2n};g)\nn\\
&& + \sum^{2n}_{j=1} \lra{a_0,d_\la a_1\clips,
[\dq(\la),a_j]\clips,d_\la a_{2n};g}_{2n}\;.\nn
\eeq
Each term in the first sum 
exists as a consequence of Proposition VII.12.iii,
while each term in the second sum occurs in the sum for 
$L_{2n}(\lambda)$; hence it too is defined.
We rewrite the first sum using the identity (V.69).  Here we must
add and subtract
$$
\sum^n_{k=0}\lra{a_0,d_\la a_1\clips,d_\la a_k,d_\la \dq(\la)\clips,d_\la
a_{2n};g}_{2n+1}\;.
$$
Therefore 
\beq
 (\part h)_{2n} (a_0\clips,a_{2n};g)& =& +\sum^{2n}_{j=1}\lra{a_0,d_\la
a_1\clips,[\dq(\la),a_j]\clips,d_\la a_{2n};g}_{2n}\nn\\
&&  -
\sum^{2n}_{j=0}\tau^\jlo_{2n+1}(a_0,a_1\clips,\dq(\la)\clips,a_{2n};g)\;.\nn
\eeq
The right side of this identity is exactly $L_{2n}(\la)$.  This completes the
proof of the proposition.

\subsection{\ Equivalence of Parallel, Radon Hyperplanes}
The basic ingredient in the definition of the cocycle $\tau^\jlo$ is
$\hx(\beta)$.  This operator is the Radon transform of the heat
kernel regularization $X(s)$ of $(n+1)$ vertices, evaluated
on the hyperplane $\beta=s_0+\cdots +s_n$.  
Until now, we restricted attention to the
hyperplane $\beta=1$. In this section, we
consider a general, parallel, hyperplane defined by $\beta>0$. 
We define the corresponding $\tau^{\jlo, \beta}$ and show that
both $\tau^{\jlo,\beta}$ and $\tau^\jlo=\tau^{\jlo,1}$ belong to the 
same equivalence class.  
This justifies our restriction earlier to the case $\beta = 1$.

Let $\{\CH,Q,\gamma,U(g),\A\}$ denote a $\Theta$-summable, 
fractionally-differentiable structure.  Let 
$X=\{a_0,da_1\clips,da_n\}$ be an
$(n+1)$-vertex set, where $a_j\in \A$. Define the expectation
$\tau^{\jlo,\beta}$ with components 
\be
\tau^{\jlo,\beta}_n (a_0\clips,a_n;g)=\beta^{-n}
\Tr \l(\gamma U(g)\hx(\beta)\r)\;,
\ee
where $\hx(\beta)$ denotes the Radon transform of the heat kernel
regularization of $X$ evaluated on the hyperplane $\sum_{j=0}^{n}=\beta$.

\mbni{Proposition VII.15.} {\it The expectation $\tau^{\jlo,\beta}\in
\C(\A)$.  Furthermore $\tau^{\jlo,\beta}\in [ \tau^{\jlo,1}]$.  In 
particular, $\tau^{\jlo,\beta}$ is a cocycle for each $\beta>0$.}

\mbni{Proof.}  The statement $\tau^{\jlo,\beta}\in \C(\A)$ follows
immediately from the estimates already proved for $X^\jlo(s)$, at least in
the case $0<\beta\le 1$.  For $\beta>1$ we require minor modification of
the constants in certain bounds; we leave this to the reader.

Note that the scaling properties of the Radon transform ensure that
$\tau^{\jlo,\beta}_n$ defined for $Q$ is identical with $\tau^{\jlo,1}_n$
defined for $\beta^{1/2}Q$.  Hence $\tau^{\jlo,\beta}_n$ is a cocycle.  In
order to demonstrate that $\tau^{\jlo,\beta}$ belongs to the same
equivalence class as $\tau^\jlo$, we need only study the family
$\tau^\jlo(\beta^{1/2})$ given by $\{\CH,\beta^{1/2}Q,\gamma,U(g),\A\}$. 
As a function of $\beta^{1/2}$, we have a linear perturbation of $Q$. Define
$Q(\beta^{1/2}) = Q+q(\beta^{1/2})$, where $q(\beta^{1/2}) =
(\beta^{1/2}-1)Q$.  Since $Q\in \T(0,1)$, and it has norm 1, we infer that 
$Q(\beta^{1/2})$ is a regular $Q$-family.  Thus Theorem VII.1 shows that
$\tau^{\jlo,\beta}=\tau^\jlo +\part H$, as long as for $|\beta^{1/2}-1|<1$. 
With redefinition of the starting point of the homotopy from $\beta=1$ to
$\beta=2$, etc., we show that $\tau^{\jlo,\beta}$ is cohomologous for all
$\beta>0$.  This completes the proof.

\section{\ \ \  End Points}
We often encounter a family $\{\CH,Q(\la), \gamma,U(g),\A\}$ which is a
regular family for $\la$ in the interior of a set $\La$, but for which we
lack some relevant information  $\la$ tends to the
boundary.  In fact, this often arises in the case that 
$\La$ is an interval and
$\la$ tends to one endpoint of $\La$.  
We consider here a case of such a
phenomenon.  For simplicity, let us assume $\la\in \La=(0,1]$,
with $\la=0$ the singular endpoint.

\setcounter{equation}{0}
\subsection{\ \ End Point Regularization}
As an example, we replace the energy operator $Q(\la)^2$ by 
the regularized energy operator $H(\ep,\la)$,
\be
H(\el) = Q(\la)^2 +\ep^2 Z^\ast Z\;.
\ee
Here $\ep$ is a real, non-zero parameter and $Z^\ast Z\ge 0$ is an
operator chosen so that it regularizes $Q(\la)^2$.  This means that 
if $\beta>0$ and $\ep>0$ are fixed, then
\be
\Tr \l(e^{-\beta H(\el)} \r) 
\ee
is bounded uniformly in $\la$ as $\la\to 0$.

Suppose that $Z^\ast Z$ commutes with the representation of ${\fg }$, namely for
all $g\in {\fg }$,
$$
U(g) Z^\ast Z = Z^\ast ZU(g)\;,
$$
and also suppose that $\gamma Z^\ast Z=Z^\ast Z\gamma$.  Then
in defining the heat kernel regularizations $\hat X$ of sets of
vertices, or in defining expections, 
we can replace $\exp(-s_jQ(\la)^2)$ wiht $\exp(-s_jH(\el))$.
We obtain 
\be
\lra{\hat X;g}_n(\el) = \lra{x_0,x_1\clips, x_n;g}_n(\el)\;.
\ee
Similarly, we arrive at a regularized JLO cochain $\tau^\jlo(\el)$,
by using regularized expectations in place of expectations (IV.16) or
(V.62).  In particular
\be
\tau^\jlo_n (a_0\clips,a_n;g)(\el) 
= \Tr \l(\gamma U(g) \hat X^{\jlo}(\el)\r)
\ee
where $X^\jlo = \{a_0,d_\la a_1\clips, d_\la a_n\}$.  In 
general, for $X=\{x_0\clips,x_n\}$, one can  define
\be
\hat X (\el) = \int x_0 e^{-s_0 H(\el)} x_1 e^{-s_1 H(\el)} \cdots
x_ne^{-s_nH(\el)} \delta (s_0 +\cdots +s_{n-1}) ds_0\cdots
ds_n\;.
\ee
Given $a\in \A^{\fg }$ and $a^2=I$,
the regularized cochain $\tau^\jlo(\el) = \{\tau^\jlo_n (\el)\}$
yields the regularized pairing defined as 
\be
\frak Z^{H(\el)} (a;g) = {1\over \sqrt\pi} \int e^{-t^2} 
\Tr\l(\gamma U(g) ae^{-H(\el)+itd_\la a}\r)dt\;.
\ee
This pairing 
$\fz^{H(\el)}(a;g)$ converges as $\ep\to 0$
to $\fz^{Q(\la)}(a;g)$ of (I.28).  
However, $\tau^\jlo(\el)$ is not a cocycle.  Nor is the
pairing (VIII.6) an invariant function of $\la$ or of $\ep$.  
For $\ep\ne 0$, the pairing
$\fz^{H(\el)}(a;g)$ depends on both $\ep$ and $\la$.  

\subsection{\ \ Exchange of Limits}
In certain examples, we have studied the dependence of
$\fz^{H(\el)}$ on $\ep$ and $\la$ in detail, at least in a
neighborhood of $\ep=\la=0$, see \cite{J2}.  In these examples we have
shown that $\fz^{Q(\la)}$ can be recovered from $\fz^{H(\ep,0)}$.  
Also we choose the regularizing factor $Z^\ast Z$ to be sufficiently simple 
so that we can evaluate $\fz^{H(\ep,0)}$ in closed form.  On the other hand,
we are interested in knowing $\fz^{Q(\la)} = \fz^{H(0,\la)}$ for
$\la>0$, where it is constant.  The important fact is that 
the function $\fz^{H(\el)}$ is {\it not} jointly continuous in 
$(\el)$ in the unit square $0\leq\ep\leq 1$, $0\leq\la\leq 1$ 
at the point $(0,0)$.

However, another fact saves the day; it is our ability to
prove that while $\fz^{H(\el)}(a,g)$ is not jointly continuous in 
$(\ep, \la)$ in the unit square and for every value of $g\in {\fg }$, 
this function {\it is} jointly continuous 
in $(\ep, \la)$ for {\it almost every} value of $g\in {\fg }$.
We establish this continuity in the examples, with the aid of a new 
expansion that we name the {\it holonomy expansion}.  As a consequence of
the expansion, we obtain bounds on
\be
{d\over d\ep} \fz^{H(\el)} \hensp{and} {d\over d\la} \fz^{H(\el)}
\ee
of the form
\be
\l| {d\over d\ep} \fz^{H(\el)} \r| \le M \ep \;,\quad \l| {d\over d\la}
\fz^{H(\el)} \r| \le M\ep^2\;,
\ee
for $0\le \el$, $0<\ep+\la$, and sufficiently close to $(\el)=(0,0)$. 
Thus we obtain,
\be
\l| \fz^{H(\ep,0)} - \fz^{H(0,\la)} \r| \le O(\ep^2)\;,
\ee
as $\ep\to 0$.  We combine this information with two other facts:
(i) in the examples, the explicit form of $\fz^{H(\ep,0)}$ has a
pointwise limit as $\ep\to 0$ for almost all $g\in {\fg }$.  (ii) In
Corollary VII.2, we established the  {\it a priori} 
continuity of $\fz^{H(0,\la)}$ as a function of $g\in {\fg }$.  These two
pieces of information ensure that for almost all $g\in {\fg }$,
\beq
\fz^{Q(\la)} (a,g)  = \lim_{\la\to 0} \lim_{\ep\to 0} \fz^{H(\el)}
(a,g) &=& \lim_{\ep\to 0} \lim_{\la\to 0}\fz^{H(\el)} \nn\\
&=& \lim_{\ep\to 0}\fz^{Q(0)^2+\ep^2 Z^\ast Z} (a,g)\;.
\eeq
The right side of (VIII.10) can be evaluated, and extends by
continuity to all $g\in {\fg }$.

\setcounter{equation}{0}
\section{\ \ Split Structures}
We define  {\it splitting} of $Q$ as a decomposition 
\be
Q={1\over \sqrt 2} (Q_1+Q_2)\;,
\ee
such that also 
$$
Q^2 = \frac12 (Q_1^2 +Q^2_2)\;.
$$
A splitting is associated with corresponding derivatives on $\A$ given by
\be
da = {1\over \sqrt 2} (d_1a+d_2a)\;, \quad d_j a=[Q_j,a]\;.
\ee
In \S IX.1 we specify this assumption in  more detail.  Clearly a splitting
into a sum of many parts is possible, but we concentrate here on a slitting
in two.

As in earlier sections, we assume
\be
\Tr \l(e^{-\beta Q^2}\r)<\infty \;,\quad \hensp{for} \beta>0\;.
\ee
However, we do not assume that $\Tr \l(e^{-\beta Q_j^2}\r)$
exists for the individual $Q_1$ or $Q_2$.  
In addition, while we assume that the group
$U(g)$ of unitary symmetries commutes with $Q_1$, we do not assume that
$Q$ or $Q_2$ are necessarily invariant.
For these reasons, the resulting 
framework will be different from the equivariant framework studied earlier. 

Within this revised setting we generalize the cochain $\tau^\jlo$ to a
cochain $\tau^{\{Q_j\}}$ defined on a suitable algebra $\A$. 
Letting $\A^{\fg }$ denote the pointwise-${\fg }$-invariant part of $\A$,
we obtain for $a\in \A^{\fg }$ and $a^2=I$ the following expression for a pairing:
\be
\fz^{\{Q_j\}} (a,g) = \frac1{\sqrt\pi} \int^\infty_{-\infty} e^{-t^2}
\Tr \l(\gamma U(g)ae^{-Q^2 +itd_1a}\r)dt\;.
\ee
While this formula bears a close resemblence to (VI.4), the
resulting pairing $\fz^{\{Q_j(\la)\}}$ in general is not an invariant.
If $Q_j(\la)$ depends on a parameter $\la$, then $\fz^{\{Q_j\}}(a;g)$ remains
a function of $\la$.  However, we describe a
special family of variations $Q_j(\la)$ and algebra $\A$, such that
(IX.4) {\it is} an invariant.  Within this class of variations, we find that,
as previously, there is a cochain $h(\la)$ such that 
\be
{d\over d\la} \tau^{\{ Q_j(\la)\}}=\part h^{\{Q_j(\la)\}}\;.
\ee

\subsection{\ \ A $Q_1$-Invariant Splitting}
We split the self-adjoint operator $Q$ into a sum of self-adjoint
operators $Q_1$ and $Q_2$, as in (IX.1).  Each operator $Q_1$, $Q_2$,
and $Q$ is odd under $\gamma$, 
\be
Q_j \gamma + \gamma Q_j = 0\;,\quad j=1,2\;.
\ee
Furthermore, the decomposition (IX.3) of $Q^2$ has the
interpretation that $Q_1$ and $Q_2$ generate {\it independent}
translations.  Algebraically,
\be
Q_1Q_2 +Q_2Q_1=0\;.
\ee
Recall that the $Q$ is assumed to be essentially self-adjoint on the domain 
$\D$; such a domain is called a core for a symmetric operator.
The core is invariant if $Q\D\subset \D$.  
If $\D$ is a common, invariant core
for $Q_1$ and $Q_2$, then products are defined on $\D$ and on this domain
\be 
[Q_1,Q^2_2]=0=[Q_2,Q^2_1]\;.
\ee

\mbni{DefinitionIX.1.} {\it The self-adjoint operator $Q$ splits into
the sum of two independent parts, if there is a common, invariant
core for $Q_1$, $Q_2$, and $Q$, such that the bounded functions of
$Q_1$ and $Q_2$ commute with the bounded functions of $Q^2_1$
nd of $Q^2_2$.}
\medbreak

Part of the definition that $Q$ splits into a sum of two independent parts, 
involves the assumption that the spectral projections of $Q_1$ commute with 
those of $Q^2_2$.  Hence the unbounded, self-adjoint operators $Q_1$ and 
$Q^2_2$ commute.  

Complementing $H$ defined in
(IX.3), we define the self-adjoint operator\footnote{The notation
$H$ and $P$ is suggestive of energy and momentum.  In fact, 
this is no accident, as such
examples arise in space-time supersymmetry. In that case, $Q_1$ and
$Q_2$ are generators of symmetries arising from two
space-time directions, and $Q^2_1 = H+P$ and
$Q^2_2=H-P$, with $H$ and $P$ being the energy and momentum
operators.  Combined with the equation for independence of
$Q_1$ and $Q_2$, this is called the algebra for space-time supersymmetry (in
a two-dimensional space-time).  The condition (IX.9--10),
supplemented by $0\le H=Q^2$, can be interpreted as a
restriction of special relativity for the energy-momentum to lie
in (or on) the positive cone.}
\be
P = \frac12 \l( Q^2_1 -Q^2_2\r)\;.
\ee
Note that $P$ and $H$ commute.  Furthermore as a consequence of (IX.3,9),
we infer
\be
\pm P\le H\;,
\ee
and the joint spectrum of $H$ and $P$ lies in a cone.  

Let us define the Sobolev spaces
$\CH_\alpha = \D((H+I)^{\alpha/2})$, which are Hilbert spaces with
inner product defining a norm $\Vrt{f}_{\CH_\alpha} =
\Vrt{(H+K)^{\alpha/2}f}$.  The space $\T(\beta,\alpha)$  of
bounded linear transformations from $\CH_\alpha$ to
$\CH_\beta$ is a Banach space with norm $\Vrt{T}_{(\beta,\alpha)}
= \Vrt{(H+I)^{\beta/2}T(H+I)^{-\alpha/2}}$.  Thus (IX.9) can be
interpreted as saying $P\in \T(-1,1)$ with
\be
\Vrt{P}_{(-1,1)} \le 1\;.
\ee

In the equivariant case, we also are interested in the unitary
group of symmetries $U(g)$ acting on $\CH$.  We assume as in
earlier sections that
\be
\gamma U(g) = U(g)\gamma \hensp{and} Q^2 U(g) = U(g) Q^2 
\ee
for all $g\in {\fg }$. 
\mbni{Definition IX.2.} {\it A splitting (IX.1) is $Q_1$-invariant, if
(IX.9) holds and also for all $g\in {\fg }$,
\be
Q_1 U(g) = U(g)Q_1\;.
\ee
}
\medbreak

Note that a $Q_1$-invariant splitting has the property
\be
Q^2_2 U(g) = U(g) Q^2_2\;.
\ee
Let us define two parameter abelian representation
\be
V(t,x)=e^{itH+ixP}\;.
\ee
The relation (IX.6) ensures that for all $(t,x)\in \R^2$,
\be
\gamma V(t,x) = V(t,x)\gamma\;.
\ee

\mbni{Definition IX.3.} {\it The representation $V(t,x)$ is
$U(g)$-{\rm invariant} if $V(t,x)U(g)=U(g)V(t,x)$ for all $g\in {\fg }$
and all
$(x,t)\in \R^2$.}

\subsection{\ \ Observables}
We define a new algebra $\A$ of observables, suitable for a
$Q_1$-invariant splitting of $Q$.  First we define an interpolation
space $\T^{(1)}_{\beta,\alpha}$ based on the $d_1$ derivative.  In
particular, let
\be
\T^{(1)}_{\beta,\alpha} = \{ b:b\in \B(\CH),\hensp{and}
R^{\beta}_1(d_1b) R^\alpha_1\in \B(\CH)\}\;,
\ee
where $R_1=(Q^2_1+I)^{-1/2}$.  We give $\T^{(1)}_{\beta,\alpha}$
the norm
\be
\Vrt{b}_{\T^{(1)}_{\beta,\alpha}}=\Vrt{b}+c_{\alpha+\beta}
\Vrt{R^\beta_1(d_1b) R^\alpha_1 }\;.
\ee
Here $c_{\alpha+\beta}$ is defined in (V.77). Clearly
$\T^{(1)}_{\beta,\alpha}$ is invariant under the action of ${\fg }$
defined by conjugation with $U(g)$, as a consequence of
(IX.11--12).
Define the (spatial) translate $b(x)$ of an element $b\in \B(\CH)$
by
\be
b(x) = e^{ixP} be^{-ixP} = V(0,x) b V(0,x)^\ast \;.
\ee

\mbni{Definition IX.4.}  {\it The zero-momentum subalgebra
$\B(\CH)_0$ of $\B(\CH)$ consists of all elements $b\in \B(\CH)$
such that $b(x)=b$ for all $x\in\R$.}

We remark that $b\in \B(\CH)$ is an element of $\B(\CH)_0$ if and
only if
\be
Pb=bP\;.
\ee
We assume that $\A$ is a Banach-subalgebra of the interpolation
space
\be \B(\CH)_0 \cap \T^{(1)}_{\beta,\alpha} \;,
\ee
which we call the zero-momentum subalgebra of
$\T^{(1)}_{\beta,\alpha}$.  Here $\alpha,\beta$ satisfy $0\le
\alpha,\beta$, with $\alpha+\beta<1$.  
Assume the pointwise, $\gamma$-invariance of $\A$, namely
$a=a^\gamma$ for $a\in \A$.  Furthermore assume
\be
\Vrt{a}_{\T^{(1)}_{\beta,\alpha}} \le \vv{a}\;.
\ee
Also let $\A^{\fg }\subset \A$ denote the pointwise-${\fg }$-invariant
subalgebra of $\A$.
\medbreak
The analysis of the interpolation spaces $\B(\CH)_0\cap
\T^{(10}_{\beta,\alpha}$ follows step by step the analysis of the
interpolation spaces $\T_{\beta,\alpha}$ in \S V.  In order to carry
this out, we note the following:
\mbni{Lemma IX.4.} {\it Let $b\in \B(\CH)_0$.  Then as a bilinear
form on $\D(Q^2)\times \D(Q^2)\subset \D(Q^2_1)\times
\D(Q^2_1)$,
\be
d^2 b=d^2_1 b\; . 
\ee
}

The proof of this lemma is an elementary consequence of $[P,b] =
0$.  Thus estimates on $d^2 b$ can be reduced to estimates on
$d^2_1 b$.  Also
\be
R=(Q^2 +I)^{-1/2} = (Q^2_1 +Q^2_2 + I)^{-1/2} \le \l(Q^2_1
+I\r)^{-1/2} = R_1\;.
\ee
As a result, we can reduce all estimates in the proof of this
statement to estimates of same form as
those of \S V.  In particular, $\B(\CH)_0\cap
\T^{(1)}_{\beta,\alpha}$ is a Banach algebra, and we can prove
analogs of Proposition V.5 (with $R_1$ replacing $R$ and $d_1$
replacing $d$), as well as Corollary V.16, Proposition V.7, and
Corollary V.8.
\subsection{\ \ The Cochain $\tau^{\{Q_i\}}$ and Invariants}
Let $Q$ in (IX.1) denote a $Q_1$-invariant, splitting of $Q$.  Let
$$
\A\subset \B(\CH)_0 \cap \T^{(1)}_{\beta,\alpha}\;, \quad 0\le
\alpha,\beta\;,\quad \alpha+\beta<1\;,$$
denote the zero-momentum algebra of observables, with
fractional $Q_1$-derivatives, as in \S IX.2.  Let
\be
\{ \CH,Q,Q_i,\gamma,U(g),\A\}
\ee
denote a $Q_1$-invariant, split, fractionally-differentiable
structure, generalizing Definition VI.1 to the $Q_1$-invariant,
split case.  Here we also assume the hypothesis of $\Theta$-summability
for $\exp (-\beta Q^2)$, $\beta>0$.

Define a cochain $\tau{\{Q_i\}}$ on $\A$ with components,
\be
tau^{\{Q_i\}}_n (a_0\clips, a_n) =
\lra{a_0,d_1a_1\clips,d_1a_n;g}_n\;,
\ee
where the $(n+1)$-linear expectation $\lra{\;\;\clips,\;\;}_n$ in (IX.26) is
defined in (IV.16).  The results of \S IX.2 allow us to establish the
following Propositions.

\mbni{Proposition IX.5.} {\it {\rm a.} Let
$\{\CH,Q,Q_i,\gamma,U(g),\A\}$ be as in (IX.25), and let $\tau{\{Q_i\}}$
be as in (IX.26).  Then $\tau^{\{Q_i\}}\in \C(\A)$.  There exists 
a constant $m<\infty$ such that
\be
\vv{\tau^{\{Q_i\} }_n}  \le m^{n+1} \l({1\over n!} \r)^{\frac12 +
\l({1-\alpha-\beta\over 2} \r)} \Tr \l(e^{-Q^2/2}\r)\;.
\ee

{\rm b.} Let $a\in \Mat_n(\A^{\fg })$ satisfy $a^2=I$.  Then
\be
\fz^{\{Q_i\}}(a;g) = \lra{\tau^{\{Q_i\}},a} = 
\frac1{\sqrt\pi} \int^\infty_{-\infty}
e^{-t^2} \Tr \l(\gamma U(g)ae^{-Q^2 +itd_1a}\r) dt
\ee
is the natural pairing of cochains in $\C(\A)$ with elements 
$a\in \Mat_n(\A^{\fg })$ satisfying $a^2=I$.  Here $\Tr$ denotes both the
trace in $\CH$ and the trace in $\Mat_n(\A)$.

{\rm c.} The cochain $\tau^{\{Q_i\}}$ is a cocycle,
\be
\part \tau^{\{Q_i\}} = 0\;.
\ee
}
\medbreak

The proof of Proposition IX.5 again follows the proofs in \S VI. 
We can also generalize the results of \S VII--VIII.  In order to
understand the parameter dependence of $\tau^{\{Q_i(\la)\}}$ on
a parameter $\la$, we consider in particular the following common
case, where the parameter $\la$ is called a {\it coupling constant}:
\mbni{Definition IX.6.} {\it We say that the splitting 
$Q(\la)=\frac1{\sqrt 2} (Q_1(\la)+Q_2(\la)$ depends parametrically on a 
coupling constant $\la$, if $Q(\la)$ depends on $\la$, but
$P=\frac12 (Q_1(\la)^2-Q_2(\la)^2)$ is independent of $\la$.}
\medbreak
As far as the analytic bounds are concerned, for a regular, linear
deformation we retain the assumptions formulated in \S VII.2.b. 
We also suppose that, as in \S VII.2.c--d, the symmetry group
$U(g)$ and the algebra $\A$ are both independent of $\la$. 
Alternatively, in the case of a regular deformation, we retain
Definition VII.7 of \S VII.3.  We can then extend Theorem VII.1 and
its corollary to the case of $Q_1$-invariant, split structures,
following our preceeding work.  We summarize this result:
\mbni{Theorem IX.7.} a. {\it Let $\{\CH,Q(\la),
Q_i(\la),\gamma,U(g),\A\}$ be a regular family (in the sense
explained above) of $\Theta$-summable, split, $Q$-invariant,
fractionally-differentiable structures, depending on a 
coupling constant $\la$ in the sense of Definition VII.6.  Then the
family of cocycles $\{\tau^{\{Q_i(\la)\}}\}\subset \C(\A)$ is
continuously differentiable in $\la$ as a function $\La \to \C(\A)$. 
There is a continuous family of cochains $h(\la)$ in $\C(\A)$ such
that for all $\la\in \La$,
$$
{d\over d\la} \tau^{\{Q_i(\la)\}} = \part h(\la)\;.
$$
Furthermore the pairing of $\tau^{\{Q_i(\la)\}}$ with $a\in
\Mat_n\{\A^{\fg }\}$ satisfying $a^2=I$ and given by (IX.27), namely
\be
\fz{\{Q_i\}}(a,g) = \lra{\tau^{\{Q_i(\la)\}} ,a}\;,
\ee
is independent of $\la\in \La$.

{\rm b.} Suppose that instead of the $\la$-dependence in $Q_i(\la)$ 
arises from a coupling constant, we substitute the condition that
$Q_1(\la)$ commutes with $a$, and that $Q_2(\la)$ is independent of 
$\la$.  Then $Z(a,g)$ is $\la$-independent for $\la\in \La$.}

\subsection{\ \ An Example of a Split Structure}
We mention here an example which we analyze elsewhere by
these methods \cite{J3}.  The splitting arises from space-time
supersymmetry, as suggested in footnote 6 or whatever.  In our
example we assume the existence of $N=2$ supersymmetry.  This
means that there exist four, pairwise, mutually-independent
self-adjoint operators $Q_1,Q_2$, $\tilde Q_1$, and $\tilde Q_2$
such that for a given $H$ and $P$,
\be
Q^2_1 = \tilde Q_1^2 = H+P\;, \quad Q^2_2 = \tilde Q^2_2 = H-P\;.
\ee
We take for the group symmetry ${\fg }$ the group $U(1)\times U(1)$
constructed in the following manner:  one factor $U(1)$ has the
form $e^{i\theta J}$, where the generator $J$ commutes with
$\gamma$, with $H$, and with $P$.  Furthermore $J$ commutes with
$Q_1$ and $\tilde Q_1$.  On the other hand,
\be
Q_2(\theta) = e^{i\theta J} Q_2 e^{-i\theta J} = Q_2 \cos \theta
+\tilde Q_2 \sin \theta\;.
\ee
Thus
$$ Q_2(\theta)^2 = H-P = Q^2_2
$$
and $J$ commutes with $Q_2(\theta)^2$.  The other $U(1)$ has the
form  $e^{itP}$; this is clearly a symmetry of both $Q_1$ and
$Q_2$, as well as with $\gamma$.  We take
\be
U(\tau,\theta) = e^{i\tau P + i\theta J}\;.
\ee

In this example, we study \cite{J3} the equivariant index 
$\fz^{\{Q_i\}}(I; g)$, namely
\be
\fz(\tau-i, \theta) = \Tr \l(\gamma U(\tau,\theta) e^{-Q^2}\r)\;,
\ee
where $g=(\tau, \theta)$.  This index was considered in the physics 
literature \cite{W2, KYY}. We can justify its evaluation in a certain class 
of non-linear quantum field theories (Wess--Zumino field theory 
examples arising form polynomial,
quasi-homogeneous superpotentials, which also satisfy 
an elliptic estimate) as a product of modular
forms \cite{J2, J3}.  The modular symmetry in $(\tau,\theta)$ is a 
hidden symmetry of the original example.

\end{document}